\documentclass[a4paper,11pt]{article}

\usepackage{diagbox}
\usepackage{amssymb,amsmath,amsfonts}
\usepackage{float}
\usepackage{relsize}
\usepackage{graphicx}
\usepackage{tikz}

\usetikzlibrary{matrix,shapes, arrows, shadows}

\tikzstyle{vertex} = [circle, draw, fill=blue!20, scale=1,auto=left]
\tikzstyle{vert} = [circle, draw, fill=blue!20, scale=.8,auto=left]
\tikzstyle{line} = [draw]

\newcommand{\midarrow}{\tikz \draw[-triangle 90] (0,0) -- +(.1,0);}

\usepackage{upgreek}

\def\be{\begin{equation}}
\def\ee{\end{equation}}
\def\bea{\begin{eqnarray}}
\def\eea{\end{eqnarray}}

\begin{document}

\begin{titlepage}
\date{\today}       \hfill

\begin{center}

\vskip .2in
{\LARGE \bf   Properties of RG interfaces for 2D boundary flows }\\
\vspace{5mm}

\today
 
\vskip .250in

\vskip .3in
{\large Anatoly Konechny}

\vskip 0.5cm
{\it Department of Mathematics,  Heriot-Watt University\\
Edinburgh EH14 4AS, United Kingdom\\[10pt]
and \\[10pt]
Maxwell Institute for Mathematical Sciences\\
Edinburgh, United Kingdom\\[10pt]
}
E-mail: A.Konechny@hw.ac.uk
\end{center}

\vskip .5in
\begin{abstract} \large
We consider RG interfaces for boundary RG flows in two-dimensional QFTs. Such interfaces are 
particular boundary condition changing operators linking the UV and IR conformal boundary conditions. We refer to them as RG operators. 
In this paper we study their general properties putting forward     a number of conjectures. 
We conjecture that an RG operator is  always a conformal primary such that the OPE of this operator with its conjugate must contain 
the perturbing UV operator when taken in one order and the leading irrelevant operator (when it exists) along which the flow enters the IR fixed point, when taken in the other order. We support our conjectures by perturbative calculations for flows between nearby fixed points, by a non-perturbative variational method inspired by the variational method proposed by J.~Cardy for massive RG flows, and by numerical results 
obtained using boundary TCSA. The variational method has a merit of its own as it can be used as a first approximation in charting the global structure of the space of boundary RG flows. We also  discuss the role of the RG operators in the transport of states and  local operators. Some of our considerations can be generalised to two-dimensional bulk flows, clarifying some conceptual issues related to  the RG interface put forward by D.~Gaiotto for bulk $\phi_{1,3}$ flows.

\end{abstract}

\end{titlepage}

\renewcommand{\thepage}{\arabic{page}}
\setcounter{page}{1}
\large 
\noindent\hrulefill
\tableofcontents 
\noindent\hrulefill

\section{Introduction }
\renewcommand{\theequation}{\arabic{section}.\arabic{equation}}
Since the dawn of quantum field theory (QFT) two-dimensional models provided us with numerous exact non-perturbative 
results describing various phenomena observed in nature and also providing us with a theoretical laboratory to gain 
insights about QFTs in general. An important ingredient in this success is the infinite-dimensional conformal symmetry 
which allowed practitioners  to solve exactly various  families of conformal  field theories (CFTs),  like minimal models, that describe 
renormalisation group (RG) fixed points. In Wilson's approach to QFT one describes a generic renormalisable QFT as a 
perturbation of a fixed point QFT. Having  such a good control over fixed points one wants next to describe RG flows originating 
from them. This was pursued from the early days of CFT and a variety of non-perturbative results were obtained for flowing 2D models, mostly for integrable perturbations. A generic perturbation of a 2D CFT is not integrable so such methods have a very limited range of applicability. Nevertheless, besides the results for integrable models,  we also have a number of general results constraining RG flows  such as  the celebrated c-theorem of Zamolodchikov \cite{c}.   While it may be hopeless to expect being able to solve a generic non-integrable 2D QFT one can aspire to describe the space of all flows originating from a fixed UV CFT in terms of its topology and geometry. Some ideas to this extent, which use the c-theorem and Morse theory, were put forward in \cite{Vafa}. Morse theory and 
other topological tools were considered in the context of RG flows more recently in \cite{Gukov1}, \cite{Gukov2}. 
 Besides trying to use  known topological methods in describing spaces of RG flows  it seems to be important to us to develop new intrinsic quantum field theoretical tools which would facilitate the description of such spaces. This paper focusses on RG interfaces 
 -- objects which in our opinion can be developed into exactly such a tool. 
 
 An RG interface, in any  dimension, is a codimension one object  on one side of which we have a UV fixed point QFT and on the other 
 we have the IR  fixed point QFT that describes the end point of a particular RG flow. If the flow is massive the IR theory is trivial and then the interface corresponds to a boundary condition in the UV theory. To construct such an interface we consider perturbing the UV theory by a given relevant operator that generates the flow, on a half space. The perturbation is then renormalised by including the usual bulk counterterms away from the interface (the hyperplane that separates the perturbed region from the unperturbed one) and possibly by additional counterterms localised on the interface. The renormalised theory is then allowed to flow to the far infrared. In that region 
 all energy scales are sent to infinity and we should end up with a scale-invariant interface separating the UV and IR fixed point QFTs. 
 In two-dimensions the scale invariance is enhanced to the conformal symmetry and therefore the RG interface is a conformal (one-dimensional) interface. The RG interfaces were introduced and  investigated by I. Brunner and D. Roggenkamp in \cite{BR}  (the main idea was also spelled out in \cite{FQ}). We find this quite an appealing feature of RG interfaces in 2D that they must respect the infinite dimensional conformal symmetry while at the same time they must carry information about the RG flow that produced them. 
 From now on we will focus on flows in 2D QFTs. 
 
 If we put an RG defect on a unit circle while inserting at the origin an operator of the UV CFT and at infinity an operator of
 the IR theory we obtain a pairing (a number) assigned to these operators. It was conjectured by D. Gaiotto in \cite{Gaiotto} that this pairing is related to  RG mixing coefficients. In \cite{Gaiotto} a conformal interface between neighbouring A-series minimal models was constructed that 
 was conjectured to be the RG interface for the RG flow triggered by the $\phi_{1,3}$ primary that was shown to link the two CFTs 
 \cite{Zamolodchikov_pert}. The unitary A-series minimal models ${\cal M}_{m}$ have the central charge 
 \be
 c_{m} = 1 - \frac{6}{m(m+1)} \, , \quad m\in {\mathbb Z} \, , \enspace m\ge 3 \, .
 \ee
 The perturbation by $\phi_{1,3}$ primary results in a flow: ${\cal M}_{m} \to {\cal M}_{m-1}$. 
In \cite{Gaiotto} the RG pairings were calculated for certain fields and matching  to the known RG mixing coefficients was demonstrated in the leading order in the $m\to \infty$ limit.  These calculations were later extended and checked at  higher order \cite{Poghossian} and for descendant operators \cite{PP}. We find that the original idea of relating the pairing at hand to RG mixing coefficients is quite interesting yet puzzling in certain aspects. Thus, while we expect only operators of nearby scaling dimensions to mix 
 in the $m\to \infty$ limit, Gaiotto's interface provides non-vanishing pairing of operators of arbitrary difference in  conformal weights already in the next to leading order.  We will provide some  comments on this  in sections \ref{sec_op_map} and  \ref{pairing_sec} that hopefully  clarify the situation. 
 
 Despite the presence of Virasoro symmetry constructing conformal interfaces is quite hard in general and we have very few examples of them even for such  well studied CFTs as  Virasoro minimal models.     The difficulty stems from the fact that a conformal interface by virtue of the folding trick \cite{AO} is equivalent to a conformal boundary condition in a tensor product of the two theories (one conjugated). And the latter very rarely happens to be a rational CFT.  
 
 There are two particular cases of RG flows where the situation is better. For massive RG flows the interface can be described as a conformal boundary condition in the UV CFT, and if the latter is rational then we know all maximal symmetry preserving conformal boundary conditions. One can thus study the mapping between all massive RG flows and conformal boundary conditions originating from 
 a given rational CFT. 
 The mapping between massive RG flows and conformal boundary conditions was determined for  the critical Ising theory in \cite{AK_Ising} using the numerical TCSA  method. In \cite{Cardy_var} such a mapping was put forward for all minimal models on a basis of a certain variational method. We will review this method in section \ref{sec:VAR} where we develop a similar method for boundary flows.
 
 The second situation for which we often know all possible RG  interfaces is the case of boundary RG flows. 
 We consider a boundary CFT (BCFT for brevity) that is a 2D CFT on the upper  half plane equipped with a conformal boundary condition. 
 If we have a boundary relevant operator $\psi(\tau)$ we can perturb the boundary condition by this operator  and generate an RG flow 
 in the space of boundary conditions. Such flows always end up at a different BCFT in the far infrared. Similarly to the bulk case 
 we can perturb the boundary only on a half-line. Renormalising this theory results in having a point-like interface or, equivalently, a boundary condition  changing operator that links the unperturbed conformal boundary condition to the perturbed flowing one. 
 We denote such an operator as  $\hat \psi^{[\lambda, 0]}$ where the unperturbed boundary condition is located to the left 
 of the operator and the perturbed one specified by the coupling $\lambda$ is located to the right\footnote{We use the ordering conventions of \cite{Runkel} for boundary operators and OPE coefficients. Note that they are different from those accepted in \cite{Lewellen}, that is a common reference on BCFTs.} (see Fig. \ref{bcco}). We can also perturb on a half line extending to negative infinity
 with the corresponding operator denoted as $\hat \psi^{[0, \lambda]}$. We will refer to these operators as interface operators.
 
 \begin{center}
\begin{figure}[H]
\centering
\begin{tikzpicture}[scale=1.4]
\filldraw[fill=gray!30!white,draw=white] (-3,0) rectangle (3,3);
\draw[blue,very thick,dashed](-3,0)--(0,0);
\draw[red,very thick] (0,0)--(3,0);
\draw (0,0) node {$\bullet$} ;
\draw (0.1,-0.3) node {$\hat \psi^{[\lambda, 0]}$};
\draw (0,1.5) node {{\small Bulk CFT}};
\draw (-2,-0.3) node {{\small Unperturbed b.c.}};
\draw (2.1, -0.3) node {{\small Perturbed b.c}} ;
\end{tikzpicture}
\caption{The boundary condition changing operator linking perturbed and unperturbed boundary conditions}
\label{bcco}
\end{figure}
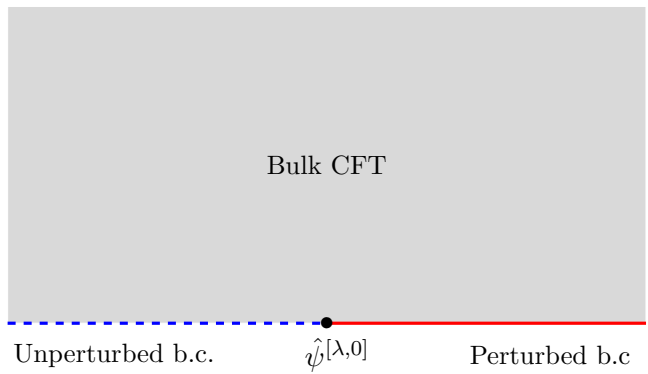
\end{center}
Allowing the boundary condition to flow to the far infrared we obtain an RG interface or RG operator
$\hat \psi^{[{\rm ir}, {\rm uv}]} $ linking the UV conformal boundary condition (on the left) to the IR conformal boundary condition (on the right). If $\hat \psi^{[0, \lambda]}$ does not mix with other operators this construction  gives a unique operator $\hat \psi^{[{\rm ir}, {\rm uv}]} $ up to normalisation. As $\hat \psi^{[0, \lambda]}$ in the UV has dimension zero it will not have any other operators to mix with if we start with an irreducible boundary condition for which  the identity operator is the only operator of dimension zero. If we start with a superposition of elementary boundary conditions and the perturbation breaks the ground state degeneracy then operators of dimension near zero may emerge and $\hat \psi^{[0, \lambda]}$ can mix with them. Such Chan-Paton symmetry breaking boundary flows were considered in \cite{Graham} and \cite{FK} (see appendix A of  \cite{FK} in particular).   It would be interesting to investigate the RG operators in this case, but in this paper we will only consider the flows in which   $\hat \psi^{[0, \lambda]}$ only mixes with itself and therefore $\hat \psi^{[{\rm ir}, {\rm uv}]} $ is uniquely defined by the flow.

For maximal  symmetry preserving conformal boundary conditions in rational CFTs not only we know how to construct  all such boundary conditions, we also know all boundary-condition changing operators linking them \cite{Cardy}, \cite{BPPZ}, \cite{BPPZ2}. On the other hand the space of all boundary RG flows is quite rich even for such well studied  theories like the tricritical Ising model \cite{Affleck}.   
This suggests  an interesting problem of mapping all boundary RG flows in terms of  their RG operators. For boundary flows 
triggered by $\psi_{1,3}$ operators  in Cardy boundary conditions of the A-series minimal models the RG operators were 
proposed in \cite{RG_boundary1}. They were conjectured to be certain primary operators (or rather linear combinations of primaries between irreducible components) and the conjecture was checked  via perturbative computations  comparing the RG mixing coefficients with RG pairings similarly to \cite{Gaiotto}. Another example of an RG operator was worked out in \cite{AKB} for the boundary magnetic field flow in the critical Ising model. As that flow is solvable via a Bogolyubov transformation the RG operator was determined exactly 
and is given up to normalisation by the boundary condition changing primary of weight $1/16$.

Besides a purely descriptive potential, i.e. breaking the space of all RG flows into domains labelled by RG operators, one would hope to use RG interfaces as a tool to constrain the space of flows. Such constraints could look like superselection rules which would tell you that a flow with a given starting point BCFT and a given perturbation which triggered it cannot end up with certain BCFTs in the IR because there would be no suitable RG operators between the two theories. It is precisely this type of rules that we put forward in this paper. 
We make the following conjectures about the RG operators $\hat \psi^{[{\rm ir}, {\rm uv}]} $.
\vspace{0.5cm}

\noindent {\bf Conjecture 1.} Any RG operator $\hat \psi^{[{\rm ir}, {\rm uv}]} $ is always a conformal primary. 
\vspace{0.8cm}

\noindent {\bf Conjecture 2.}  For a flow triggered by a relevant primary field $\psi^{\rm uv}$ the OPE of  the corresponding RG operator 
$\hat \psi^{[{\rm ir}, {\rm uv}]} $ 
with its conjugate must contain $\psi^{\rm uv}$, that is 
\be \label{OPE1}
\hat \psi^{[{\rm uv}, {\rm ir}]}(\tau)  \hat \psi^{[{\rm ir}, {\rm uv}]} (0) = C_{\rm uv} \tau^{\Delta_{\rm uv} - 2\hat \Delta} \psi^{\rm uv}(0) + \dots 
\ee
where $\tau>0$, $C_{\rm uv}\ne 0$, $\hat \Delta$ is the dimension of the RG field, and $\Delta_{\rm uv}$ is the dimension of $\psi^{\rm uv}$. (The omitted terms in the above OPE may include singular terms.) For RG flows triggered by a linear combination of different primaries we propose that at least one of the  primary operators present in the perturbation must appear in the OPE in (\ref{OPE1}).
\vspace{0.8cm}

\noindent {\bf Conjecture 3.} If a boundary RG flow with an RG operator $\hat \psi^{[{\rm ir}, {\rm uv}]} $ arrives at the IR BCFT 
via a leading irrelevant operator $\psi^{\rm ir}$ it must appear in the OPE of $\hat \psi^{[{\rm ir}, {\rm uv}]} $ with its conjugate, that is 
\be \label{OPE2}
\hat \psi^{[{\rm ir}, {\rm uv}]}(\tau)  \hat \psi^{[{\rm uv}, {\rm ir}]} (0) = C_{\rm ir} \tau^{\Delta_{\rm ir} - 2\hat \Delta} \psi^{\rm ir}(0) + \dots 
\ee
where $C_{\rm ir}\ne0$.
\vspace{0.5cm}

In the main body of the  paper we provide arguments in support of these conjectures, both perturbative and non-perturbative. 
Perturbation theory is applied to boundary flows between two nearby fixed points. 
Among the non-perturbative arguments we use a new variational method which we mentioned above. Besides supporting our 3 conjectures the variational method goes beyond that  
 in allowing one to differentiate candidate fixed points for which there are candidate RG operators each satisfying the conditions stated in the conjectures. 
   In the main body of the paper we start out in sections \ref{section_states} and  \ref{sec_op_map} by discussing the generalities of mappings of states and operators by means of the RG interfaces (often one refers to these as  transport of states and operators). 
   We keep the discussion there fairly general switching between bulk and boundary perturbations. In section \ref{nearby_fp} 
   we present explicit perturbative calculations related to RG operators for a boundary flow between nearby fixed points. Our 
   calculations there are model independent, valid for any nearby fixed points. To the best of our knowledge such a general analysis has not been done before. 
   In section \ref{sec:VAR}  we develop a new variational method for boundary flows based on  RG operators.  In 
   section \ref{sec:TCSA} we present some numerical TCSA results in which we check numerical data against component ratios in the 
   vacuum vector and the first excited state as predicted by the RG operators. In section \ref{sec:conclusions} we offer some concluding remarks pointing at a number of open questions.

 \section{Mappings  of states} \label{section_states}
 \setcounter{equation}{0}
 
 Here we consider some formal aspects of RG operators starting with mappings of states between the perturbed and unperturbed theories. 
 To that end we  consider operator quantisation in which the Euclidean time runs parallel to the boundary of the upper half plane (to the right). The constant time slices are thus semi-infinite intervals. (Later we will also consider quantisation on a strip with some fixed boundary condition on one end and the perturbation appearing on the other end.)
 If we have an insertion of $\hat \psi ^{[\lambda, 0]}(0)$ then to the left of the insertion we have the state space of the UV BCFT 
  which we denote as ${\cal H}^{0}$ and to the right of the insertion we have the state space of the perturbed theory: ${\cal H}^{\lambda}$. Similarly for an insertion of $\hat \psi ^{[ 0,\lambda]}(0)$  we have the two spaces swapped. 
  We can thus consider the matrix elements of the form $\langle u| \hat \psi ^{[\lambda, 0]}(0) |v\rangle$ and 
 $\langle v| \hat \psi ^{[0,\lambda]}(0) |u\rangle$ where $|u\rangle \in {\cal H}^{\lambda}$, $|v\rangle \in {\cal H}^{0}$. 
  These pairings define linear mappings 
  \be \label{maps}
  \hat \psi^{[0,\lambda]}: {\cal H}^{\lambda} \to ({\cal H}^{0})^{'} \, , \quad  \hat \psi^{[\lambda,0]}: {\cal H}^{0} \to ({\cal H}^{\lambda})^{'} 
  \ee
  where ${\cal H}'$ stands for the anti-dual space. 
  If our theory is invariant under the reflections $\tau\to -\tau$ preserving the boundary then the two mappings are related to each other by conjugating by the reflection  charge. If the images of mappings (\ref{maps}) belong to the state spaces themselves (that are  naturally identified as  subspaces of the anti-dual spaces) then we have a mapping (or transport) of states between the two theories. In this case one may wonder if there are kernels in these mappings, for example not every UV state may have an image inside the deformed theory state space. Alternatively it may also happen that the element of the anti-dual space has infinite norm. 
   For example this happens when we consider an improper Bogolyubov transformation\footnote{In QFT Haag's theorem proves that for interacting theories in Poincare-invariant space-time this is inevitable. The proof however relies on the invariance under spatial translations which is not applicable when our spatial slice is a half-line.} for which the transformed vacuum, although has finite overlaps with all states in the original Fock space, itself has an infinite norm (see \cite{Berezin} for a detailed discussion). In line with that 
   terminology we will  say that (\ref{maps}) defines proper mappings between state spaces when the images lie in the corresponding state spaces and say that we have an improper mapping when we merely get an element in the anti-dual space. 
   
We can distinguish between the two cases using the OPE of $\hat \psi$ with its conjugate. 
  Let $|0\rangle$ and $|0\rangle_{\lambda}$ denote the vacua in the UV BCFT and the perturbed theory respectively. 
  We can represent the norm squared of the image of perturbed  vacuum in the unperturbed state space as a limit of two-point 
  function
\be\label{norm}
 \|\hat \psi^{[0,\lambda]} |0\rangle_{\lambda}\|^2 = \lim_{\epsilon \to 0} \langle \hat \psi^{[\lambda,0]}(-\frac{\epsilon}{2}, 0) 
\hat \psi^{[0,\lambda]}(\frac{\epsilon}{2}, 0) \rangle_{\lambda} 
\ee
that is illustrated  on  Figure  \ref{Fig_norm}. 
\begin{center}
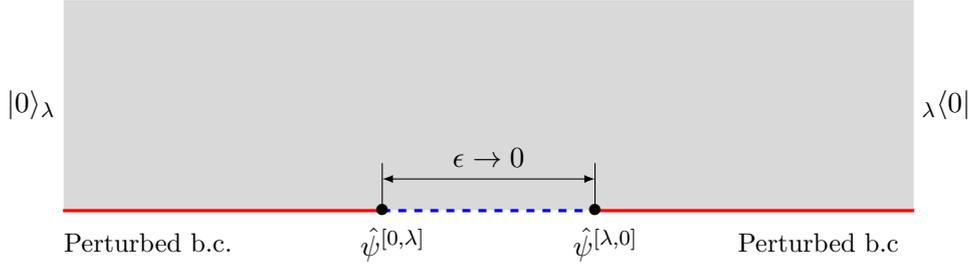
\begin{figure}[H]
\centering
\begin{tikzpicture}[>=latex,scale=1.4]
\filldraw[fill=gray!30!white,draw=white] (-4,0) rectangle (4,2);

\draw[blue,very thick,dashed](-1,0)--(1,0);
\draw[red,very thick] (-4,0)--(-1,0);
\draw[red,very thick] (1,0)--(4,0);
\draw (-1,0) node {$\bullet$} ;
\draw (1,0) node {$\bullet$} ;
\draw (4.3,1) node {${}_{\lambda}\langle0|$};
\draw (-4.3,1) node {$|0\rangle{}_{\lambda}$};
\draw (1.1,-0.32) node {$\hat \psi^{[\lambda, 0]}$};
\draw (-0.9,-0.32) node {$\hat \psi^{[0,\lambda]}$};
\draw (-3.2,-0.3) node {{\small Perturbed b.c.}};
\draw (3.1, -0.3) node {{\small Perturbed b.c}} ;
\draw (-1,0) -- (-1, 0.45);
\draw (1,0)--(1,0.45);
\draw[<->] (-1, 0.3) -- (1,0.3);
\draw (0,0.5) node {$\epsilon \to 0$};
\end{tikzpicture}
\caption{Norm squared of the perturbed theory vacuum represented as a short-distance limit of a two-point function of the RG operator}
\label{Fig_norm}
\end{figure}
\end{center}
Similarly we can swap the operators $\hat \psi^{[\lambda, 0]}$, $\hat \psi^{[0,\lambda]}$ and obtain a representation of 
the norm squared of $|0\rangle$ as represented in the perturbed theory state space. If $\lambda$ is not at the value of the infrared fixed point the short distance behaviour is governed by  the UV fixed point. We can find the short distance asymptotic behaviour from 
the Callan-Symanzik equation which has the form 
\be \label{CS1}
\left( \epsilon\frac{\partial}{\partial \epsilon} + \beta(\lambda) \frac{\partial}{\partial \lambda} + 2\hat \gamma \right)
 \langle \hat \psi^{[\lambda,0]}(-\frac{\epsilon}{2}, 0) 
\hat \psi^{[0,\lambda]}(\frac{\epsilon}{2}, 0) \rangle_{\lambda} =0
\ee
where $\beta=\beta(\lambda)$ is the beta function and $\hat \gamma=\hat \gamma(\lambda)$ is the anomalous dimension function for 
the interface operators.
 A general solution to (\ref{CS1}) can be written as 
\be \label{sd_int}
 \langle \hat \psi^{[\lambda,0]}(-\frac{\epsilon}{2}, 0) 
\hat \psi^{[0,\lambda]}(\frac{\epsilon}{2}, 0) \rangle_{\lambda}=\exp\left( \int\limits_{\bar \lambda}^{\lambda}\frac{2\hat \gamma}{\beta}d\lambda' \right) \langle \hat \psi^{[\lambda,0]}(-\frac{\mu^{-1}}{2}, 0) 
\hat \psi^{[0,\lambda]}(\frac{\mu^{-1}}{2}, 0) \rangle_{\bar \lambda}
\ee
where $\mu$ is the RG scale, $\bar \lambda=\bar \lambda(\epsilon)$ is the value of the running coupling constant at distance $\epsilon$ and 
$\lambda=\bar \lambda(\mu^{-1})$. For strictly relevant or asymptotically free perturbations the effective coupling goes to zero at short distances and thus the asymptotic behaviour  of the integral in (\ref{sd_int}) is determined by the region near $\lambda=0$. 
In section \ref{section_RGop} we calculate $\hat \gamma$ at the leading order in perturbation theory.Using the result (see formula (\ref{hat_gamma})) it is easy to check  that both for the relevant case with $\beta(\lambda)\approx -y\lambda$ and 
the asymptotically free case with $\beta(\lambda) \approx - D\lambda^2$, $D>0$ 
 the integral in (\ref{sd_int}) approaches a finite value. This indicates  that whenever the UV behaviour is governed by a UV fixed point boundary perturbations always give rise to proper mappings between states. We hope to be able to strengthen this argument in  future work 
 investigating the convergence of perturbation theory using methods similar to \cite{CF}.

The situation changes if  $\lambda$ is at the infrared fixed point and we have a pairing between states in the UV and IR  BCFTs. 
In this case we expect $\hat \psi$ and its conjugate to be a scaling operator (or at any rate to be a linear combination of scaling operators) and the two point function is then always singular at short distances. The mappings (\ref{maps}) between the UV and IR BCFT states are thus always improper. 

Next we would like to formally demonstrate  that the mappings (\ref{maps}) give the expected representation of the energy eigenstates.  For the sake of the general discussion we could stay on the plane but the picture is particularly nice (and more familiar) on the strip when we have a discrete state space. We start with a BCFT 
on a strip of width $L$. 
Let $w=\tau+i\sigma$, $\tau \in {\mathbb R}$, $\sigma\in [0,L]$ be  the complex coordinate on the strip, then it is related to the 
coordinate on the upper half plane $z$ by  the standard conformal mapping $w=\frac{L}{\pi}{\rm Log}(z)$. 
Suppose that we have the same UV BCFT boundary condition on both ends.
The Hamiltonian 
for the $\tau$-translation on the strip $H_{0}$ is then related to the scaling generator via 
\be
H_{0} = \frac{\pi}{L}( L_{0} - \frac{c}{24})  \, .
\ee
Perturbing the strip on the lower end: $\sigma=0$ according to (\ref{bare_ps}), (\ref{DeltaS}) results in having a perturbed (bare)
Hamiltonian 
\be \label{perturbed_bare_H}
H_{\lambda}^{0} = H_{0} - \lambda_{0} \psi_{0}(0) 
\ee
where $\psi$ is the operator on the strip. 
The renormalised Hamiltonian $H_{\lambda}$ generates translations of renormalised operators in the perturbed theory. 
Consider now a single insertion of the renormalised interface operator: $\hat \psi^{[0,\lambda]}(\tau)$. Its derivative can be written  as 
\be \label{id_1}
\partial_{\tau} \hat \psi^{[0,\lambda]}(\tau) = H_{0}\hat \psi^{[0,\lambda]}(\tau)-\hat \psi^{[0,\lambda]}(\tau)H_{\lambda} 
\ee
that holds for any boundary-condition changing operator. 
Let $|E_{i}\rangle_{0}\in {\cal H}^{0}$ be the eigenstates of $H_{0}$ with eigenvalues $E_{i}$ and 
$|{\cal E}_{I}\rangle_{\lambda}\in {\cal H}^{\lambda}$ be the eigenstates of $H_{\lambda}$ with eigenvalues ${\cal E}_{I}$.
Applying  identity  (\ref{id_1}) taken at $\tau=0$ to an eigenvector $|{\cal E}_{I}\rangle_{\lambda}$ we can rewrite it as 
\be \label{id_2}
H_{0} \hat \psi^{[0,\lambda]}(0)|{\cal E}_{I}\rangle_{\lambda} - \partial_{\tau} \hat \psi^{[0,\lambda]}(0)|{\cal E}_{I}\rangle_{\lambda}   =
 {\cal E}_{I} \hat \psi^{[0,\lambda]}(0)|{\cal E}_{I}\rangle_{\lambda}  
\ee
On the other hand if we differentiate the perturbative series defining the insertion of $\hat \psi^{[0,\lambda ]}(\tau)$
we formally obtain 
\be
\frac{\partial}{\partial \tau} {\rm T}{\rm exp}\left(\lambda \int\limits_{-\infty}^{\tau}\psi(t)dt\right) = 
\lambda \psi(\tau) {\rm T}{\rm exp}\left(\lambda \int\limits_{-\infty}^{\tau}\psi(t)dt\right) \, .
\ee
We can give this formula a precise meaning if we assume that we started with a regularised expression in which all insertions 
were separated and then removed the regularisation and added counterterms both for the collisions happening away from $t=\tau$ 
renormalising the perturbing operator and coupling, and for the collisions happening at $t=\tau$ that define a new composite 
boundary-condition changing operator. This procedure gives rise to  
\be
\partial_{\tau} \hat \psi^{[\lambda, 0]}(0) = \lim_{\epsilon \to\, + \,0} \Bigl[ \lambda \psi(\epsilon) \hat \psi^{[\lambda, 0]}(0)  
+ \mbox{Counterterms}\Bigr] \, .
\ee 
Substituting this into (\ref{id_2}) we obtain 
\be \label{id_3}
\lim_{\epsilon \to\, + \,0}  \Bigl[H_{0} -   \lambda \psi(\epsilon) - \mbox{Counterterms}  \Bigr] \hat \psi^{[0,\lambda]}(0) |{\cal E}_{I}\rangle_{\lambda} =  {\cal E}_{I} \hat \psi^{[0,\lambda]}(0)|{\cal E}_{I}\rangle_{\lambda}    \, .
\ee
Comparing with (\ref{perturbed_bare_H}) the left hand side here can be interpreted as the action of the renormalised perturbed Hamiltonian operator 
  $H_{\lambda}$ written as an operator in the unperturbed theory space ${\cal H}^{0}$. Perturbative calculations show that 
  for strongly relevant operators with $\Delta>1/2$ the ground state energy divergence is the same as the power divergence coming from the OPE of $\psi_{0}$ with 
  $\hat \psi_{0}$ and, being subtracted, precisely gives the perturbative ground energy shift in the left hand side of (\ref{id_3}).

  To summarise, we see that when renormalisation effects are properly taken into account the operator $\hat \psi^{[0,\lambda]}(0)$ 
  maps the perturbed theory energy eigenstates into the eigenstates of the perturbed Hamiltonian acting in the unperturbed state space. 
  This property is rather formal and does not give a recipe for constructing these states in ${\cal H}^{0}$. To obtain an explicit  expression 
 for the vacuum of the perturbed theory  we can start from the  two point function on a strip
  \be 
   \langle X|\hat \psi^{[0,\lambda]}(0)\hat \psi^{[\lambda,0]}(-T)|Y\rangle
   \ee
   where $|X\rangle, |Y\rangle$ are energy eigenstates from ${\cal H}^{0}$.
   Taking the limit $T\to \infty$ and inserting a complete set of energy eigenstates $|E^{\lambda}_{i}\rangle\in {\cal H}^{\lambda}$ between the two interface operators we obtain 
   \be
    \langle X|\hat \psi^{[0,\lambda]}(0)\hat \psi^{[\lambda,0]}(-T)|Y\rangle \sim \sum_{i} e^{-T(E_{i}^{\lambda}-E_{Y})} 
    \langle X|\hat \psi^{[0,\lambda]}(0)|E_{i}^{\lambda}\rangle \langle E_{i}^{\lambda}| \psi^{[\lambda,0]}(0)|Y\rangle 
   \ee
    Comparing this with perturbation theory representation of this correlation function  we see that, assuming that 
    $\langle E_{0}^{\lambda}| \psi^{[\lambda,0]}(0)|Y\rangle \ne 0$ 
    the vacuum of the perturbed theory as a state in ${\cal H}^{0}$ can be expressed as 
  \be
  \hat \psi^{[0,\lambda]}(0)|0\rangle_{\lambda} = \frac{1}{\langle E_{0}^{\lambda}| \psi^{[\lambda,0]}(0)|Y\rangle } \lim_{T\to \infty} e^{T(E_{0}^{\lambda}-E_{0})} 
  {\rm T}{\rm exp}\left(\lambda\int\limits_{-T}^{0}\psi(t)dt \right) |Y\rangle \, .
  \ee
 Here the normalisation of this state is fixed by the normalisation of $ \hat \psi^{[0,\lambda]}$ and it does not automatically have the unit norm. As we are not expecting the orthogonality catastrophe (which happens due to divergent norm of the perturbed vacuum state), at least for small $\lambda$ we can take $|Y\rangle$ to be the unperturbed vacuum $|0\rangle$.  At the infrared fixed point the situation with normalisation changes, as we discussed above we expect the perturbed vacuum to have infinite norm. Fixing the normalisation of 
 $ \hat \psi^{[0,\lambda]}$ to be such that this field has a finite two-point function at the fixed point gives a representation of the vacuum state as an element in the anti-dual state space $({\cal H}^{0})'$.
 
  Another approach to constructing the perturbed theory  energy eigenstates is by using the Gell-Mann-Low formula based on the adiabatic switching  of the interaction. Under some assumptions this approach can be also applied to excited energy eigenstates. 
 At the level of perturbation theory such constructions, that use the interface operator, should be of course equivalent to the usual Rayleigh-Schrodinger perturbation theory for the Hamiltonian eigenvalues and eigenvectors.

 \section{Mappings  of operators} \label{sec_op_map}
 \setcounter{equation}{0}
 
 Besides the mappings of states between the two theories discussed in the previous section the interface operator also enters into 
 the mappings between local operators. Formally speaking (ignoring the divergences and assuming that the mappings are proper) once we have the mappings between states:
\be
 \hat \psi^{[0,\lambda]}: {\cal H}^{\lambda} \to {\cal H}^{0} \, , \quad  \hat \psi^{[\lambda,0]}: {\cal H}^{0} \to {\cal H}^{\lambda}  
 \ee
 we can can sandwich any operator acting in the unperturbed theory $A: {\cal H}^{0}\to {\cal H}^{0}$ by the interface operator to obtain an operator in the perturbed theory
 \be
 \hat \psi^{[\lambda,0]}  A\hat \psi^{[0,\lambda]}:  {\cal H}^{\lambda}   \to  {\cal H}^{\lambda}  \, .
 \ee
  However for local operators this composition is in general singular and the naive composition should be replaced by the following procedure. Consider a local operator $\psi_{a}^{0}(\tau)$ in the UV BCFT.  
  To construct from it an operator in the perturbed theory we first surround it by the interface operators inserted at $\tau-\epsilon$ and $\tau+\epsilon$ as depicted on Figure \ref{Fig_operator}. We then take the limit $\epsilon \to 0$ and subtract divergences by adding counterterms. 
  
  \begin{center}
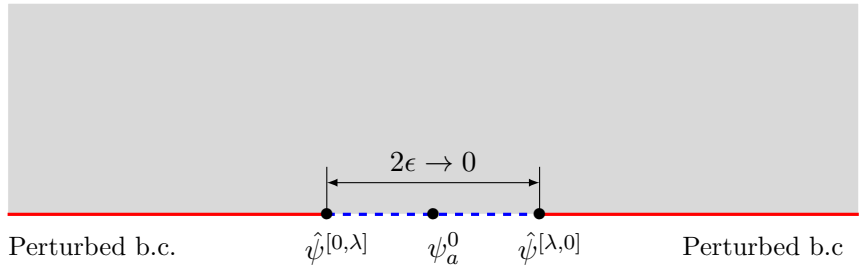
\begin{figure}[H]
\centering
\begin{tikzpicture}[>=latex,scale=1.4]
\filldraw[fill=gray!30!white,draw=white] (-4,0) rectangle (4,2);

\draw[blue,very thick,dashed](-1,0)--(1,0);
\draw[red,very thick] (-4,0)--(-1,0);
\draw[red,very thick] (1,0)--(4,0);
\draw (-1,0) node {$\bullet$} ;
\draw (1,0) node {$\bullet$} ;
\draw (0,0) node {$\bullet$};
\draw (1.1,-0.32) node {$\hat \psi^{[\lambda, 0]}$};
\draw (-0.9,-0.32) node {$\hat \psi^{[0,\lambda]}$};
\draw (0.1,-0.32) node {$\psi_{a}^{0}$};
\draw (-3.2,-0.3) node {{\small Perturbed b.c.}};
\draw (3.1, -0.3) node {{\small Perturbed b.c}} ;
\draw (-1,0) -- (-1, 0.45);
\draw (1,0)--(1,0.45);
\draw[<->] (-1, 0.3) -- (1,0.3);
\draw (0,0.5) node {$2\epsilon\to 0$};
\end{tikzpicture}
\caption{A renormalised local operator in the perturbed theory can be obtained by surrounding a local operator in the UV theory by two interface operators, taking the limit when all 3 operators are at the same point, and subtracting divergences.}
\label{Fig_operator}
\end{figure}
\end{center}
We can write then 
\be\label{psi_lambda}
[\psi_{a}]^{\lambda}(\tau) = \lim\limits_{\epsilon\to 0} \left[ \hat \psi^{[\lambda,0]}(\tau + \epsilon) \psi_{a}^{0}(\tau)\hat \psi^{[0,\lambda]}(\tau-\epsilon)  + \mbox{ Counterterms } \right]
\ee
where $[\psi_{a}]^{\lambda}$ denotes a renormalised operator in the deformed theory. While the key ingredient in this construction is the seed UV operator $\psi_{a}^{0}$ we also have some arbitrariness in how we choose the counterterms. 

Similarly, we can consider bulk perturbations\footnote{In this section we consider in parallel  the two cases of bulk and boundary perturbations as conceptually the issues involved are very similar and also because we want to make connection to prior work \cite{Gaiotto}.} in which the RG interface is one-dimensional. To construct a local operator in the perturbed 
theory we surround an operator $\phi_{i}^{\rm UV}$ in the UV theory  by the deformation interface putting it on a circle of radius $\epsilon$. The resulting object can be expanded in terms of renormalised operators. If we already have some basis of renormalised operators $\tilde \phi_{i}^{\lambda}$ available we can write this expansion as 
\be \label{bulk_exp}
{\cal D}^{\lambda}_{\epsilon}\phi_{i}^{\rm UV}(0)= \sum_{j} C_{i}^{j}(\epsilon, \lambda) \tilde \phi^{\lambda}_{j}(0)\, .
\ee
Shrinking the circle and subtracting divergences we obtain a local operator in the perturbed theory (see Figure \ref{Fig_operator_bulk}). 

  \begin{center}
\begin{figure}[H]
\centering
\begin{tikzpicture}[>=latex,scale=1.4]
\filldraw[fill=red!10!white,draw=white] (-4,0) rectangle (4,4);
\filldraw[blue!10!white,draw=black, thick] (0,2) circle (0.7);
\draw (0,2) node {$\bullet$};
\draw (0.1,2.3) node {$\phi_{i}^{\rm UV}$};
\draw (2.8,3.5) node {{\small perturbed theory}};
\draw (0,2) --(0.7,2);
\draw (0.33,1.88) node {$\epsilon$};
\draw[->] (-1,1) to[out=0,in=-90]  (-0.3,1.8);
\draw (-2.2,1) node {{\small unperturbed theory }};
\draw (-2.4,0.75) node {{\small inside the circle}};
\end{tikzpicture}
\caption{A renormalised local bulk operator in the perturbed bulk theory can be obtained by surrounding a local operator in the UV theory by the deformation interface put on a circle of radius $\epsilon$ which is sent to zero and subtracting divergences.}
\label{Fig_operator_bulk}
\end{figure}
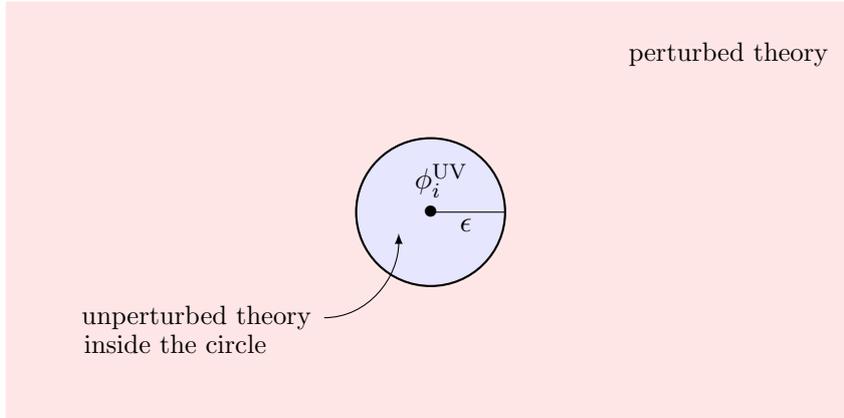
\end{center} 

We can understand this procedure from the point of view of standard renormalisation theory. To construct a local composite 
operator in the deformed theory we usually start by taking an operator $\phi_{i}^{\rm UV}$ in the UV theory and regularising the perturbation series expansion that involves this operator. This amounts to regularising the collisions of the perturbing operators between themselves and away from the insertion as well as the collisions of perturbing operators with the insertion. For example we can take a
  hard disc  regularisation in which we cut out discs of radius $\epsilon$ around the insertions of the perturbing operators and around the insertion of  $\phi_{i}^{\rm UV}$ and do not allow the centers of the discs to be closer than $2\epsilon$. When the regularisation is removed the divergences are cancelled by counterterms added to the action and by counterterms added to $\phi_{i}^{\rm UV}$. Here we have a single cutoff for both types of collisions and it is removed in a single step. The above procedure illustrated on Figures \ref{Fig_operator} and \ref{Fig_operator_bulk} is a modification of the standard scheme in which one uses two regulators: one cuts out discs of radius $\epsilon'$ around the perturbing operator insertions and of radius $\epsilon$ around $\phi_{i}^{\rm UV}$ not allowing the discs to overlap. 
  One then first takes $\epsilon'\to 0$, adds action counterterms and constructs the deformation interface which now surrounds $\phi_{i}^{\rm UV}$ in the part regulated and part renormalised perturbation series. Next one takes $\epsilon$ to zero adding the counterterms to $\phi_{i}^{\rm UV}$. It may be hard to prove renormalisability theorems in this mixed approach but formally this is a legitimate procedure because everything in the construction is local. The same picture goes for the boundary case with the sole modification being that discs are replaced by symmetric intervals around the insertions and the surrounding interface degenerates into two points where the interface operators are inserted. 
  
  Returning to the boundary case, if we have some basis of local boundary  operators $\tilde \psi_{b}^{\lambda}$ available in the deformed theory then we can use the 
  OPE expansion to write 
 \be \label{expansion_hathat}
 \hat \psi^{[\lambda,0]}(\tau + \epsilon) \psi_{a}^{0}(\tau)\hat \psi^{[0,\lambda]}(\tau-\epsilon) = \sum_{b} C_{a}^{b}(\epsilon, \lambda) 
 \tilde \psi_{b}^{\lambda}(\tau) 
 \ee
where $C_{a}^{b}(\epsilon, \lambda)$ are OPE coefficients. In section \ref{nearby_fp} we present explicit perturbative calculations for these coefficients for the case 
of nearby fixed points.
 One can single out the coefficients that are singular 
when $\epsilon \to 0$ and subtract them. The remaining terms after $\epsilon \to 0$ will give an expansion of $[\psi_{a}]^{\lambda}$ defined in 
(\ref{psi_lambda}): 
\be \label{expansion}
[\psi_{a}]^{\lambda} ={ \sum_{b}} [C_{a}^{b}](0, \lambda) 
 \tilde \psi_{b}^{\lambda}(\tau) 
 \ee
 where the square brackets around $C_{a}^{b}$ indicate the subtraction being done when taking $\epsilon \to 0$. It is also possible to include finite counterterms 
 in the subtraction.
If in (\ref{expansion}) we choose $\lambda$ to be at the value $\lambda=\lambda_{*}$ specifying the infrared fixed point and choose 
$ \tilde \psi_{b}^{\lambda_{*}}$ to be a basis of scaling fields at that fixed point, the coefficients we obtain 
are called RG mixing coefficients in \cite{Gaiotto}. If the renormalisation scheme (i.e. the counterterms) defining $[\psi_{a}]^{\lambda}$ is 
fixed these coefficients are unambiguously defined.

 Similarly to the above, for the case of bulk RG flows we can expand a UV theory operator surrounded by the deformation defect 
 as in (\ref{bulk_exp})  (see Figure  \ref{Fig_operator_bulk}).  Shrinking the circle and  adding counterterms 
 we obtain a renormalised operator 
 \be \label{expansion2}
[\phi_{i}]^{\lambda} ={ \sum_{j}} [C_{i}^{j}](0, \lambda) 
 \tilde \phi_{i}^{\lambda}(\tau) \, . 
 \ee
 Setting $\lambda=\lambda_{*}$ - the infrared fixed point and choosing $\tilde \phi_{i}^{\lambda_{*}}= \phi_{i}^{\rm IR}$ to be a basis of CFT scaling operators we obtain the RG mixing coefficients $[C_{i}^{j}](0, \lambda_{*})$.
 
  For the case of bulk $\phi_{1,3}$ flows between nearby minimal models the RG mixing coefficients  were 
calculated in \cite{Zamolodchikov_pert}.  In \cite{Gaiotto} these coefficients are compared to the so called RG pairings 
which are defined as pairings of states via the RG interface. For the bulk flows the RG interface is  one-dimensional 
it can be specified via the folding trick by a boundary state $\langle {\rm RG}| \in ({\cal H}^{\rm UV}\otimes {\cal H}^{\rm IR})^{*}$. 
If $|i,{\rm UV}\rangle \in {\cal H}^{\rm UV}$ and $|j, {\rm IR}\rangle\in  {\cal H}^{\rm IR}$ then the pairing between these two states is defined as the overlap
$\langle {\rm RG}|(| i, {\rm UV}\rangle \otimes |  j,{\rm IR}\rangle)$. This pairing of states specifies a pairing of operators via the usual state-operator correspondence. We will refer to this pairing as  Gaiotto's pairing. Equivalently we can define this pairing intrinsically in terms of operators by taking an operator $\phi_{i}^{\rm UV}$ in the UV CFT
and inserting it at the origin in the complex plane. We then place the RG interface on a unit circle surrounding $\phi_{i}^{\rm UV}$ 
as depicted on Figure \ref{Fig_operator_bulk} so that outside of the unit circle we have the IR CFT. We can insert an operator 
$\phi_{j}^{\rm IR}$ and calculate the correlation function to pick up the value of the pairing or, equivalently, we can expand the 
UV operator surrounded by the interface in terms of local operators  of the IR CFT as in (\ref{bulk_exp}). If 
the operators $\phi^{\lambda}_{i}$ taken at the IR fixed point $\lambda=\lambda_{*}$ are scaling operators then 
 the Gaiotto's pairing coefficients are given by $C_{i}^{j}(1, \lambda_{*})$ where we set\footnote{We also assume that at the fixed point the RG scale is set to one: $\mu=1$.} $\epsilon=1$. Note that  expansion (\ref{bulk_exp}) taken at $\lambda=\lambda_{*}$ contains operators of arbitrarily high conformal weights. Thus the pairing coefficients $C_{i}^{j}(1, \lambda_{*})$ are defined for pairs of UV and IR operators which
  have arbitrarily large difference in conformal dimension. Explicit calculations show that for the interface proposed in 
 \cite{Gaiotto} such pairings are non-vanishing already
 at the first order in $1/m$. This presents an apparent problem with Gaiotto's interpretation \cite{Gaiotto} of these coefficients as the RG mixing coefficients $C_{i}^{j}(0, \lambda^{*})$  as in any RG flow only operators of dimensions smaller or equal than the UV dimension of the seed operator $\phi_{i}^{\rm UV}$ get admixed in counter terms so at the IR fixed point we expect an upper bound on the conformal weights involved. Moreover for flows between nearby fixed points in the perturbative scheme of \cite{Zamolodchikov_pert} we expect only operators of nearby conformal dimensions to mix 
that is the differences between conformal dimensions should go to zero as $m\to \infty$ ($c\to 1$). It seems that for the Gaiotto's 
prescription to work we only need to apply it to the coefficients  $C_{i}^{j}(1, \lambda_{*})$ taken for operators of nearby conformal dimensions.

If instead of the radius of surrounding defect being $1$ we take it to be $\epsilon$ we can write a small $\epsilon$  expansion of $\phi_{i}^{\rm UV}$ 
surrounded by the RG interface as 
\be \label{exp_eps}
{\cal D}^{\rm RG}_{\epsilon }\phi_{i}^{\rm UV}(0)= \sum_{j} C_{i}^{j}(1, \lambda_{*}) \epsilon^{\Delta_{j}-\Delta_{i}} \phi^{\rm IR}_{j}(0) \, .
\ee
We see that the contribution of the higher dimensional operators is suppressed if we take $\epsilon\to 0$. The leading contributions will come from operators of dimensions close to $\Delta_{i}$ (or smaller if it mixes with more relevant operators). We can imagine then somehow truncating this expansion by retaining only  the leading terms at $\epsilon\to 0$. Thus the Gaiotto's pairing coefficients 
$C_{i}^{j}(1, \lambda_{*})$ are recovered from the OPE coefficients $C_{i}^{j}(\epsilon, \lambda)$ by first taking $\lambda=\lambda_{*}$ and then the limit $\epsilon \to 0$. It is not clear then why these coefficients should match with the 
RG mixing coefficients $[C_{i}^{j}](0, \lambda^{*})$ which are obtained essentially using the opposite order of limits:  by first taking $\epsilon$ to zero, subtracting the divergences and then taking the coupling to the IR fixed point value.
Given that at the IR fixed point the short distance behaviour jumps discontinuously there is no reason to believe both expansions have matching coefficients. The RG mixing coefficients however are scheme dependent and Gaiotto's proposal, which was tested 
in the Zamolodchikov's scheme up to the next to leading order and also for descendants in \cite{Poghossian}, \cite{PP}, may 
well be true in a particular scheme. In section   \ref{pairing_sec} we discuss a  scheme which would guarantee the matching 
and which at higher orders may differ from the Zamolodchikov's scheme.

\section{Nearby fixed points} \label{nearby_fp}
\setcounter{equation}{0}

\subsection{Some generalities}
In this section we consider flows triggered by a single boundary primary operator $\psi^{0}$ with dimension 
$0<\Delta<1$ such that the anomalous dimension\footnote{We limits ourselves to the strictly relevant case 
as perturbation theory can be safely applied to 
calculate OPE coefficients \cite{GM}.} $y=1-\Delta$ is a small number. A prototype model of this situation is 
a flow in the A-series Virasoro models triggered on a Cardy boundary condition by a $\psi_{1,3}$ boundary field perturbation. 
Such flows were studied by perturbative methods in \cite{RRS} and non-perturbatively in \cite{LSS}, \cite{GW}. 
In the context of RG interfaces they were considered in \cite{RG_boundary1}. 
We will consider the leading order perturbative results trying to keep the discussion  model independent as 
much as possible. The general approach is that we would like to be able to use perturbation theory in $y$ to compute 
approximations to  quantities like correlation functions and anomalous dimensions at the IR fixed point. 
To have a reliable approximation we need to choose a renormalisation scheme in which   correlation functions do not have any poles in $y$ in the limit $y\to 0$. 
As the perturbing operator becomes marginal for $y\to 0$ such poles signify the emergent logarithmic divergences in 
the perturbation theory integrals. On dimensional grounds such divergences can only occur in the mixings of operators 
whose dimensions become the same in the $y\to 0$ limit. In this subsection we focus on the perturbing operator $\psi^{0}$ 
while operator mixings are considered in sections \ref{sec_nonderiv}, \ref{sec:derivatives}.

We assume that the perturbing primary operator is normalised so that its two- and three-point functions are 
\be
\langle \psi^{0}(\tau)\psi^{0}(0)\rangle_{0} =\frac{1}{|\tau|^{2\Delta}}\, , \quad 
\langle  \psi^{0}(\tau_1)\psi^{0}(\tau_2)\psi^{0}(0)\rangle_{0} = \frac{C}{|\tau_1-\tau_2|^{\Delta} |\tau_1|^{\Delta}|\tau_2|^{\Delta}}
\ee
where $\langle ...\rangle_{0}$ stand for correlators in the UV BCFT and $C\ne 0$ is the OPE coefficient in the OPE 
\be
\psi^{0}(\tau)\psi^{0}(0) = \frac{1}{\tau^{2\Delta}} + \frac{C}{\tau^{\Delta}}\psi^{0}(0) + \dots \, , \quad \tau>0 
\ee
  where among the omitted terms  other singular terms may be present, however they should not contain any operators, other than 
  $\psi^{0}$ itself, 
  whose dimensions become marginal for $y\to 0$. This ensures that in the perturbative scheme  the  operator $\psi^{0}$ only 
  mixes with itself under the RG action.
 
We define the bare perturbation series for correlation functions according to 
\be \label{bare_ps}
\langle \psi^{0}_{n}(x_n) \dots \psi_{1}^{0}(x_1) \rangle_{\lambda} = {\cal N}^{-1}\langle e^{\Delta S} \psi_{n}^{0}(x_n) \dots \psi_{1}^{0}(x_1) \rangle_{0}
\ee
where 
\be \label{DeltaS}
\Delta S = \lambda_{0} \int \! d\tau \psi^{0}(\tau) \, , 
\ee
$\lambda_{0}$ is the bare coupling, $\psi_{i}^{0}$ are boundary operators,   $\langle ...\rangle_{\lambda}$ denotes the  correlators in the deformed theory, and
\be
{\cal N} = \langle e^{\lambda_{0} \int \! d\tau \psi^{0}(\tau)}  \rangle_{0} \, 
\ee 
is the normalisation factor. The divergences in the perturbative series  generated by expanding the exponent 
come from the regions of integration in which one or more copies of $\psi^{0}$ collide with themselves away from 
the external fields $\psi_{i}^{0}(\tau_i)$ or at those insertions. In the first case we cancel the divergences via counterterms 
in the perturbation action $\Delta S$ while in the latter by counterterms defining composite operators in the perturbed theory.

In addition to divergent terms there are also finite contributions which 
are singular in the $\Delta \to 1$ limit. We would like to subtract such terms as well. 
Let us consider the two-point function of the perturbing operator $\psi^{0}$. 
The leading order correction is 
\bea
&& \langle \psi^{0}(x) \psi^{0}(0)\rangle_{\lambda} = \frac{1}{|x|^{2\Delta}} + \lambda_{0} \frac{C}{|x|^{3\Delta -1}} \int\limits_{-\infty}^{\infty}dt \frac{1}{|t|^{\Delta}|1-t|^{\Delta}} + \dots \nonumber \\
&& = \frac{1}{|x|^{2\Delta}} +  \lambda_{0} \frac{C}{|x|^{3\Delta -1}}I(\Delta) + \dots
\eea
where the integral at hand converges to the value  
\be
I(\Delta) =  \frac{\sqrt{\pi}\Gamma^{2}(1/2-\Delta/2)\Gamma(\Delta - 1/2)}{\Gamma^{2}(\Delta/2)\Gamma(1-\Delta)}
\ee
 for $\Delta>1/2$, which is the assumption, but it has a pole at $\Delta\to 1$:
\be
I(\Delta) \sim \frac{4}{1-\Delta}   \, , \quad \Delta \to 1\, .
\ee
We can subtract this pole by introducing a renormalised operator 
\be 
\psi^{\lambda} = Z^{-1/2}\psi^{0}
 \ee
  and choosing the $Z$-factor to be of the form 
\be \label{Zf}
Z = \mu^{2(\Delta-1)} + \lambda_{0}C\tilde I(\Delta) \mu^{3(\Delta-1)} + {\cal O}(\lambda_{0}^2)
\ee
where $\tilde I(\Delta)$ is any function that has a simple pole at $\Delta=1$ with the same residue as $I(\Delta)$ and 
$\mu$ is a subtraction scale which has units of mass. 
For example we could choose a variant of minimal subtraction in which we subtract just the poles at every  order in perturbation. 
Instead we will follow another scheme fixed by setting 
\be \label{RGscale}
\langle \psi^{\lambda}(\mu^{-1})\psi^{\lambda}(0)\rangle = Z^{-1}\langle \psi^{0}(\mu^{-1})\psi^{0}(0)\rangle = \mu^2 
\ee 
that ties  the scale $\mu$ and the  coupling to the two-point function. This is essentially the scheme used in  
\cite{Zamolodchikov_pert} (see also \cite{CL}
for a  thorough discussion). 
For this choice of scheme we have $\tilde I(\Delta)=I(\Delta)$ in (\ref{Zf}). 

The renormalised operator $\psi^{\lambda}$ is linked to the renormalised dimensionless coupling $\lambda$ via the action principle: 
\be 
\frac{\partial}{\partial \lambda} \langle  \dots \rangle_{\lambda} = \langle \int\! d\tau \psi^{\lambda}(\tau) \dots \rangle_{\lambda} 
\ee
that by virtue of (\ref{bare_ps}), (\ref{DeltaS}) gives 
\be
\frac{\partial \lambda_{0}}{\partial \lambda} = Z^{-1/2}  \, .
\ee
Using (\ref{Zf}) with $\tilde I(\Delta)=I(\Delta)$ we obtain 
through the order $\lambda^2$
\be
\lambda_{0} = \mu^{1-\Delta}(\lambda - \frac{C}{4}I\lambda^2) 
\ee
that gives 
\be
\mu \frac{\partial \lambda}{\partial \mu} = - (1-\Delta)\lambda -  \frac{C}{4}I(1-\Delta) \lambda^2 
\ee
and therefore in the chosen scheme the beta function is 
\be \label{beta}
\beta(\lambda) = -y\lambda - D\lambda^2  \, , \quad  \mbox{where } y=1-\Delta\, , \enspace D= \frac{C}{4}I(\Delta)(1-\Delta)\, .
\ee
We assume from now on that in the limit  $y\to 0$ the coefficient $D$ tends to a non-zero constant. Under this assumption 
the beta function (\ref{beta}) has a zero corresponding to an infrared fixed point at 
\be \label{fpt}
\lambda_{*} = -\frac{y}{D} \, .
\ee
Under our assumptions this value is small which is the basis for calling this fixed point perturbative or "nearby".

Besides the renormalised perturbing operator $\psi^{\lambda}$ we are also interested in renormalised composite operators 
which we will denote $\psi_{i}^{\lambda}$ and which satisfy the Callan-Symanzik equation 
\be \label{CS_gen1}
(\mu \frac{\partial}{\partial \mu}  + \sum_{i=1}^{N} \hat \Gamma_{i} + \beta(\lambda) \frac{\partial}{\partial \lambda}) 
\langle \psi_{1}^{\lambda}(\tau_{1})  \psi_{2}^{\lambda}(\tau_{2}) \dots \psi_{N}^{\lambda}(\tau_{N})  \rangle_{\lambda} = 0
\ee
where $\hat \Gamma_{i}$ is an operator that acts as 
\be \label{gamma_general}
\hat \Gamma_{i} \psi_{j}^{\lambda} = \delta_{ij} \sum_{k}\gamma_{ik}\psi_{k}^{\lambda} \, 
\ee
where  $\gamma_{ij}$ is the matrix of anomalous dimensions.
For the perturbing operator it has only one entry
\be
 \gamma \equiv \gamma_{\psi \psi} =  \partial_{\lambda}\beta \, .
\ee
At the infrared fixed point (\ref{fpt}) the value of the scaling dimension of $\psi$ is 
\be
1+ \gamma(\lambda_{*}) =  \partial_{\lambda}\beta(\lambda_{*}) =1 + y\, 
\ee
that means that the flow approaches the infrared fixed point along an operator of marginally irrelevant dimension: 
$\Delta_{\rm ir}=1+y$.  

Trading the scale change for a dilation transformation we obtain another form of the Callan-Symanzik equation 
\be\label{CS_gen2} 
\Bigl[ \sum_{i=1}^{N} \tau_{i}\frac{\partial}{\partial \tau_{i}}  + \sum_{i=1}^{N}(1+\hat \Gamma_{i})  
+ \beta(\lambda) \frac{\partial}{\partial \lambda} \Bigr] 
\langle \psi_{1}^{\lambda}(\tau_{1})\psi_{2}^{\lambda}(\tau_{2})  \dots \psi_{N}^{\lambda}(\tau_{N})\rangle = 0 \, .
\ee

\subsection{The interface operator} \label{section_RGop}
Before we discuss the RG operator for the flow between $\lambda=0$ and $\lambda=\lambda_{*}$ 
we first discuss the bare interface operator $\hat \psi^{[\lambda, 0]}_{0}$ for the deformation at hand (see Fig. \ref{bcco}). 
This operator can be constructed perturbatively by expanding $e^{\Delta S}$ on a half line extending to the right of 
the insertion point. 
Similarly the conjugate operator $\hat \psi^{[ 0,\lambda]}_{0}$ is obtained by the same expansion on a half line extending to 
the left of the insertion point. To construct the renormalised operators   $\hat \psi^{[\lambda, 0]}$, $\hat \psi^{[ 0,\lambda]}$ 
we consider the two point function $\langle \hat \psi_{0}^{[0,\lambda]}(\tau/2)\hat \psi_{0}^{[\lambda,0]}(-\tau/2)\rangle_{\lambda}$. 
This  two-point function can be
equivalently described as a perturbation of the BCFT on the interval: $[-\tau/2,\tau/2]$ as 
depicted on Figure \ref{Figpp}.

\begin{center}
\begin{figure}[H]
\centering
\begin{tikzpicture}[scale=1.4]
\filldraw[fill=gray!30!white,draw=white] (-3.8,0) rectangle (3.8,3);
\draw[blue,very thick,dashed](-3.8,0)--(-1.4,0);
\draw[red,very thick] (-1.4,0)--(1.4,0);
\draw[blue,very thick,dashed](1.4,0)--(3.8,0);
\draw (-1.4,0) node {$\bullet$} ;
\draw (1.4,0) node {$\bullet$} ;
\draw (-1.3,-0.3) node {$\hat \psi^{[\lambda, 0]}$};
\draw (1.5,-0.3) node {$\hat \psi^{[ 0,\lambda]}$};

\draw (0,1.5) node {{\small Bulk CFT}};
\draw (-3.1,-0.2) node {{\small Unperturbed}};
\draw (0,-0.2) node {{\small Perturbed}};
\draw (3.1, -0.2) node {{\small Unperturbed }} ;
\end{tikzpicture}
\caption{Two point function of the boundary condition changing operator is equivalent to perturbing the boundary condition on an interval}
\label{Figpp}
\end{figure}
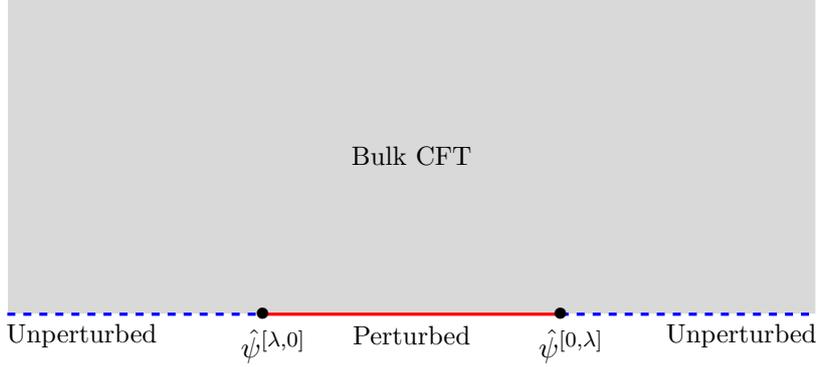
\end{center}
For the leading order correction we obtain 
\bea \label{hat2pt}
&&\langle \hat \psi^{[0,\lambda]}_{0}(\tau/2)\hat \psi_{0}^{[\lambda,0]}(-\tau/2)\rangle_{\lambda} = 
\langle \exp\left( \lambda_{0} \!\! \int\limits_{-\tau/2}^{\tau/2}\!\! dt \, \psi^{0}(t) \right)\rangle_{0} \nonumber \\
&&=1 + 
\frac{\lambda_{0}^2}{2!}\int\limits_{-\tau/2}^{\tau/2}\!\! dt\!\!  \int\limits_{-\tau/2}^{\tau/2}\!\! dt'\, 
 \langle \psi^{0}(t) \psi^{0}(t')\rangle_{0} + \dots 
\eea
This integral diverges in the region $t\to t'$ as a power (as we assume that $1/2<\Delta$).   
Subtracting this divergence we obtain 
\be
\langle \hat \psi^{[0,\lambda]}_{0}(\tau/2)\hat \psi_{0}^{[\lambda,0]}(-\tau/2)\rangle_{\lambda} =
1 + \frac{\lambda_{0}^2}{2}\frac{|x|^{2(1-\Delta)}}{(1-2\Delta)(1-\Delta)}  + {\cal O}(\lambda_{0}^2) \, .
\ee 
The remaining finite contribution has a pole at $\Delta=1$ which we subtract via a counterterm present in the $Z$-factor for 
the renormalised operators: 
\be  \hat \psi^{[0,\lambda]}= \hat \psi^{[0,\lambda]}_{0} \hat Z^{-1/2}\, , \qquad 
 \hat \psi^{[\lambda,0]}= \hat \psi^{[\lambda,0]}_{0} \hat Z^{-1/2}\, . 
 \ee
  Similarly to  (\ref{RGscale}) we adopt a renormalisation scheme in which 
\be
\langle \hat \psi^{[0,\lambda]}(\mu^{-1}/2) \hat \psi^{[\lambda,0]}(-\mu^{-1}/2)\rangle_{\lambda} = 1 
\ee
that gives in terms of the renormalised coupling  
\be
\hat Z = 1 + \frac{\lambda^2}{2(1-2\Delta)(1-\Delta)}  + {\cal O}(\lambda^3) \, .
\ee
The renormalised two-point function satisfies the Callan-Symanzik equation 
\be\label{CShat}
\left( \tau \frac{\partial}{\partial \tau} + \beta(\lambda) \frac{\partial}{\partial \lambda} + 2\hat \gamma \right)\langle \hat \psi^{[0,\lambda]}(\tau/2)\hat \psi^{[\lambda,0]}(-\tau/2)\rangle_{\lambda} =0
\ee
where 
\be \label{hat_gamma}
 \hat \gamma = \frac{\mu}{2}\frac{d \ln \hat Z}{d\mu} =  \frac{\lambda^2}{2(2\Delta-1)}  + {\cal O}(\lambda^3) 
\equiv \frac{a\lambda^2}{2} + {\cal O}(\lambda^3)  \, , \quad a=\frac{1}{2\Delta-1} \, .
\ee
(Note that in (\ref{CShat}) we treat $\hat \psi$ fields as fields of classical dimension zero as opposed to the perturbing field 
that has classical dimension 1. This is consistent as $\hat \psi_{0}$ is obtained by integrating $\psi^{0}$.)
At the fixed point (\ref{fpt}) we obtain in the leading order in $y$ the scaling dimension of $\hat \psi^{[{\rm ir}, {\rm uv}]}$ :
\be\label{hat_dim}
\hat \Delta =  \frac{a\lambda^2_{*}}{2} + {\cal O}(y^3) = \frac{y^2}{2D^2} + {\cal O}'(y^3)
\ee
and the same for the conjugate operator $\hat \psi^{[{\rm uv}, {\rm ir}]}$. This can be compared with the $c\to 1$ expansion 
of the dimension for RG operator conjectured in \cite{RG_boundary1}. We find a precise match of the leading term with 
(\ref{hat_dim}). 
 
It is interesting to note that we get the same leading value if we calculate the lowest dimension of a boundary condition changing 
primary field linking the UV and the IR BCFTs. This can be demonstrated for example by using perturbation theory on a strip perturbed on one edge and finding the shift in the vacuum energy. It may be thus true for ``short'' RG flows we are looking at here, that the RG operator is always a primary of lowest dimension, but it is hard to see any general argument as to why this may be true for generic flows.
 
 We would like next to calculate the OPE coefficients in (\ref{OPE1}) and (\ref{OPE2}) in the leading order in $y$. 
 For the coefficient $C_{\rm uv}$ we start from the same setup as depicted on Figure  \ref{Figpp} but with an additional insertion 
 of $\psi^{0}$ at  a point on the boundary: $z\gg \tau$. At the leading order we can set $\hat Z=1$ and calculate 
 \bea
 &&\langle  \psi^{0}(z)  \hat \psi^{[0,\lambda]}_{0}(\tau/2)\hat \psi_{0}^{[\lambda,0]}(-\tau/2) \rangle_{\lambda} =
 \lambda\mu^{y}\!\! \int\limits_{-\tau/2}^{\tau/2}\!\! dt \langle \psi^{0}(t) \psi^{0}(z)\rangle + \dots  \nonumber \\
 && = \frac{\lambda \mu^{y}}{ 2\Delta -1} \left( \frac{1}{(z-\tau/2)^{2\Delta-1}} -   \frac{1}{(z+ \tau/2)^{2\Delta-1}} \right) + \dots 
 \sim \frac{\lambda \mu^{y} \tau}{z^{2\Delta}} 
 \eea
 where in the last step we retained the leading asymptotic term for $z/\tau \to \infty$. Setting $\lambda=\lambda_{*}=-\frac{y}{D}$ 
 and comparing  with  (\ref{OPE1}) we find, using  (\ref{hat_dim}), that 
 \be\label{Cuv_val}
 C_{\rm uv} = \lambda_{*} +  {\cal O}(y^2) =-\frac{y}{D} + {\cal O}'(y^2) \, . 
 \ee
 
 The coefficient $C_{\rm ir}$ can be also calculated in the same simple-minded manner, by swapping the perturbed and unperturbed regions on Figure \ref{Figpp} and inserting at $z\gg \tau$ the renormalised operator $\psi$. A more instructive way to do  
 calculate $C_{\rm ir}$, which leads to the same result, is by expanding the RG resummed two-point function 
 $$\langle \hat \psi^{[0,\lambda]}(\tau/2)\hat \psi^{[\lambda,0]}(-\tau/2)\rangle_{\lambda}$$ 
 near the infrared fixed point 
  $\lambda=\lambda_{*}$. The RG resummation\footnote{It should be noted that, although small in the $y\to 0$ limit, 
  there are also non-perturbative contributions to the two-point function via the one-point functions. We should regard the RG resummed result which is based on perturbative calculations as the part of the two-point function coming from the identity operator in the OPE. 
  It is known \cite{GM} that for relevant perturbations OPE coefficients can be computed perturbatively. } is done by solving the Callan-Symanzik equation (\ref{CShat}) 
  with the beta function (\ref{beta}), $\hat \gamma$ given by the leading term in (\ref{hat_gamma}) and the 
  initial condition taken as the leading order value, that is just 1. 
  The result can be expressed in various ways. An expression particularly adapted to
  expansion in the  infrared  is  
 \be \label{2p_resum}
 \langle \hat \psi^{[0,\lambda]}(\tau/2)\hat \psi^{[\lambda,0]}(-\tau/2)\rangle_{\lambda}  = 
 \frac{1}{|\tau\mu|^{ay^2/D^2}} \exp\left( 
\frac{ay}{D^2}(f(Y)-f(\xi)) \right) 
\ee
 where 
\be
Y=-\left(\frac{\lambda - \lambda^{*}}{\lambda}\right)|\mu \tau|^{-y} \, , \qquad \xi = -\left(\frac{\lambda - \lambda^{*}}{\lambda}\right) \, ,
\ee
\be
f(x)=\frac{1}{1+x}-\ln(1+x) \, .
\ee
 It is easy to see that in the short distance  limit: $\tau \to 0$ ($\lambda \ne \lambda_{*}$) this function is non-singular while for the 
 large distance limit $\tau \to \infty$ ($\lambda \ne 0$) it 
 decays as ${\rm Const}\cdot \tau^{-2\hat \Delta}$. 
 
The infrared fixed point dominates when $|Y|<1$, this is when we can use conformal perturbation theory near the corresponding BCFT.
 The leading correction to the two-point function is given by an integral of the three point function:
 \be\label{ir_pert1}
  \langle \hat \psi^{[0,\lambda]}(\tau/2)\hat \psi^{[\lambda,0]}(-\tau/2)\rangle_{\lambda}= 
\frac{b}{\tau^{2\hat \Delta}} + (\lambda-\lambda_{*})\! \int\limits_{-\tau/2}^{\tau/2}\!\! dt\, \langle  \hat \psi^{[0,\lambda]}(\tau/2) 
\psi^{\rm ir}(t) \hat \psi^{[\lambda,0]}(-\tau/2)\rangle_{*} + \dots  
\ee
 where $b$ is a normalisation constant and $\langle \dots \rangle_{*}$ stands for a correlator in the IR BCFT.
 The three-point function at hand is 
 \be 
 \langle  \hat \psi^{[0,\lambda]}(\tau/2) 
\psi^{\rm ir}(t) \hat \psi^{[\lambda,0]}(-\tau/2)\rangle_{*} = \frac{C_{\rm ir}}{\tau^{2\hat \Delta - \Delta_{\rm ir}} 
(\tau/2-t)^{\Delta_{\rm ir}}(t+\tau/2)^{\Delta_{\rm ir}}} \, .
 \ee
 Substituting this into (\ref{ir_pert1})  we obtain  
  \be\label{ir_pert2}
  \langle \hat \psi^{[0,\lambda]}(\tau/2)\hat \psi^{[\lambda,0]}(-\tau/2)\rangle_{\lambda}= 
\frac{b}{\tau^{2\hat \Delta}} +  (\lambda-\lambda_{*})  \frac{C_{\rm ir}}{\tau^{2\hat \Delta+\Delta_{\rm ir}-1}} 
\int\limits_{0}^{1} \frac{dx}{(1-x)^{\Delta_{\rm ir}}x^{\Delta_{\rm ir}}}  + \dots 
\ee
As $\Delta_{\rm ir}=1+y>1$ the integral at hand has power divergences. Subtracting them via analytic continuation to Euler's 
beta function we obtain a renormalised value 
 \be\label{ir_pert3}
  \langle \hat \psi^{[0,\lambda]}(\tau/2)\hat \psi^{[\lambda,0]}(-\tau/2)\rangle_{\lambda}= 
\frac{b}{\tau^{2\hat \Delta}} +  (\lambda-\lambda_{*}) \frac{C_{\rm ir}}{\tau^{2\hat \Delta+\Delta_{\rm ir}-1}} 
\frac{\Gamma^2(1-\Delta_{\rm ir})}{\Gamma(2-2\Delta_{\rm ir})} + \dots 
\ee
Comparing this with the first two terms in the expansion of  (\ref{2p_resum}) in powers of $Y$ and $\xi$ we obtain 
\be \label{Cir_val}
b=1 + {\cal O}(y) \, , \quad C_{\rm ir} = -\lambda_{*} + {\cal O}(y^2) = \frac{y}{D} + {\cal O}'(y^2) \, . 
\ee

At this stage we can summarise what our calculations say about the three conjectures spelled out in the introduction. 
The first conjectured fact, that the RG operator is a conformal primary, in the perturbative case comes practically for 
free as the anomalous dimension of $\hat \psi^{[\rm ir, \rm uv]}$ is a series in powers of $y$ that is assumed to be infinitesimally small. 
Thus its scaling dimension $\hat \Delta$ is perturbatively less than 1 that in a unitary theory implies that it is a primary. 

In (\ref{Cuv_val}), (\ref{Cir_val}) we also got non-vanishing values of the two OPE coefficients: $C_{\rm uv}$ and $C_{\rm ir}$ in accord with conjectures 2 and 3. If the perturbative series in powers of $y$ has a non-zero radius of convergence then both constants, 
which are believed not to have any non-perturbative contributions for strictly relevant perturbations, will be non-vanishing at least for a generic value of $y$. 

\subsection{Renormalisation of composite operators}  \label{sec_nonderiv}
Let us consider  a basis of quasi-primary  operators $\psi_{i}^{0}$ in the UV BCFT with scaling dimensions $\Delta_{i}$. 
We assume that the BCFT at hand is symmetric under reflections preserving the boundary: $\tau \to -\tau$ with a charge $R$ 
that is an anti-unitary operator. We assume further that the operators $\psi_{i}^{0}$ transform as 
\be
(\psi_{j}^{0}(\tau))^{\dagger} \equiv R\, \psi_{j}^{0}(-\tau) R^{-1} = \eta_{j} \psi_{j}^{0}(-\tau)
\ee 
where $\eta_{j}=\pm 1$ are the charges. We will refer to operators with charge $\eta_{j}=1$ as reflection symmetric and 
to operators of charge $\eta_{j}=-1$ as reflection anti-symmetric operators. For example, in the $A$-series Virasoro unitary minimal 
models for a boundary condition changing operator $\psi_{i}^{[a,b]}$ we have $(\psi_{i}^{[a,b]})^{\dagger}(0) = \psi_{i}^{[b,a]}(0)$
so that the operator $\psi_{i}^{[a,b]} - \psi_{i}^{[b,a]}$ has charge $-1$ and $\psi_{i}^{[a,b]} + \psi_{i}^{[b,a]}$ has charge 1. 
Any derivative applied to an operator flips the charge so that the charge of $\partial^{n}\psi_{i}^{0}$ is $(-1)^{n}\eta_{i}$. We choose the operators $\psi_{i}^{0}$ to be normalised
 so that 
\be
\langle \psi_{i}^{0}(\tau) (\psi_{j}^{0})^{\dagger}(0)\rangle_{0} = \frac{\delta_{ij}}{\tau^{2\Delta_{i}}} \, , \quad \tau>0 \, .
\ee
We define OPE coefficients as
\be
\psi_{i}^{0}(\tau) \psi_{j}^{0}(0) = \sum_{k}\sum_{n=0}^{\infty}\tau^{\Delta_{k} + n-\Delta_{i}-\Delta_{j}} D_{ij}^{k(n)}\partial^{n}\psi_{k}^{0}(0)  \quad \tau>0
\ee
where 
\be \label{derivative_OPE}
 D_{ij}^{k(n)} = D_{ij}^{k} \frac{\Gamma(n-\Delta_{j}+\Delta_{i}+\Delta_{k})\Gamma(2\Delta_{k})}{n!\Gamma(\Delta_{i}+\Delta_{k}-\Delta_{j})\Gamma(2\Delta_{k}+n)}
\ee
and $D_{ij}^{k}$ are the OPE coefficients for the quasiprimary fields. With our choice of basis  $D_{ij}^{k}$ are real in a unitary BCFT. 
In general the cyclic  symmetry of the 3-point function coefficients implies
\be
D_{ij}^{k} = D_{ki}^{j}\eta_{j}\eta_{k}\, .
\ee
 If the boundary condition is reflection symmetric 
the OPE coefficients must also satisfy 
\be  \label{D_symm}
D_{ij}^{k} = \eta_{i}\eta_{j}\eta_{k}D_{ji}^{k}\, .
\ee
For nearby fixed points we assume that $D_{\psi\psi}^{\psi}$ is non-vanishing so that $\eta_{\psi}=1$.  
Denote the OPE coefficients standing at the perturbing operator $\psi^{0}$ as $D_{ij}^{\psi}$,  $D_{ji}^{\psi}$. 
We see from (\ref{D_symm}) that for operators of the same reflection charge the coefficients $D_{ij}^{\psi}$ are symmetric 
and for the operators of opposite charge they are anti-symmetric. 

Assuming that at least one of  $D_{ij}^{\psi}$,  $D_{ji}^{\psi}$  is non-vanishing the leading order correction to the two-point function is 
\be  \label{2pt_bares}
\langle \psi_{j}^{0}(\tau) (\psi_{i}^{0})^{\dagger}(0)\rangle_{\lambda} = \frac{\delta_{ij}}{\tau^{2\Delta_{i}}} + \frac{\lambda_{0}}{\tau^{\Delta_{i} + \Delta_{j}-y}} 
\eta_{i}T_{ij} 
\ee
where 
\bea \label{IJ}
&& T_{ij}=D_{ji}^{\psi}(I_{ij} + I_{ji}) + D_{ij}^{\psi} J_{ij} \, , \nonumber \\
&& I_{ij} = \frac{\Gamma(2\Delta-1)\Gamma(1-\Delta + \Delta_{i} - \Delta_{j})}{\Gamma(\Delta + \Delta_{i} - \Delta_{j})} \, ,\nonumber \\
&& J_{ij} = \frac{\Gamma(1-\Delta - \Delta_{i}+ \Delta_{j})\Gamma(1-\Delta - \Delta_{j}+ \Delta_{i})}{\Gamma(2-2\Delta)} \, 
\eea
where it is assumed that only power divergences may be present which are minimally subtracted by analytic continuation in dimensions.

We  observe that $I_{ij}$ and $J_{ij}$ 
have poles when 
$y+\Delta_{i}-\Delta_{j}=-n$ or $y+\Delta_{j}-\Delta_{i}=-n$ where $n$ is a non-negative integer, assuming $0<\Delta<1$. 
If one of such equalities takes place then the above expressions are not valid as there is a logarithmic divergence in the integral at hand. 
Such cases are called resonances. We  assume that no exact resonance is present. However for nearby fixed points, when we 
treat $y$ as a small parameter, there are in general values of $\Delta_{i}$, $\Delta_{j}$ and $n$ when the combinations at hand  vanish in the limit $y\to 0$. When $\Delta_{i}-\Delta_{j} = {\cal O}(y)$ then both $I_{ij}$ and $J_{ij}$ have poles which 
need to be subtracted by counterterms to ensure that perturbation theory in $y$ is applicable at the IR fixed point. 
To have a well defined operator at the IR fixed point one needs to subtract such poles by finite counterterms.
Interestingly if 
 $\Delta_{j}-\Delta_{i}+n \sim C(1-\Delta)$, $C\ne 0$ where 
$n$ is a positive integer, then the poles are absent. This is explained by the fact that the OPE coefficients 
$D_{\psi i}^{j (n)}$ given in (\ref{derivative_OPE}) vanish in this limit.
The same happens in the situation when  $\Delta_{i}-\Delta_{j}+n \sim C'(1-\Delta)$ where $C'\ne 0$.

We express the renormalised operators $\psi_{i}^{\lambda}$ as linear combinations of the bare operators $\psi_{j}^{0}$ 
\be 
 \label{zij}
\psi_{i}^{\lambda}(\tau) = \sum_{j}Z_{ij} \mu^{1-\Delta_{j}} \psi_{j}^{0}(\tau)
\ee
where the matrices $Z_{ij}$ are functions of the renormalised coupling $\lambda=\mu^{-y}\lambda_{0}$ only. 
The reflection action is then defined on the renormalised fields via the action on the bare fields. 
Note that, more generally, derivatives of $\psi_{j}^{0}$ can be present in the expansion (\ref{zij}). In this subsection we focus on the simplest case when 
this does not take place while mixings with derivatives are discussed in section \ref{sec:derivatives}.

The Callan-Symanzik equation for the renormalised operators holds in the form of  (\ref{CS_gen1}), (\ref{gamma_general}) with 
the matrix of anomalous dimensions
\be
\gamma_{ij} = - \sum_{k}  (\mu \frac{d}{d\mu} Z_{ik}\mu^{1-\Delta_{k}}) \mu^{\Delta_{k}-1} Z^{-1}_{kj}  \, .
\ee
At the leading 
order in the coupling we can write
\be \label{Zc1}
Z_{ij}=\delta_{ij} + \lambda z_{ij} \, 
\ee
The renormalised two-point function at the linear order in $\lambda$ is 
\be \label{2pt_ren_ij}
 \langle \psi_{j}^{\lambda}(\tau) (\psi_{i}^{\lambda})^{\dagger}(0)\rangle_{\lambda} = \mu^2 \Bigl[ \frac{\delta_{ij}}{(\tau \mu)^{2\Delta_{i}}} 
 + \lambda \Bigl(  \frac{\eta^{i}T_{ij}}{(\tau\mu)^{\Delta_{i} + \Delta_{j}-y}}   + \frac{z_{ij}}{(\tau \mu)^{2\Delta_{j}}} 
 + \frac{z_{ji}}{(\tau \mu)^{2\Delta_{i}}}  
 \Bigr) \Bigr]\, . 
\ee

  Imposing an analogue of the Zamolodchikov's condition
\be \label{Zcond}
\langle \psi^{\lambda}_{i}(\mu^{-1})(\psi_{j}^{\lambda})^{\dagger}(0)\rangle_{\lambda} = \mu^2 \delta_{ij} \, , 
\ee
we arrive at the leading order at the condition
\be \label{Zc2} 
z_{ij} + z_{ji} = -\eta_{i}T_{ij} \, .
\ee
Note that since $\eta_{\psi}=1$ the symmetry property (\ref{D_symm}) implies that $\eta_{i}T_{ij}$ is symmetric  under the interchange of $i$ and $j$. We fix the antisymmetric part of $Z_{ij}$  by requiring that 
$\gamma_{ij}$ is symmetric. This gives  
\be \label{Zam_scheme_Z0}
 z_{ij} =  -\frac{ \eta_{i}T_{ij}(y + \Delta_{i} - \Delta_{j})}{2y}  \, ,  
\ee  
\bea \label{Zam_scheme_gamma0}
\gamma_{ij} &=& (\Delta_{i}-1)\delta_{ij} + \lambda z_{ij}(y+ \Delta_{j}-\Delta_{i}) 
   \nonumber \\&=& (\Delta_{i}-1)\delta_{ij} -\lambda \eta_{i}T_{ij}\frac{y^2 - (\Delta_{i}-\Delta_{j})^2}{2y} \, .
   \eea

From now on we focus on the case of nearby fixed points with $y \ll 1$. As poles in $y$ can come due to mixings of operators 
whose UV dimensions become the same in the limit $y\to 0$ in the Zamolodchikov's scheme only mixings between such operators 
are allowed in the construction of renormalised operators (\ref{zij}). 
We consider first the simplest case of having a subset  of operators $\psi_{i}^{0}$ with 
$$\Delta_{i}-\Delta_{j} = {\cal O}(1-\Delta)\, $$
which only mix between themselves. 
For example, in the case of boundary RG flows triggered in the minimal models ${\cal M}_{m}$ by a $\psi_{1,3}$ operator, 
the primaries $\psi_{r,r+1}$, $\psi_{r,r-1}$ form such a subset provided $r\ll m$. 
At the leading order in the small $y$ expansion   we obtain 
\be
T_{ij} \sim  t_{ij}\equiv \frac{2(D_{ij}^{\psi}+ D_{ji}^{\psi})}{(1-\Delta)\left[1 - \left(\frac{\Delta_{i}-\Delta_{j}}{1-\Delta}\right)^2 \right]} \, ,
\ee
\be \label{Zam_scheme_Z}
 z_{ij} =  -\frac{ \eta_{i}t_{ij}(y + \Delta_{i} - \Delta_{j})}{2y} =  
- \frac{\eta_{i}(D_{ij}^{\psi} + D_{ji}^{\psi})}{y-\Delta_{i} + \Delta_{j}} \, , 
\ee  
\bea \label{Zam_scheme_gamma}
\gamma_{ij} &=& (\Delta_{i}-1)\delta_{ij} + \lambda z_{ij}(y+ \Delta_{j}-\Delta_{i}) 
   \nonumber \\&=&(\Delta_{i}-1)\delta_{ij} - \lambda \eta_{i}(D_{ij}^{\psi} + D_{ji}^{\psi}) \, .
\eea
Here we assume that $D_{ij}^{\psi}+ D_{ji}^{\psi}$ tends to a non-vanishing constant in the limit 
$y \to 0$. Since this tensor is  symmetric in $i,j$ this implies that mixings of operators with opposite reflection charges vanish at the order $y^{1}$ and may start only from the order $y^2$.

If we set $\lambda=\lambda^{*} = -\frac{y}{D}$ in the above expression for $\gamma_{ij}$ we get a numerical matrix $\gamma^{*}_{ij}$. 
Since it is symmetric it can be diagonalised by an orthogonal transformation. Let $\xi_{i}^{p}$ be the  eigenvectors 
labelled by the upper index $p$ so that 
\be
\sum_{j}\gamma^{*}_{ij} \xi_{j}^{p}= (\Delta_{p}^{\rm ir}-1) \xi_{i}^{p}  \, .
\ee
Here $\Delta_{p}^{\rm ir}$
are the scaling  dimensions of fields at the IR fixed points and the corresponding eigenvectors $\xi^{i}_{p}$
express the IR scaling fields in terms of renormalised 
operators which are built in the perturbed theory from the UV operators $\psi_{i}^{0}$ using the RG scheme fixed by (\ref{Zcond}). 
Namely, in our notation we can write 
\be
\chi_{p}^{\rm ir} = \sum_{i}\xi_{i}^{p}\psi^{\rm ir}_{i}
\ee 
where $\chi_{p}^{\rm ir}$ are the scaling fields. Assuming that the scaling fields are chosen so that\footnote{This is essentially the Zamolodchikov's condition (\ref{Zcond}) for $\mu=1$ which is the standard choice at fixed point CFTs.} 
\be
 \langle \chi_{p}^{\rm ir}(1) (\chi_{q}^{\rm ir})^{\dagger}(0)\rangle =  \delta_{pq} 
\ee
the eigenvectors must satisfy the orthonormality conditions
\be
\sum_{i} \xi_{i}^{p}\xi_{i}^{q} = \delta_{pq}\, , \qquad   \sum_{p} \xi_{i}^{p}\xi_{j}^{p} = \delta_{ij} \, . 
\ee
The coefficients $\xi_{i}^{p}$ are the RG mixing coefficients considered by Zamolodchikov in \cite{Zamolodchikov_pert} and by Gaiotto in \cite{Gaiotto}.

\subsection{Mixings with derivatives} \label{sec:derivatives}
Another interesting case of operator mixing arises  when 
\be
 \Delta_{j}-\Delta_{i}+ n = {\cal O}(y)\, ,  \enspace y=1-\Delta\, , \enspace n\in {\mathbb N}
\ee 
for a pair of operators $\psi_{i}^{0}$, $\psi_{j}^{0}$. 
For the $\psi_{1,3}$ flows this happens for example for operators of the form $\psi_{i}^{0}=\psi_{r,r+2}$ or 
$\psi_{i}^{0}=\psi_{r,r-2}$  and $\psi_{j}^{0}=\psi_{r,r}$ when  $n=1$.

To make the discussion more concrete consider a situation when we have a group of operators $\psi_{I}^{0}$ with 
the capital letter indices $I,J,K$ belonging to some subset of indices ${\cal A}$ such that 
\be
\Delta_{I}-\Delta_{J} = {\cal O}(y)\enspace \forall I,J\in {\cal A}\, .
\ee
 And suppose that there is another group of operators $\psi_{a}^{0}$ labelled by small case letters from the beginning of the alphabet that belong to some subset ${\cal B}$, for 
which 
\be \label{DeltaJJ}
\Delta_{a}-\Delta_{b} = {\cal O}(y) \enspace \forall a,b \in {\cal B} \, , 
\ee
\be \label{DeltaiJ}
\Delta_{a}-\Delta_{I} + n = {\cal O}(y) \enspace \forall a\in {\cal B}\, , \forall I\in {\cal A} \, .
\ee
for some fixed $n\in {\mathbb N}$. The last condition means that the operators $\psi_{I}^{0}$ and $\partial_{\tau}^{n}\psi^{0}_{a}$ 
have nearby dimensions and thus can mix under the RG flow\footnote{This is a model situation. In real models like $\psi_{1,3}$ flows 
for large enough dimensions the situation is more complicated as derivatives of different order of different groups of operators can all mix together.}. 
As we noted in our comments after formula (\ref{IJ}) the renormalised two-point functions 
$\langle \psi_{I}(\tau) \psi_{J}(0)\rangle_{\lambda}$,  $\langle \psi_{I}(\tau) \psi_{a}(0)\rangle_{\lambda}$
do not develop any additional singularities under (\ref{DeltaiJ}). It is due to this fact that, although we can introduce mixings with derivatives to satisfy the Zamolodchikov's condition (\ref{Zcond}), such mixings are suppressed at the leading order unless 
some other degenerations take place like $\Delta_{a} \sim 0$.

Following Zamolodchikov we treat operators built on derivatives of bare operators $\partial^{n}\psi^{0}_{a}$ on equal footing with 
operators built from quasiprimaries, namely allowing them to mix with all other operators as long as their dimensions are nearby. 
The advantage of this approach is that we get manifestly symmetric anomalous dimension matrices (at least at the leading order 
in $y$). The disadvantage is that we loose a simple characterisation of derivatives of renormalised operators as with the basis chosen they 
are embedded in some non-trivial way into the space of all renormalised operators. It was noted in \cite{CL} 
that off-criticality there is no characterisation of Virasoro descendants, however one can always single out the derivatives of other operators. We will discuss an alternative to Zamolodchikov's approach later. 

To treat the operators built from derivatives uniformly with the other operators we normalise the bare operators by introducing the operators 
\be \label{psi_an}
\psi_{a,n}^{0} =  \frac{1}{\sqrt{(2\Delta_{i})_{2n}}} \partial_{\tau}^{n}\psi^{0}_{a}(\tau) 
\ee
where 
\be
(a)_{n}= \frac{\Gamma(a+n)}{\Gamma(a)} 
\ee
is the Pochhammer symbol.
 The conjugate operator to $\psi_{a,n}^{0}$ is defined as 
 $$
 (\psi_{a(n)}^{0} )^{\dagger} = \eta_{a} (-1)^{n}\psi_{a(n)}^{0} 
 $$ so that 
\be
\langle \psi_{a,n}^{0}(\tau) (\psi_{b,n}^{0})^{\dagger}(0)\rangle_{0} = \delta_{ab} \, .
\ee
The corresponding renormalised operators are then linear combinations 
\be
\psi_{a,n}^{\lambda} = \sum_{b} Z^{(n)}_{ab} \mu^{1-\Delta_{b}-n}\psi_{b,n}^{0} + \sum_{J} Z_{aJ}^{(n)}\mu^{1-\Delta_{J}}\psi_{J}^{0}\, , 
\ee
\be
\psi_{I}^{\lambda} = \sum_{J} Z_{IJ}\mu^{1-\Delta_{J}}\psi_{J}^{0} + \sum_{b} Z^{(n)}_{Ib} \mu^{1-\Delta_{b}-n}\psi_{b,n}^{0} \, . 
\ee
where at the leading order in the coupling
\bea
&& Z_{IJ} = \delta_{IJ} + \lambda z_{IJ} \, , \quad Z^{(n)}_{ab} = \delta_{ab} +  \lambda z^{n}_{ab} \, , \nonumber \\
&&  Z^{(n)}_{Ib} = \lambda z^{n}_{Ib}\, , \quad Z^{(n)}_{bI} = \lambda z^{n}_{bI} \, .
\eea
We impose the Zamolodchikov's condition (\ref{Zcond}) that now reads as 
\be \label{Z1}
\langle \psi_{I}^{\lambda}(\mu^{-1})(\psi_{J}^{\lambda})^{\dagger}(0)\rangle_{\lambda} = \mu^2 \delta_{IJ}\, , \quad 
\langle \psi_{a,n}^{\lambda}(\mu^{-1})(\psi_{b,n}^{\lambda})^{\dagger}(0)\rangle_{\lambda} = \mu^2\delta_{ab} \, , 
\ee
\be \label{Z2}
\langle \psi_{I}^{\lambda}(\mu^{-1})  \psi_{a,n}^{\lambda}(0)\rangle_{\lambda} = 0 \, .
\ee
Using (\ref{2pt_bares}), (\ref{IJ}) we find that conditions (\ref{Z1}), (\ref{Z2}) are satisfied at the leading order in the coupling if we set  
$z_{IJ}$ as in (\ref{Zam_scheme_Z}) (with the replacement of indices) and 
\be \label{z111}
 z_{ab}^{n}= -\eta_{a}(-1)^{n}T_{ab}^{(n)}\frac{[y+\Delta_{a}-\Delta_{b}]}{2y} \, , \quad T_{ab}^{(n)}=(-1)^{n}T_{ab}  \frac{(\Delta_{a} + \Delta_{b} -y)_{2n}}{\sqrt{(2\Delta_{a})_{2n}(2\Delta_{b})_{2n}}} \, , 
\ee 
\be \label{z222}
z_{aI}^{n} = - \eta_{a}(-1)^{n}T_{aI}^{(n)} \frac{[y+ \Delta_{a} + n - \Delta_{I}]}{2y} \, , \quad z_{Ia}^{n} = - \eta_{a}T_{aI}^{(n)} \frac{[y+ \Delta_{I}  - \Delta_{a}-n]}{2y}
\ee
where 
\be
T_{aI}^{(n)}  = T_{aI} \frac{(\Delta_{I} + \Delta_{a} - y)_{n}}{\sqrt{(2\Delta_{a})_{2n}}} 
\ee 
and $T_{ab}$ and $T_{aI}$ are given by (\ref{IJ}). 
Assuming that $\Delta_{a}$ and all OPE coefficients have finite limits when $y\to 0$  we can expand  the above expressions in $y$. Using (\ref{IJ}),  (\ref{Zam_scheme_gamma}), (\ref{DeltaJJ}), and (\ref{DeltaiJ}) 
we obtain  the leading order  corrections to the anomalous  dimensions matrix elements
\bea
&&\gamma_{IJ} = (\Delta_{I}-1)\delta_{IJ}  -\lambda \eta_{I} (D_{IJ}^{\psi} + D_{JI}^{\psi}) \,  , \nonumber \\
&&\gamma_{a,n;\, b,n} = (n+\Delta_{a} -1)\delta_{ab} - \lambda\eta_{a} (D_{ab}^{\psi} + D_{ba}^{\psi}) 
 \, , \nonumber \\
&&\gamma_{a,n;\, I} = \gamma_{I;\, a,n}\nonumber \\
&&=-\lambda \eta_{a}(D_{aI}^{\psi} + (-1)^{n}D_{Ia}^{\psi}) 
\frac{\sqrt{(2\Delta_{a}+n)_{n}}}{n\sqrt{(2\Delta_{a})_{n}}}(y + \Delta_{I}-n - \Delta_{a})  \, .
\eea 
We see from the last expression that  $\gamma_{a,n;\, I} = {\cal O}(\lambda y)$ and at the leading order 
mixings with derivatives are negligible. For the boundary $\psi_{1,3}$ flows this happens for example for the operators 
$\psi_{r,r+3}$ that could mix with first derivatives of $\psi_{r,r+1}$ and $\psi_{r,r-1}$ but do not mix with derivatives of $\psi_{r,r}$. 
We can conclude from the above analysis that  the allowed mixings with derivatives are suppressed at least by the first order in $y$. 

The case $\Delta_{a} = {\cal O}(y)$ is special as we get additional singularities in the anomalous dimension matrices. For 
the $\psi_{1,3}$ flows $\Delta_{r,r} = {\cal O}(y^2)$ but $D_{(r,r)(1,3)}^{(r,r\pm 2)}$ vanish as ${\cal O}(y)$ so that the 
elements $\gamma_{a,n;\, b,n}$, $\gamma_{a,n;\, I}$ have a finite limit. However, the second order correction in the coupling  
taken at the fixed point $\lambda=\lambda_{*}$ also contributes at the same order. It is hard to proceed in a model independent fashion 
in this case. For the  boundary $\psi_{1,3}$ flow that starts from the Cardy $(2,2)$ boundary condition the leading order analysis of 
mixings with derivatives was done in \cite{GRW}, \cite{RG_boundary1}  via different methods. 

Note that in this approach the derivative operators are not manifest, rather we have an embedding of the form 
\be
\mu^{n}\partial^{n}_{\tau}\psi_{a}^{\lambda} = \sum_{b} {\cal M}_{ab} \psi^{\lambda}_{b,n} + \sum_{J}{\cal N}_{aJ}\psi_{J}^{\lambda}
\ee
where at the leading order 
\be
{\cal M}_{ab} = \sqrt{(2\Delta_{a})_{2n}}\left( \delta_{a,b} - \lambda\Bigl[ z_{ab}^{n} - z_{ab}\sqrt{\frac{(2\Delta_{b})_{2n}}{(2\Delta_{a})_{2n}}}\Bigr] \right)\, , 
\ee
\be
{\cal N}_{aJ} = -\lambda \sqrt{(2\Delta_{a})_{2n}} z_{a,J}^{n} \, .
\ee
There is however a different approach one could adapt to treating the derivative operators. Instead of 
introducing operators $\psi_{i,n}^{\lambda}$ which are built on derivatives of bare operators but are allowed to mix with 
non-derivative operators we can consider the operators  $\partial_{n}\psi_{i}^{\lambda}$ which are manifest derivatives. 
They can get admixed to operators $\psi_{I}^{\lambda}$ but not the other way around. We can fix a renormalisation scheme 
for $\psi_{I}^{\lambda}$ by demanding that they are orthogonal to the fields $\partial_{n}\psi_{i}^{\lambda}$. In this scheme 
the matrix of anomalous dimensions is not symmetric but the derivative fields are manifest. The leading order calculations are 
quite similar to the ones we reproduced above for the Zamolodchikov's scheme. The components of the  anomalous dimension matrix  between operators $\phi_{I}$ and the derivatives $\partial_{n}\psi_{i}^{\lambda}$ are generically ${\cal O}(y)$ unless $\Delta_{i} \sim 0$ 
which is the case when one needs the second order integrals to be taken into account. 
At the IR fixed point, in this scheme  one still should 
be able to diagonalise the matrix of anomalous dimensions and have the   derivatives of scaling fields decoupled from quasiprimaries. 
If $\Delta_{i}$ remains finite in the limit $y\to 0$ this holds at the leading order due to the mentioned suppression of mixings with derivatives. 
 For the $\psi_{1,3}$ boundary flows the analysis of derivative fields 
is complicated by the fact that the IR fixed point  in general is a superposition of irreducible components with 
 the $\psi_{r,r}$ fields becoming the identity fields on these components. Their derivatives are identically zero but with  divergent rescalings they produce reflection odd primary fields at the IR fixed point which act between different boundary irreducible components (see \cite{GRW}, \cite{RG_boundary1}). 
 It would be also interesting to analyse mixings of derivatives in a similar scheme for the bulk $\phi_{1,3}$ flows, but this is outside of the scope of this paper.

\subsection{Transport of composite operators via the interface} \label{operator_pert}


We will next calculate the expansion coefficients $C_{i}^{j}(\epsilon, \lambda)$ introduced in (\ref{expansion_hathat}) 
taking for the basis $\tilde \psi_{b}^{\lambda} $   a set of composite operators  $\psi_{i}^{\lambda}$ and their derivatives. 
We assume that the operators $\psi_{i}^{\lambda}$ are constructed from the quasiprimary fields $\psi^{0}_{i}$.
Allowing mixings with derivatives we write the most general expression for renormalised operators\footnote{Here we are not 
splitting the set for $i,j$- indices into two sets ${\cal A}$ and ${\cal B}$ as we did in section \ref{sec:derivatives} although those can be easily introduced if one wants to work out more details.}
\be \label{zij2}
\psi_{i}^{\lambda}(\tau) = \sum_{j} Z_{ij} \mu^{1-\Delta_{j}} \psi_{j}^{0}(\tau) 
+ \sum_{j}\sum_{n=1}^{\infty}Z_{ij}^{(n)} \mu^{1-\Delta_{j}-n} \psi_{j,n}^{0}(\tau)   \, .
\ee
At the leading 
order in the coupling we can write
\be \label{Zc1_2}
Z_{ij}^{(n)}=\delta_{ij}^{(n)} + \lambda z_{ij}^{(n)} \, 
\ee
where 
\be
 \delta_{ij}^{(n)} = 
 \left \{
\begin{array}{l@{\qquad}l}
\delta_{ij}   &\mbox{ if } n=0
\\[1ex]
0 &\mbox{ otherwise} 
\end{array}
\right . 
\ee
and for the time being we won't specify $z_{ij}^{(n)}$ keeping our scheme general. 
Note that at the leading order in $\lambda$ we can replace $\psi_{j,n}^{0}$ in (\ref{zij2}) by 
$[(2\Delta_{j})_{2n}]^{-1/2}\partial^{n}\psi_{j}^{\lambda}$.
The OPE  we are after can be written as 
\be \label{OPE_gen_form}
 \hat \psi^{[\lambda,0]}(\tau + \epsilon) \psi_{i}^{0}(\tau)\hat \psi^{[0,\lambda]}(\tau-\epsilon) = \sum_{j} \sum_{n=0}^{\infty} 
 C_{i}^{j(n)}(\epsilon, \lambda)  \partial_{\tau}^{n} \psi_{j}^{\lambda}(\tau)   \, .
\ee
To calculate  $C_{i}^{j(n)}(\epsilon, \lambda)$ we start by considering a 4-point function of bare operators: 
\be
\langle \psi_{k}^{0}(x)  \hat \psi^{[\lambda,0]}(\tau + \epsilon) \psi_{i}^{0}(\tau)\hat \psi^{[0,\lambda]}(\tau-\epsilon) \rangle 
= \frac{\delta_{ik}}{(x-\tau)^{2\Delta_{i}}} + \lambda_{0} \Bigl[ \int\limits_{-\infty}^{\tau-\epsilon}\!\!  dt + \int\limits_{\tau+\epsilon}^{\infty} \!\!dt\Bigr] 
\langle \psi^{0}(t)  \psi_{i}^{0}(\tau) \psi_{k}^{0}(x)\rangle + \dots
\ee
where $\epsilon \ll \tau \ll x$. The integrals in the above expression can be expressed via Gaussian hypergeometric functions:
\bea
&& \langle (\psi_{k}^{0})^{\dagger}(x)  \hat \psi^{[\lambda,0]}(\tau + \epsilon) \psi_{i}^{0}(\tau)\hat \psi^{[0,\lambda]}(\tau-\epsilon) \rangle 
= \frac{\delta_{ik}}{(x-\tau)^{2\Delta_{i}}} + \frac{\lambda_{0}\eta_{k}}{(x-\tau)^{\Delta_{i} + \Delta_{k}-y}}  \Bigl(
T_{ik} 
 \nonumber \\
&& +  \frac{1}{\alpha-1}\left(\frac{\epsilon}{x-\tau}\right)^{1-\alpha}\Bigl[
D_{ik}^{\psi}\, {}_{2}F_{1}\left(\beta,1-\alpha,2-\alpha; -\frac{\epsilon}{x-\tau}\right) \nonumber \\
&& +D_{ki}^{\psi}\, {}_{2}F_{1}\left(\beta,1-\alpha,2-\alpha; \frac{\epsilon}{x-\tau}\right)
\Bigr] \Bigr) \nonumber \\
&&
\eea
where $\alpha = \Delta + \Delta_{i} - \Delta_{k}$, $\beta=\Delta + \Delta_{k}-\Delta_{i}$. 

To extract the OPE coefficients $C_{i}^{j(n)}(\epsilon, \lambda)$ we contract the above expression and its normalised derivatives with respect to $x$ with the matrices 
$$
Z_{jk}^{(n)}\mu^{1-\Delta_{k}}=(\delta_{jk}^{(n)} + \lambda z_{jk}^{(n)})\mu^{1-\Delta_{k}-n} \, , 
$$
 set $\lambda_{0}=\lambda \mu^{y}$, and expand in  powers of $\frac{\epsilon}{(x-\tau)}$. 
Amputating the two-point functions (\ref{2pt_ren_ij}) and its derivatives we obtain from that expansion 
\be \label{CvsC}
C_{i}^{j(n)}(\epsilon, \lambda) =( \delta_{ij}^{(n)} + \lambda \tilde C_{i}^{j(n)}(\epsilon)) \mu^{\Delta_{i} -1-n} 
\ee
where 
\be \label{Cij0_gen}
\tilde C_{i}^{j(0)}(\epsilon) = -z_{ij} - \frac{\eta_{j}(D_{ij}^{\psi}+D_{ji}^{\psi}) (\epsilon \mu)^{1-\Delta - \Delta_{i}+ \Delta_{j}}}{1-\Delta - \Delta_{i} + \Delta_{j}} \, , 
\ee
\be \label{Cijn_odd}
\tilde C_{i}^{j(n)}(\epsilon) =-\frac{z_{ij}^{(n)}}{\sqrt{(2\Delta_{j})_{2n}}} + (\epsilon \mu)^{y+\Delta_{j}-\Delta_{i}+n} 
\frac{\eta_{j}(D_{\psi i}^{j(n)}-D_{i\psi}^{j (n)}) }{ 1-\Delta-\Delta_{i} + \Delta_{j} + n}
\ee
when 
$n$ is odd, and 
\be \label{Cijn_even}
\tilde C_{i}^{j(n)}(\epsilon) = - \frac{z_{ij}^{(n)}}{\sqrt{(2\Delta_{j})_{2n}}} - (\epsilon \mu)^{y+\Delta_{j}-\Delta_{i}+n} 
\frac{\eta_{j}(D_{\psi i}^{j(n)}+D_{i\psi}^{j(n)}) }{1-\Delta-\Delta_{i} + \Delta_{j} + n}
\ee
when $n$ is even. Here $D_{\psi i}^{j(n)}$ and $D_{i \psi }^{j(n)}$ are the OPE coefficients of derivatives (\ref{derivative_OPE}).
We note that the above expressions are valid at the leading order in the coupling.
In particular they apply in the case when the dimensions $\Delta_{i}$ and $\Delta_{j}$ are not close, as would be the case 
e.g. for $\psi_{1,3}$ flows whith $\psi_{i}^{0}=\psi_{1,3}$ and $\psi_{j}^{0}=\psi_{1,5}$. 
We note that when $\eta_{i}=\eta_{j}$ only even order derivatives are present while when $\eta_{i}=-\eta_{j}$ only 
odd order derivatives. 

We note that the expression for the non-derivative coefficients (\ref{Cij0_gen}) has a pole for quasiprimary operators of nearby 
conformal dimension, that is when  $\Delta_{i}-\Delta_{j} = {\cal O}(y)$. To have a scheme in which the operators 
are well behaved at the IR fixed point we need to subtract such poles by a suitable choice of $z_{ij}$.   
Zamolodchikov's scheme (\ref{Zam_scheme_Z}) subtracts  the pole as well as ensures that the anomalous dimension 
matrix $\gamma_{ij}$ is symmetric. Substituting (\ref{Zam_scheme_Z}) into (\ref{Cij0_gen}) we obtain at the leading order in $y$
\be \label{C0_Zam}
\tilde C_{i}^{j(0)}(\epsilon) = \eta_{j}(D_{ij}^{\psi} + D_{ji}^{\psi}) \frac{1-(\epsilon \mu)^{1-\Delta - \Delta_{i}+\Delta_{j}}}{1-\Delta - \Delta_{i}+\Delta_{j}} \, .
\ee
In the case when the difference $\Delta_{i}-\Delta_{j}$ does not go to zero when $\Delta \to 1$ 
we choose $z_{ij}=0$ in keeping with the rule that only operators of nearby dimensions mix. 

There are 3 special cases of OPE coefficients in (\ref{OPE_gen_form}) where the answer is particularly simple. 
In these cases $z_{ij}^{(n)}=0$ for $n>0$. 

\noindent {{\bf 1)} $ \psi_{i}^{0}=\psi^{0}$, $\psi_{j}^{\lambda}=1\, .$}  \\
In this case there are no derivative terms and we get from (\ref{C0_Zam}) in the leading order in $1-\Delta$:
\be
C_{\psi, 1}^{1} = \frac{2\lambda} {\epsilon \mu} 
\ee
which is consistent with $C_{\rm uv}=\lambda_{*}$ that we found in section \ref{section_RGop}.

\noindent {{\bf 2)} $\psi_{i}^{0}=1$, $\psi_{j}^{\lambda}=\psi^{\lambda}$\, .}  \\
In this case we get the following terms in the OPE
\be
\hat \psi^{[\lambda, 0]}(\epsilon)  \hat \psi^{[0,\lambda]} (-\epsilon) = \sum_{k=0}^{\infty} -\frac{2\lambda}{(2k+1)!}\epsilon^{2k+1}
\partial_{\tau}^{2n}\psi^{\lambda}(0) + \dots \, .
\ee
If we set here $\lambda=\lambda_{*}$ and $\psi^{\lambda} = \psi^{\rm ir}$ we obtain the leading term with 
$C_{\rm ir} = -\lambda_{*}$ (as found before) plus the derivative contributions with coefficients  matching those expected from the conformal symmetry.  

\noindent {{\bf 2)} $\psi_{i}^{0}=\psi^{0}$, $\psi_{j}^{\lambda}=\psi^{\lambda}$\, .}  \\

\bea \label{psi_to_psi}
&&\hat \psi^{[\lambda, 0]}(\epsilon) \psi^{0}(0) \hat \psi^{[0,\lambda]} (-\epsilon) = \mu^{y}[1 + 2\lambda D\frac{(1-(\epsilon \mu)^{y})}{y}]\psi^{\lambda}(0)  \nonumber \\
&&  -\lambda \mu^{y}  D  (\epsilon\mu)^{y}      \sum_{k=0}^{\infty} \frac{\epsilon^{2k}}{k(2k+1)!} \partial_{\tau}^{2k}\psi^{\lambda}(0) + \dots 
\eea
where the omitted terms contain contributions from other quasiprimaries. Setting here $\lambda=\lambda_{*}$ and 
$\psi^{\lambda}=\psi^{\rm ir}$ we obtain an expansion that at the leading order in $1-\Delta$ should match with what is expected from 
conformal symmetry. We checked that it does indeed match.  As this involves studying the relevant  four point function the details are a bit tedious and we omit them.

As discussed in section \ref{sec_op_map} to obtain a renormalised operator $[\psi_{i}]^{\lambda}$ from a seed operator $\psi_{i}^{0}$ 
we need to take the limit $\epsilon \to 0$ in (\ref{OPE_gen_form}) and subtract the divergences. 
The divergent terms come from the operators $\partial_{\tau}^{n}\psi^{\lambda}_{j}$ for which 
\be \label{div_cond}
1-\Delta + \Delta_{j} - \Delta_{i} < -n 
\ee
and $D_{ij}^{\psi}+D_{ji}^{\psi}\ne 0$. (Recall that we assume there are no resonances hence no equality sign in (\ref{div_cond}).)
This inequality can hold whenever $\Delta_{i}$ is sufficiently larger than $\Delta_{j}$. It can also hold for two operators of 
nearby dimension if  $n=0$ and $\psi_{j}^{0}$ 
is a more relevant operator than $\psi_{i}^{0}$ with dimensions satisfying\footnote{In the explicit example 
of $\psi_{1,3}$ boundary flows this happens for e.g. mixings between $\psi_{r,r+1}$ and $\psi_{r,r-1}$ operators. }: 
$\Delta_{j} - \Delta_{i} < -y$. Subtracting the power divergences and then taking $\epsilon$ to zero is equivalent to simply dropping 
the $\epsilon$-dependent terms in (\ref{Cij0_gen}), (\ref{Cijn_odd}), (\ref{Cijn_even}). In such a minimal subtraction scheme however 
the operators are not adapted to a perturbative treatment in $y$  that is signified by the poles in $y$ which are still present in correlation functions.
If we choose to add to $\psi^{0}_{i}$ exactly the same finite counterterms  as in the scheme defining the operators $\psi^{\lambda}_{i}$ (and include the relevant powers of $\mu$) we would recover the operators $\psi^{\lambda}_{i}$ themselves that is 
\be \label{ren_ren}
[\psi_{i}]^{\lambda} = \psi_{i}^{\lambda} \, 
\ee
at the leading order in $\lambda$.
At higher orders  the $\hat Z$ factor will be important and will contribute to the counterterms to be added 
if one wants (\ref{ren_ren}) to hold for a chosen basis $\psi_{i}^{\lambda}$ e.g. for the one specified by the Zamolodchikov's conditions.

 We can also calculate the operator product expansion for derivatives of UV operators sandwiched between two interface operators. This can be done either by calculating the integral of the relevant three-point function or by differentiating the OPE in (\ref{OPE_gen_form}). 
The details are a bit tedious and we omit them here. 

\subsection{Gaiotto's pairing and scaling eigenvectors} \label{pairing_sec}

We can use our leading order calculations of the OPE coefficients $C_{i}^{j(n)}(\epsilon, \lambda)$ from the 
previous section to obtain the leading order Gaiotto's pairing coefficients. We discussed the latter in section 
\ref{sec_op_map}. Although they were defined in \cite{Gaiotto} for the case of bulk RG interface they can be 
easily generalised to the case of boundary flows. For the case of $\psi_{1,3}$ boundary flows in the minimal models 
the Gaiotto's pairings have been previously considered in \cite{RG_boundary1}.   
Here we consider flows between nearby fixed points assuming  for simplicity that no mixings with derivatives are present as in section  \ref{sec_nonderiv}.   The Gaiotto's 
pairing coefficients are then given as\footnote{Here we choose to keep $\mu$ generic rather than setting it equal to 1 as we 
did in section \ref{sec_op_map}.} 
 \be
 A_{i}^{p}= \sum_{j}\mu^{1-\Delta_{i}}C_{i}^{j}(\mu^{-1},\lambda_{*})\xi^{p}_{j} 
 \ee
 where $\xi^{p}_{j}$ are the normalised eigenvectors of the anomalous dimension matrix $\gamma^{*}_{ij}$ which we call 
 following \cite{Gaiotto} the RG mixing coefficients.
In the Zamolodchikov's scheme at the leading order in $y$ the coefficients $C_{i}^{j}$ are given in   (\ref{CvsC}), (\ref{C0_Zam}) and we see 
substituting $\epsilon=\mu^{-1}$ that 
\be \label{new_gauge_cond}
C_{i}^{j}(\mu^{-1},\lambda_{*})  = \delta_{i}^{j} \mu^{\Delta_{i}-1}
\ee
and thus 
\be \label{A}
 A_{i}^{p} = \xi^{p}_{i} \, .
\ee
This means that at the leading order in $y$, in the Zamolodchikov scheme, the Gaiotto's pairing coefficients 
(taken for the mixing fields) match with the RG mixing coefficients. This was demonstrated in \cite{Gaiotto} for the 
bulk $\phi_{1,3}$ flows and the proposed RG interface constructed algebraically. The main point in that calculation was a check 
of that algebraic construction.
 Here we are interested in understanding better the correspondence between the pairings and the RG mixing coefficients. 
 Formula (\ref{A})  shows that  both quantities match  in the leading order using Zamolodchikov's scheme  for any flow between  nearby fixed points. We have  also checked that an analogue of  (\ref{A}) holds at the leading order for  the case of mixings with first order derivative operators considered in section \ref{sec:derivatives}.
 
  Looking at formulae (\ref{CvsC}),    (\ref{Cij0_gen}) which are written for a general scheme we see that we can take the condition 
  (\ref{new_gauge_cond}), where the indices $i,j$ are those of operators of nearby dimension, as an alternative to the Zamolodchikov's 
  conditions (\ref{Zcond}). It may happen that at higher orders this scheme differs from Zamolodchikov's scheme. In particular the matrix $\gamma_{ij}$ may not be  symmetric but at the IR fixed point 
  it must be nonetheless diagonalisable by some linear transformation which is not necessarily orthogonal. This is because RG schemes differ by a coordinate transformation at least perturbatively, order by order in the coupling.  In the scheme that adopts (\ref{new_gauge_cond}) it is guaranteed though that the RG mixing coefficients are given by Gaiotto's pairing. Thus, at least in principle such a scheme should always exist  for nearby fixed points.

\section{Variational method} \label{sec:VAR}
\setcounter{equation}{0}
\subsection{Cardy's variational ansatz for bulk massive flows}
In this section we will introduce a variational method related to RG interfaces. The idea is quite similar to the variational 
ansatz introduced in \cite{Cardy_var} so we start with a brief review of that construction. 

Let us consider perturbations of 2D bulk CFTs by terms of the form
\be \label{bulk_S}
\Delta S =-\sum_{i} \lambda^{i} \int\! \phi_{i}(x_1,x_2) d^{2}x 
\ee
where  $\phi_{i}$ are bulk relevant operators with scaling dimensions $\Delta_{i}$ and the correlation functions are deformed as in\footnote{We put a minus sign in front of 
(\ref{bulk_S}) to match the conventions in \cite{Cardy_var}.}  (\ref{bare_ps}). 
Putting the perturbed theory on a cylinder with circumference $R$, the compact coordinate $\sigma$, and the Euclidean time coordinate $\tau$,  we can write the perturbed Hamiltonian as  
\be
H=H_{0} + \sum_{i}\lambda_{i} \int\!  \phi_{i}(0,\sigma) d\sigma 
\ee
where $H_{0}$ is the UV CFT Hamiltonian.  

For massive  flows in the 2D bulk theories only the vacuum state or a finite dimensional space of degenerate vacua survive 
in the far infrared. The RG interface in this case is a conformal boundary condition and the vacuum state is given by 
a boundary state in the UV CFT. For degenerate vacua we have a superposition (direct sum) of irreducible boundary conditions. 
Any conformal boundary state $|B\rangle\!\rangle$ has an infinite norm. However for the perturbed theories (\ref{bulk_S}), before they 
reach the trivial fixed point in the far infrared,  
the vacuum $|0\rangle_{\lambda}$ may have a finite norm. We are not aware of any general result which would tell us when this is the case, but certainly for a number of concrete examples (e.g.  the massive free fermion) that is the case. 
Assuming that the norm of the perturbed vacuum, that is of its image in the UV CFT state space, is finite we can use a variational method to approximate it. Using  that in the far infrared the vacuum is given by some conformal boundary state $|B\rangle\!\rangle$ 
Cardy proposed the following simple ansatz for a variational trial state 
\be \label{C_ansatz}
|\tau, B\rangle = e^{-\tau H_{0}} |B\rangle\!\rangle \, .
\ee
This state has a finite norm which is given by the square root of the partition function evaluated on a finite cylinder of length $2\tau$ 
with the boundary conditions specified by  $|B\rangle\!\rangle$  at both ends. We can find an approximation to the vacuum state by 
minimising over $\tau$ and over the choices of $|B\rangle\!\rangle$ the variational energy average
\be
E(\tau) = \frac{1}{\langle \tau,B |\tau,B\rangle} \langle \tau, B| \left[ H_{0} + \sum_{i}\lambda_{i} \int\!  \phi_{i}(0,\sigma) d\sigma  \right]|\tau, B\rangle 
\, .
\ee
This expression simplifies drastically when $\tau \ll R$. As $\tau$ measures how close we are to the IR fixed point we can choose to 
satisfy this condition if we are after describing the vicinity of the IR fixed point. In this limit we can substitute the cylinder by an infinite 
strip and obtain\footnote{Repeating the calculations from \cite{Cardy_var} we obtained a slightly different answer. In the second term in 
\cite{Cardy_var} there is $2\tau$ standing in the denominator rather than $4\tau$ that we obtained (see formula (35) of that paper).} 
\be \label{bulk_var}
E_{B}(\tau)= R\Bigl[  \frac{\pi c}{24(2\tau)^2} + \sum_{i} \lambda_{i} A^{B}_{i} \left(\frac{\pi}{4\tau}\right)^{\Delta_{i}}    \Bigr]
\ee
where $c$ is the UV CFT central charge and $A^{B}_{i}$ are the values of the disc one-point functions for the operator $\phi_{i}$ inserted at the centre and the boundary condition  $|B\rangle\!\rangle$ put on the boundary. If one knows all conformal boundary states in the UV CFT one may minimise the energies (\ref{bulk_var}) in $\tau$ and then choose the smallest value among all conformal 
boundary states. The overlaps between the trial states $|\tau, B\rangle$ with different boundary conditions are suppressed exponentially when $\tau \ll R$ so that one does not need to consider their superpositions unless the vacuum becomes degenerate (see \cite{LVT} for a nice discussion of the off-diagonal terms in Cardy's ansatz). 

The minimum of $E_{B}(\tau)$ can be negative provided $\lambda^{i}A^{B}_{i} <0$ for at least one of the perturbing operators and 
it is zero if this is not the case. In the latter case the minimum lies at $\tau=\infty$ where strictly speaking we would not expect 
our simple ansatz  to be valid.  
The negative minimum always exists if $\lambda^{i}A^{B}_{i} <0$ for the most relevant of the perturbing operators. 
Considering for simplicity  a single coupling $\lambda$ it is easy to see that at the minimum of variational energy the value of $\tau$ equals
\be \tau_{\rm min} \sim  |\lambda|^{\frac{1}{\Delta-2}}
\ee
 so that in the far infrared where $\lambda \to \infty$ the parameter $\tau_{\rm min}\to 0$ as expected.

 In \cite{Cardy_var} the ansatz (\ref{C_ansatz}) was applied to massive  RG flows originating from $A$-series unitary minimal models. 
 While the method gives reasonable results for single coupling flows (in particular describing vacuum degenerations) some problems were noted in the case of several couplings, in particular for the critical Ising model perturbed by both the thermal and the magnetic operators. The difficulties might be coming from the fact that the ansatz (\ref{C_ansatz}) captures well only low weight components 
 of the vacuum vector. Namely, for the image of the true vacuum state $|0\rangle_{\lambda}$ in the CFT UV state space we can consider overlaps $ \langle \Delta |0\rangle_{\lambda}  $ 
 with normalised states of conformal weight $\Delta$. 
Such overlaps  should be approximated  well by (\ref{C_ansatz}) when $\frac{\Delta}{R}\ll   |\lambda|^{\frac{1}{2-\Delta}}$ 
with the approximate behaviour 
\be
 \langle \Delta | 0\rangle_{\lambda}  \sim e^{- \tau_{\rm \min} \Delta/R}  
\ee
while for very high weight components when $\frac{\Delta}{R}\gg   |\lambda|^{\frac{1}{2-\Delta}}$  we expect a different  behaviour 
 as it should be governed by the UV perturbation theory. For example for the mass perturbation of the massless 
fermions in the NS sector (Ising model in the disordered phase) the exact vacuum state written in terms of fermion creation operators is
\be
|0\rangle_{m} = {\cal N} {\rm exp}\left(\frac{i}{m}\sum_{n=0}^{\infty} (\omega_{n}-\frac{2\pi |n+1/2|}{R} )a_{n+1/2}^{\dagger}\bar a_{n+1/2}^{\dagger}\right)  |0\rangle
\ee
where 
\be
\omega_{n} = \sqrt{m^2 + \left( \frac{2\pi(n+1/2)}{R}\right)^2   } \, ,
\ee
${\cal N}={\cal N}(mR)$ is the normalisation factor, and  $m<0$ is the mass coupling. When $n \ll |m| R$  the corresponding components in $|0\rangle_{m}$ can be approximated by 
\bea \label{low_tail}
&& {\cal N} {\rm exp}\left(-i\sum_{n=0}^{\infty} (1-\frac{2\pi |n+1/2|}{R|m|} )a_{n+1/2}^{\dagger}\bar a_{n+1/2}^{\dagger}\right)  |0\rangle 
\nonumber \\
&& \approx {\cal N} e^{-\frac{\pi}{R|m|}(L_{0} + \bar L_{0})}  {\rm exp}\left(-i\sum_{n=0}^{\infty}a_{n+1/2}^{\dagger}\bar a_{n+1/2}^{\dagger}\right)  |0\rangle 
\eea
 that up to normalisation is  of the form (\ref{C_ansatz}) with $|B\rangle\!\rangle$ being the free fermion representation of the  free spin conformal boundary condition 
and $\tau= \frac{1}{2|m|}$. When $n \gg |m| R$ we obtain instead 
\be
{\cal N} {\rm exp}\left(-i\sum_{n=0}^{\infty} \frac{|m|R}{4\pi (n+1/2)} a_{n+1/2}^{\dagger}\bar a_{n+1/2}^{\dagger}\right)  |0\rangle 
\ee
where the coupling and the weight dependences are essentially flipped. 

To summarise we see two types of discrepancies between the actual vacuum state and a trial state of the form  (\ref{C_ansatz}). 
In the low weight part of the vacuum vector while the shifted conformal boundary state dominates at large coupling there are 
corrections (they are related to approximations made to derive the leading asymptotic behaviour  in  (\ref{low_tail})) 
which presumably can be written as local perturbations of the boundary state. In the high weight tail of the vacuum the 
behaviour is essentially different from the one in  (\ref{C_ansatz}). On one hand this can be explained by  the emergence of the UV fixed point behaviour in the true vacuum which we do not put into the ansatz. On the other hand, perhaps more intuitively,  from the point of view of the high frequency modes in the UV theory the interface is not a boundary but a permeable object they can pass through. 
 It would be interesting to understand when any of these corrections are important and whether they can help 
 resolving the issues with applying the ansatz to two coupling flows in the Ising model which were noted in \cite{Cardy_var}.

\subsection{Variational ansatz for boundary flows. Perturbation on one end.}  \label{boundary_var_main}

While for the boundary flows we always flow to a non-trivial fixed point in which at least the identity tower of states is always present, 
the variational principle can still be applied to the vacuum state. As discussed in section \ref{section_states}  in the far infrared the vacuum state of the perturbed theory is represented 
by a non-normalisable state constructed using the RG operator 
\be
\hat \psi^{[\rm uv, ir]}(0)|0\rangle_{\rm ir} 
\ee
where $|0\rangle_{\rm ir}$ is the vacuum in the IR BCFT (assuming for simplicity it is non-degenerate). 
Similarly to (\ref{C_ansatz}) we can modify this state by shifting the insertion of the RG operator. Define 
\be \label{trial_b}
|\tau, \hat \psi\rangle = \hat \psi^{[\rm uv, ir]}(-\tau)|0\rangle_{\rm ir} \, , \enspace \tau>0 \, .
\ee
This state has a finite norm squared given by the two-point function 
\be
\langle \tau, \hat \psi  |\tau, \hat \psi\rangle  ={}_{\rm ir}\langle 0|\hat \psi^{[\rm ir, uv]}(\tau) \hat \psi^{[\rm uv, ir]}(-\tau)|0\rangle_{\rm ir}
\ee
We propose using states $|\tau, \hat \psi\rangle$ as variational trial states. They are labeled by a choice of IR BCFT, a choice of 
RG operator $\hat \psi$ that is a boundary condition changing operator linking the UV and IR BCFTs, and by a continuous parameter $\tau$. In section \ref{section_states} we gave a general argument based on the Callan-Symanzik equation for the RG operator two-point function that in the boundary case   the vacuum has a finite norm for relevant and marginally-relevant perturbations. This puts the use of the variational method on a firmer basis than in the  bulk case. 

To work  out the variational energy we put our theory on a strip of width $L$ as discussed in section \ref{section_states}. 
We denote by $\sigma \in [0,L]$ the coordinate across the strip and by $\tau \in {\mathbb R}$ the coordinate along the strip. 
There are two different set-ups. We can fix the boundary condition on the top edge of the strip to be some conformal boundary condition 
and consider the perturbed boundary condition on the bottom edge. Alternatively we can consider the perturbed boundary condition on both edges. The latter setup is better adopted to studying degenerations of vacua which can emerge at the IR fixed point. However 
in both cases the final equations are quite similar and we start by discussing the first set-up in which the derivation of the variational 
energy is a bit simpler. We thus fix the spectator boundary condition on the top edge throughout the discussion in this subsection. In the case of unitary rational models it is convenient to choose the Cardy boundary condition labeled by the identity state but for now we will keep the discussion general. 

We write the perturbed Hamiltonian on the strip as 
\be \label{H_pert2}
H = H_{0} - \sum_{i}\lambda^{i}\psi_{i}^{\rm uv}(0) \quad \mbox{where}\quad   H_{0} = \frac{\pi}{L}(L_{0}^{\rm uv}-\frac{c}{24}) \, .
\ee
For simplicity we assume that there are no logarithmic divergences present. The coupling constant $\lambda$ and the parameter $\tau$ are dimensionful. 
 As all insertions take place at the bottom edge where $\sigma=0$ we will suppress the sigma coordinate in operator positions.
We assume next that we have chosen the tentative end point of the flow triggered by the perturbation in (\ref{H_pert2}) and will 
refer to it as the tentative
IR BCFT,  sometimes dropping for brevity the ``tentative''.
Denote the vacuum state on the strip with the tentative IR BCFT  on the bottom edge by $|0\rangle_{\rm ir}$. It is given by the 
lowest weight boundary condition changing operator linking the IR BCFT and the spectator BCFT. 
Let $\Delta_{0}^{\rm ir}$ be its conformal weight. We further choose a tentative RG operator $\hat \psi^{[\rm uv,\rm ir]}$ about which 
at this stage we will only assume that it is a scaling operator of dimension $\hat \Delta$. This assumption is easy to justify. 
Close to the IR fixed point the interface operator $\hat \psi^{[0,\lambda]}$ can be expanded in scaling boundary condition changing operators. Continuing the RG flow we perform dilations on this superposition and the scaling operator of the smallest dimension present dominates  becoming the RG operator in the far infrared. With the choices made we consider the trial state given in (\ref{trial_b}). 
The variational energy 
\be
E_{\hat \psi}(\tau) = \frac{\langle \tau, \hat \psi| H|\tau, \hat \psi\rangle}{\langle \tau, \hat \psi| \tau, \hat \psi\rangle}
\ee
has two contributions: from $H_{0}$ and from the perturbation. For the first one we have 
\bea \label{term1_var}
&& \frac{\langle \tau, \hat \psi| H_{0}|\tau, \hat \psi\rangle}{\langle \tau, \hat \psi| \tau, \hat \psi\rangle} = 
\left(\frac{\pi}{L}\right)\frac{ {}_{\rm ir}\langle 0|\hat \psi^{[\rm ir, uv]}(\tau) L_{0}^{\rm uv}\hat \psi^{[\rm uv, ir]}(-\tau)|0\rangle_{\rm ir}     }{{}_{\rm ir}\langle 0|\hat \psi^{[\rm ir, uv]}(\tau) \hat \psi^{[\rm uv, ir]}(-\tau)|0\rangle_{\rm ir}} 
\nonumber \\
&&= \frac{ {}_{\rm ir}\langle 0|\hat \psi^{[\rm ir, uv]}(\tau) [(\pi/L)\Delta_{0}^{\rm ir} - \partial_{\tau}]\hat \psi^{[\rm uv, ir]}(-\tau)|0\rangle_{\rm ir}     }{{}_{\rm ir}\langle 0|\hat \psi^{[\rm ir, uv]}(\tau) \hat \psi^{[\rm uv, ir]}(-\tau)|0\rangle_{\rm ir}} 
\eea
where we used 
\be
\frac{\pi}{L}\left( L_{0}^{\rm uv}\hat \psi^{[\rm uv, ir]}(-\tau) - \hat \psi^{[\rm uv, ir]}(-\tau) L_{0}^{\rm ir}\right)  = -\partial_{\tau}\hat \psi^{[\rm uv, ir]}(-\tau) \, .
\ee
For $\tau \ll L$, to extract the leading behaviour  in (\ref{term1_var})  we can use the flat space OPE\footnote{Note that we are not mapping the operators to the half plane but do the OPE directly on the strip.} 
\be  \label{flat_OPE}
\hat \psi^{[\rm ir, uv]}(\tau) \hat \psi^{[\rm uv, ir]}(-\tau) = \frac{D_{\hat \psi \hat \psi}^{1}}{(2\tau)^{2\hat \Delta}} + 
 \sum_{a} \frac{D_{\hat \psi \hat \psi}^{a}}{(2\tau)^{2\hat \Delta -\Delta_{a}}}  \psi_{a}^{\rm ir}(0)
\ee 
where $\psi_{a}^{\rm ir}$ are  scaling operators in the IR BCFT of positive dimension. 
The leading contribution in (\ref{term1_var}) comes from the identity term in (\ref{flat_OPE}) and is readily evaluated to be 
\be \label{first_contr}
\frac{\langle \tau, \hat \psi| H_{0}|\tau, \hat \psi\rangle}{\langle \tau, \hat \psi| \tau, \hat \psi\rangle}  \sim \frac{\hat \Delta}{\tau} + \frac{\pi \Delta_{0}^{\rm ir}}{L} 
\ee
where strictly speaking the second term is suppressed by a factor of  $\tau/L$ but we keep it as it is important when the IR vacua become degenerate. The other subleading terms that come from the less singular terms  in the OPE (\ref{flat_OPE}) are suppressed by additional powers of $\tau/L$. 

  Similarly, the contribution coming from the interaction term 
  \be
 -\sum_{i}\lambda_{i}  \frac{\langle \tau, \hat \psi| \psi_{i}^{\rm uv}(0) |\tau, \hat \psi\rangle}{\langle \tau, \hat \psi| \tau, \hat \psi\rangle} = 
 -\sum_{i}\lambda_{i}   \frac{ {}_{\rm ir}\langle 0|\hat \psi^{[\rm ir, uv]}(\tau) \psi_{i}^{\rm uv}(0)\hat \psi^{[\rm uv, ir]}(-\tau)|0\rangle_{\rm ir}     }{{}_{\rm ir}\langle 0|\hat \psi^{[\rm ir, uv]}(\tau) \hat \psi^{[\rm uv, ir]}(-\tau)|0\rangle_{\rm ir}}  
  \ee
can be expanded using the triple OPE expansion 
\be \label{triple_flat_OPE}
\hat \psi^{[\rm ir, uv]}(\tau) \psi_{i}^{\rm uv}(0)\hat \psi^{[\rm uv, ir]}(-\tau) =  \frac{ D_{\hat \psi i \hat \psi}   4^{\Delta_{i}^{\rm uv}}}{(2\tau)^{2\hat \Delta+ \Delta_{i}^{\rm uv} }} + \sum_{a}\frac{D_{\hat \psi i \hat \psi}^{a}}{(2\tau)^{2\hat \Delta + \Delta_{i}^{\rm uv} - 
\Delta_{a}^{\rm ir}}} \psi_{a}^{\rm ir}(0)
\ee
where the coefficient $D_{\hat \psi i \hat \psi}$ that stands at the identity operator can be obtained from the three point function on the half plane: 
\be
\langle   \hat \psi^{[\rm ir, uv]}(\tau) \psi_{i}^{\rm uv}(0)\hat \psi^{[\rm uv, ir]}(-\tau)  \rangle  = \frac{ D_{\hat \psi i \hat \psi}   4^{\Delta_{i}^{\rm uv}}}{(2\tau)^{2\hat \Delta+ \Delta_{i}^{\rm uv} }}  \, .
\ee
Here we normalised the coefficient in such a way that when $\hat \psi^{[\rm ir, uv]}$ is a quasiprimary field we have 
\be \label{DDD}
D_{\hat \psi i \hat \psi} = D_{i \hat \psi}^{\hat \psi} D_{\hat \psi \hat \psi}^{1} 
\ee
where  $D_{i \hat \psi}^{\hat \psi}$  is the OPE coefficient 
in the expansion  
\be \label{needed_OPE}
 \psi_{i}^{\rm uv}(0)\hat \psi^{[\rm uv, ir]}(-\tau) = \frac{D_{i \hat \psi}^{\hat \psi}}{\tau^{\Delta_{i}^{\rm uv}}} \hat \psi^{[\rm uv, ir]}(-\tau) + \dots 
\ee 
From now on we will assume that $\hat \psi^{[\rm ir, uv]}$ is a quasiprimary and thus formula (\ref{DDD}) applies.   
Using (\ref{triple_flat_OPE}) we obtain the leading behaviour 
\be \label{second_contr}
 -\sum_{i}\lambda_{i}  \frac{\langle \tau, \hat \psi| \psi_{i}^{\rm uv}(0) |\tau, \hat \psi\rangle}{\langle \tau, \hat \psi| \tau, \hat \psi\rangle}  
 \sim -\sum_{i}\lambda_{i} D_{i \hat \psi}^{\hat \psi} 
 \left(\frac{2}{\tau}\right)^{\Delta_{i}^{\rm uv}} \, .
\ee
The corrections to this leading behaviour come from the operators of positive dimension in (\ref{triple_flat_OPE}). Their one-point functions go as negative  powers of $L$  and hence these corrections are 
suppressed by powers of $\tau/L$. Combining (\ref{first_contr})  and (\ref{second_contr}) we obtain the leading order expression for the variational energy 
\be \label{var_boundary_1end}
E_{\hat \psi}(\tau) =  \frac{\pi \Delta_{0}^{\rm ir}}{L}   + \frac{\hat \Delta}{\tau} -\sum_{i}\lambda_{i} D_{i \hat \psi}^{\hat \psi} 
 \left(\frac{2}{\tau}\right)^{\Delta_{i}^{\rm uv}} \, .
\ee
Several comments are now in order about formula (\ref{var_boundary_1end}).

We note that the two $\tau$-dependent terms here have a great similarity to Cardy's variational energy (\ref{bulk_var}) 
with the central charge in the interaction independent term replaced by the dimension of RG operator and the one-point 
functions in the interaction term replaced by the OPE coefficient (\ref{needed_OPE}). (The signs in front of the interaction terms differ due to different conventions.) Moreover, for the Virasoro minimal models the similarity goes even further as the OPE coefficients at hand  can be expressed in 
terms of Virasoro fusion matrices \cite{Runkel} which satisfy the orthogonality relations. For the minimal models the bulk one-point functions can be expressed via the modular $S$-matrix entries which also satisfy   orthogonality relations. The latter  were used  in \cite{Cardy_var} to demonstrate some general properties of the proposed variational ansatz.

 The OPE coefficient $D_{i \hat \psi}^{\hat \psi} $ in standard normalisation of the operators involved equals 
the OPE coefficient $C_{\rm uv}$ from the second conjecture that we introduced in the introduction (see formula (\ref{OPE1})). Thus, to obtain a negative 
variational energy at finite value of $\tau$ we need that coefficient to be non-vanishing at least 
for one of the perturbing operators. This supports our second conjecture from the introduction. 

The leading term expression 
(\ref{var_boundary_1end}) remains a good approximation provided that the value of $\tau$ at the minimum: $\tau_{\rm min}\ll L$. 
As $\tau_{\rm min}$ depends only on the couplings $\lambda_{i}$ we can always assume they are large enough so that this condition is satisfied. The $L$-independent terms in (\ref{var_boundary_1end}) give the flat space part of the finite size ground state energy. 
It dominates in the $L\to \infty$ limit and is independent of the choice of the spectator boundary condition.

Given the trial states $|\tau, \hat \psi\rangle$ it is natural to ask whether their linear combinations may provide better 
approximations of the vacuum. To gauge the importance of such combinations we need to consider 
 off-diagonal matrix elements of $H$ taken between the states of the form $|\tau, \hat \psi \rangle$ 
taken  for 
different IR BCFTs or for different RG operators $\hat \psi$ with the same IR BCFT\footnote{Potentially one could even look at 
different $\tau$'s used in each state}.  
To estimate such matrix elements we can use the OPE of the operators involved. For  $\tau/L\ll 1$ the leading contributions, as 
in our analysis of the diagonal terms,  
will come from the identity terms in the OPEs involved as the non-identity terms have one-point functions suppressed by 
inverse powers of $L$. For different RG operators their product does not contain the identity so there are no identity contributions from 
the $H_{0}$ part of the Hamiltonian. Also for different IR BCFTs there is no identity field available. 
The identity fields may appear in the interaction term when we have the same IR BCFT but 
different RG operators.
   We will discuss an example of such situation in section \ref{sec:offdiag}. In this case we get an $L$-independent contribution 
   to the off-diagonal terms and the corresponding matrix needs to be diagonalised. It should be also noted that the finite size suppressed off-diagonal terms may become important when the diagonal size-independent terms become equal.

The constant term in (\ref{var_boundary_1end}) which can be ignored for generic flows becomes important 
for flows leading to degenerate vacua. Consider for simplicity the case when the flow ends up in a superposition 
of two irreducible and different boundary conditions which we label as  $b_{1}$ and $b_{2}$. In section \ref{tricrit_sec} we will 
consider an explicit example of such a boundary flow in the tricritical Ising model. Such a degeneracy typically emerges due to 
a symmetry that gives equal OPE coefficients $D_{i \hat \psi}^{\hat \psi} $ and the dimensions $\hat \Delta$. 
In this case the dominant contribution to the variational energy that comes from minimising the non-constant  
terms in (\ref{var_boundary_1end}) is exactly the same for each of the trial states 
\be 
\hat \psi^{[{\rm uv}, b1]}(-\tau)|0\rangle_{b1}\, , \qquad \hat \psi^{[{\rm uv}, b2]}(-\tau)|0\rangle_{b2} \, .
\ee
The subleading constant term thus becomes important and describes the spontaneous symmetry breaking 
due to the presence of the spectator boundary condition. In particular it ensures that if the spectator is $b_{1}$ the 
first of the above trial states has lower energy because $\Delta^{\rm b_1}_{0} = 0$ and $\Delta^{\rm b_2}_{0}>0$. 
Moreover, the off-diagonal matrix elements may also lead to  important 
corrections if $\Delta_{1,2}<1$ where $\Delta_{1,2}$ is the lowest weight appearing in the operator products 
\be
\hat \psi^{[b2,{\rm uv}]}(\tau)\hat \psi^{[{\rm uv}, b1]}(-\tau)\, , \qquad \hat \psi^{[b2,{\rm uv}]}(\tau)\psi_{i}^{\rm uv}(0)\hat \psi^{[{\rm uv}, b1]}(-\tau) \, .
\ee

It should be also noted regarding the derivation of formula (\ref{var_boundary_1end}) that we assumed that there are no resonances and thus no UV logarithmic divergences. 
Such logarithms contributing to ground state energy arise in some models e.g. for $\psi_{1,3}$ boundary perturbations 
of minimal models ${\cal M}_{m}$ 
where $\Delta_{1,3} = \frac{m-1}{m+1}$ and the coupling constant dimension is $1-\Delta_{1,3}=\frac{2}{m+1}$. 
The resonance with the identity operator occurs for odd $m$ at the order $\lambda^{\frac{m+1}{2}}$. The simplest case 
is the Ising model with the free spin boundary condition perturbed by the boundary condition changing operator. 
In such cases there is an additional contribution to the flat space ground energy 
\be
E_{\rm uv} \sim \lambda^{\frac{m+1}{2}} {\rm ln}(\tau \mu)
\ee
where $\mu$ is the subtraction energy scale. As proposed in \cite{Cardy_var} one can add such a contribution to 
the main part (\ref{var_boundary_1end}) by hand. 

Consider now  a single coupling $\lambda$ case denoting the dimension of UV operator $\psi$  as $\Delta^{\rm uv}$.  Assuming that 
\be \label{min_cond}
\lambda D_{\psi \hat \psi}^{\hat \psi}  >0 
\ee
the minimum of (\ref{var_boundary_1end}) 
is achieved at a finite point 
\be
\tau=\tau_{\rm min} = \left(\frac{\hat \Delta}{\Delta^{\rm uv} 2^{\Delta^{\rm uv}} \lambda D_{\psi \hat \psi}^{\hat \psi}}  \right)^{\frac{1}{1-\Delta^{\rm uv}}} 
\ee
and is equal to 
\be \label{Emin2}
E_{\rm min}(\hat \psi) = -(1-\Delta^{\rm uv})(2\Delta^{\rm uv})^{\frac{\Delta^{\rm uv}}{1-\Delta^{\rm uv}}} |\lambda|^{\frac{1}{1-\Delta^{\rm uv}}} \left(\frac{|D_{\psi \hat \psi}^{\hat \psi}|}{\hat \Delta^{\Delta^{\rm uv}}} \right)^{\frac{1}{1-\Delta^{\rm uv}}} \, .
\ee
For a fixed UV BCFT and a fixed perturbation the values $E_{\rm min}(\hat \psi)$ need to be compared for different tentative IR fixed points and different RG operators. The method is most efficient in situations when we know all conformal boundary conditions in a given bulk CFT as well as all boundary condition changing operators between them. This is the case for example in the Virasoro minimal models. Moreover, if we can argue that the perturbation we consider results in a flow to a fixed point invariant under the maximal chiral algebra\footnote{For example it would be the case if the perturbation preserves the global symmetry group in a WZW model.} we would be in the same situation for any rational CFT. As the g-factors in the known unitary CFTs are bounded from below, the g-theorem \cite{gThm1}, \cite{gThm2}  restricts the choices of the IR fixed points to a finite number of candidates. With our conjecture 1 and   the assumptions on the bulk CFT just made, there are also finitely many candidates for the RG operator. 

In deriving (\ref{var_boundary_1end}) we assumed after formula (\ref{needed_OPE}) that $\hat \psi^{[\rm ir, uv]}$ is a quasiprimary 
as that leads to the simplest form of the variational energy. It is easy to generalise (\ref{DDD})  and (\ref{var_boundary_1end})
to a derivative field. For example if $\hat \psi^{[\rm ir, uv]}(\tau) = \partial_{\tau}\hat \chi^{[\rm ir, uv]}$ where $\hat \chi^{[\rm ir, uv]}$ 
is a quasiprimary field the OPE coefficient  $D_{i \hat \psi}^{\hat \psi} $ in (\ref{var_boundary_1end}) should be replaced by 
\be \label{derivative_D}
D_{\hat \psi i \hat \psi}^{(1)} =  D_{i \hat \chi}^{\hat \chi} \left( 1 + \frac{\Delta_{i}^{\rm uv}(\Delta_{i}^{\rm uv}-1 + 4\hat \Delta)}{2\hat \Delta(2\hat \Delta + 1)} \right) \, .
\ee
It is interesting to note that the minimal energy for a single derivative field is not manifestly larger than that for the 
underlying quasiprimary. If that was the case that would have landed additional support to our conjecture 1. However, we did check for the particular case of the tricritical Ising model considered in section \ref{tricrit_sec} that the variational energies for single derivatives are larger than those of the primaries.

\subsection{Variational ansatz for boundary flows. Perturbation on both ends.} 
Here we consider the case when we put the perturbed boundary condition on both ends of the strip. 
The perturbed Hamiltonian now is 
\be
H=H = H_{0} - \sum_{i}\lambda^{i}(\psi_{i}^{\rm uv}(0,0) + \psi_{i}^{\rm uv}(0,L))
\ee
where in the operator insertions the first coordinate is $\tau$ and the second is $\sigma$.
In this case, since we have the same boundary condition on both ends of the strip, the IR BCFT vacuum state  has weight zero.
For simplicity we first focus on the case when the IR vacuum is non-degenerate.
Its image in the UV BCFT state space is given by 
\be 
\hat \psi^{[\rm uv, ir]}(0,L)\hat \psi^{[\rm uv, ir]}(0,0)|0\rangle_{\rm ir} \, .
\ee
Requiring that the vacuum state is symmetric under the reflection $\sigma \to L-\sigma$ we write a trial state 
\be \label{trial_2_ends}
|\tau, \hat \psi\rangle_{2} = \hat \psi^{[\rm uv, ir]}(-\tau,L)\hat \psi^{[\rm uv, ir]}(-\tau,0)|0\rangle_{\rm ir} \, , \enspace \tau>0 \, 
\ee
 
The  average of the perturbed Hamiltonian in the state (\ref{trial_2_ends}) can be expressed via  certain 4- and 5-point functions on a strip. 
In the limit $\tau/L\ll 1$ these functions factorise onto contributions from the OPEs given in (\ref{flat_OPE}), (\ref{triple_flat_OPE}) 
and the analogous OPEs on the bottom end of the strip. 
The leading contributions to the variational energy again come from the identity fields in each OPE. 
For quasiprimary fields $\psi_{a}$ in standard normalisation 
we have 
\be
{}_{\rm ir}\langle 0| \psi_{a}^{\rm ir}(\tau_1,0) \psi_{b}^{\rm ir}(\tau_2,L)|0 \rangle_{\rm ir} = \delta_{a,b}\left(\frac{\pi}{2L}  \right)^{2\Delta_{a}} 
\left[{\rm cosh}\left(\frac{\pi(\tau_1-\tau_2)}{2L}  \right) \right]^{-2\Delta_{a}} 
\ee
so that 
\be \label{metric}
{}_{\rm ir}\langle 0| \psi_{a}^{\rm ir}(0,0) \psi_{b}^{\rm ir}(0,L)|0\rangle_{\rm ir} = \delta_{a,b}\left(\frac{\pi}{2L}  \right)^{2\Delta_{a}} \, . 
\ee
Similar equations can be obtained for the derivative fields. 
These equations imply that the  contributions to the average energy coming from non-identity fields in the OPEs at hand are suppressed by powers of $\tau/L$.  The leading contributions are then easily evaluated to be 
\be
E_{\hat \psi}^{\rm 2 ends}(\tau) = \frac{{}_{2}\langle \tau, \hat \psi| H |\tau, \hat \psi\rangle_{2}}{{}_{2}\langle \tau, \hat \psi| \tau, \hat \psi\rangle_{2}} = 2\left[ \frac{\hat \Delta}{\tau} -\sum_{i}\lambda_{i} D_{i \hat \psi}^{\hat \psi} 
 \left(\frac{2}{\tau}\right)^{\Delta_{i}^{\rm uv}} \right]
\ee
so that, up to the subleading constant term in (\ref{var_boundary_1end}), the leading  variational energy for the two-end perturbation is twice the 
variational energy for the one end perturbation. This of course can be expected given that the leading terms are boundary contributions to the ground state energies on the half-plane. 


\subsection{Boundary flows in tricritical Ising model} \label{tricrit_sec}

In this section we are going to illustrate the use of the variational method developed in section \ref{boundary_var_main} 
on the boundary flows in the tricritical Ising model which originate from  Cardy boundary conditions. 
The tricritical Ising model (TIM) is the Virasoro A-type   minimal model $M(5,4)$ with central charge $c=7/10$. 
The model has 6 primary states and 6 associated irreducible conformal boundary conditions given 
by the Cardy's construction \cite{Cardy}. The notation, weights of the primaries and g-factors for Cardy boundary conditions 
are collected in Table  \ref{TIM:table}. The notation used for Cardy boundary conditions reflects the underlying Blume-Capel 
lattice  model in which the spins can take three values: 0,+1,-1. The notation for $(d)$ comes from the word ``disordered''.

The space of boundary RG flows that originate from Cardy boundary conditions was put together in \cite{Affleck}. 
It was also investigated in \cite{Giokas} using mean field theory. It is 
summarised on the diagram depicted on Figure \ref{space_of_flows}. The diagram respects  a number of constraints. 
Three boundary conditions: $(+), (-), (0)$ are stable, they do not have any relevant boundary operators.  
The $g$-factors satisfy 
\be
g_{(+)}=g_{(-)} < g_{(0)} < g_{(0+)}=g_{(-0)} < g_{(-)} + g_{(+)} < g_{(d)} 
\ee
that by virtue of the g-theorem puts restrictions on the possible IR fixed points. 
 Topological defects which relate various sectors of boundary theories place more constraints. 
 The elementary topological defects $X_{i}$ are labeled by primary states $i$  and act on Cardy boundary 
 states $|j\rangle\!\rangle$ as 
 \be
 X_{i} |j\rangle\!\rangle = \sum_{k}N_{ij}^{k}|k\rangle\!\rangle
 \ee
 where $N_{ij}^{k}$ are the fusion coefficients. 

\begin{center}
\begin{table}[H]
\centering
\begin{tabular}{|c|>{\centering\arraybackslash}p{1cm}|>{\centering\arraybackslash}p{1cm}|>{\centering\arraybackslash}p{1cm}|>{\centering\arraybackslash}p{1.4cm}|>{\centering\arraybackslash}p{1.4cm}|>{\centering\arraybackslash}p{2cm}|}
\hline 
\rule{0pt}{4ex}  \rule[-3ex]{0pt}{0pt}  Kac table label & (1,1)& (2,1)& (3,1)& (1,2)& (1,3) & (2,2) \\
\hline 
\rule{0pt}{4ex}  \rule[-3ex]{0pt}{0pt}  Conformal weight & 0& 7/16& 3/2& 1/10& 3/5 & 3/80 \\
\hline 
\rule{0pt}{4ex}  \rule[-3ex]{0pt}{0pt}  Operator notation & {\bf 1} & $\sigma'$ & $\epsilon''$ & $\epsilon$ & $\epsilon'$& $\sigma$\\
\hline 
\rule{0pt}{4ex}  \rule[-3ex]{0pt}{0pt}  Cardy b.c. notation & $(-)$ & $(0)$ & $(+)$ & $(-0)$ & $(0+)$ & $ (d)$\\
\hline
\rule{0pt}{4ex}  \rule[-3ex]{0pt}{0pt}  Approx. $g$-factor & 0.51 & 0.72 & 0.51& 0.83 & 0.83 & 1.17 \\
\hline
\rule{0pt}{4ex}  \rule[-3ex]{0pt}{0pt} Boundary fields & (1,1)& (1,1), (3,1)& (1,1)& (1,1), (1,3)& (1,1), (1,3) & 
(1,1), (1,3), (1,2), (3,1)\\
\hline 
\end{tabular}
\caption{Primary fields and Cardy boundary conditions  in the Tricritical Ising Model}
\label{TIM:table}
\end{table}
\end{center}
\begin{center}
\begin{figure}[H]
\centering
\begin{tikzpicture}[scale=1.2]
\draw (0,2.5)  node {$(\!-\!)\!\oplus\!(\!+\!)$}; 
\draw (0,0) circle [radius=0.4] node {d};
\draw (0,-3.5) circle [radius=0.4] node {0};
\draw (-4,-3.5) circle [radius=0.4] node {$-$};
 \draw (-2,-2.5) circle [radius=0.4] node {--\,0};  
\draw (2,-2.5) circle [radius=0.4] node {0+}; 
\draw (4,-3.5) circle [radius=0.4] node {+};
\begin{scope}[thick, every node/.style={sloped,allow upside down}]
\draw[dashed] (0.65,2.47) to[out=0,in=90] node {\midarrow} (4,-3.1); 
\draw[dashed] (-0.65,2.47) to[out=180,in=90] node {\midarrow}  (-4,-3.1);
\draw[red] (0.4,0) to[out=0,in=90] node {\midarrow} (4,-3.1);
\draw (0.35,-0.25) to[out=-40,in=90] node {\midarrow} (2,-2.1); 
\draw[red] (-0.4,0) to[out=180,in=90] node {\midarrow} (-4,-3.1);
\draw (-0.35,-0.25) to[out=220,in=90] node {\midarrow} (-2,-2.1); 

\draw[blue] (0,0.4) -- node {\midarrow} (0,2.3); 
\draw[blue] (0,-0.4) -- node {\midarrow} (0,-3.1);

\draw[blue]  (-2.4,-2.5) to[out=180,in=45] node {\midarrow} (-3.65,-3.35); 
\draw[blue] (-1.6,-2.5) to[out=0,in=135] node {\midarrow} (-0.4,-3.5); 

\draw[blue]  (1.6,-2.5) to[out=180,in=45] node {\midarrow} (0.4,-3.5);
\draw[blue]  (2.4,-2.5) to[out=0,in=135] node {\midarrow} (3.6,-3.5);
\end{scope}
\end{tikzpicture}
\caption{The space of boundary flows in the Tricritical Ising Model. Blue arrows are the flows triggered by $\psi_{1,3}$ operators, 
red -- by $\psi_{1,2}$ operators, black -- by particular linear combinations of $\psi_{1,3}$ and $\psi_{1,2}$, and dashed -- 
 by the component identity fields.}
\label{space_of_flows}
\end{figure}
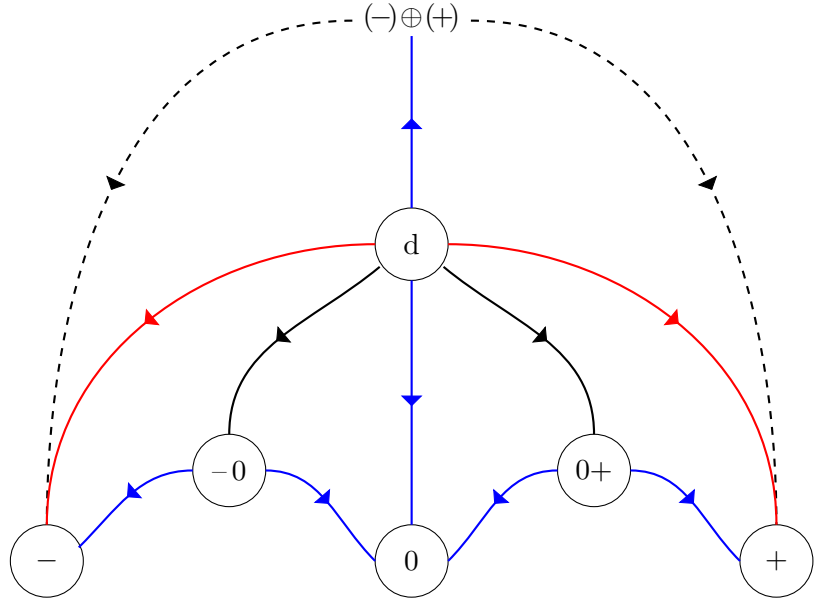
\end{center}
A theorem proved in \cite{GW} states that given an RG flow from a boundary condition $A$ to a boundary condition 
$B$, for any topological defect $X_{i}$  there exists an RG flow from $X_{i}A$ to $X_{i}B$ triggered by a boundary 
operator of the same conformal weight as in the original flow.  The group-like defect $X_{\epsilon''}$ in TIM 
realises the spin reversal symmetry. It acts by reflections about the vertical symmetry axis of the flow diagram 
on Figure \ref{space_of_flows} interchanging the $+$ and $-$ labels. The defect $X_{\sigma'}$ relates pairs of flows 
triggered by the $\psi_{1,3}$ fields with positive and negative coupling which are depicted by blue arrows on  Figure \ref{space_of_flows}. 
These flows as well as the flows triggered by the $\psi_{1,2}$ field on the disordered boundary condition $(d)$ are 
integrable. Their end points have been established using the TBA. Note in particular the flow generated by $\psi_{1,3}$ with a positive coupling 
switched on the $(d)$ boundary condition. It ends with a degenerate boundary condition that is the direct sum of the $(+)$ and 
$(-)$ Cardy boundary conditions. The flows given by the black lines are conjectural flows which are generated by some special linear combinations of $\psi_{1,3}$ and $\psi_{1,2}$ operators which should be there by continuity of the space of flows. We  focus here on the single coupling flows. We comment briefly  on  the two-coupling flows at the end of section \ref{sec:offdiag}.

\subsection{Flows triggered by the identity operators} \label{sec:id}
We first look at the case of flows from the superposition $(+)\oplus(-)$ to the components which are triggered by the identity operators: 
$1_{+}$, $1_{-}$. These flows are depicted by the dashed lines on Figure \ref{space_of_flows}. The state space of  $(+)\oplus(-)$ is 
a direct sum: ${\cal H}^{(+)}\oplus {\cal H}^{(-)}$ and the operators $1_{+}$, $1_{-}$ act as orthogonal projectors on each component. 
The most general perturbation is a superposition: $\lambda_{+}1_{+} + \lambda_{-}1_{-}$. However, since $1_{+}+1_{-}=1$ commutes with all other operators it suffices to consider a flow triggered by $1_{+}$ alone. Perturbing the Hamiltonian as 
\be
H=H_{0} - \lambda 1_{+}
\ee
we see that the eigenvectors are just the eigenvectors of $H_{0}$ but the eigenvalues of all states in ${\cal H}^{(+)}$ are shifted 
by $-\lambda$. For $\lambda>0$ the energies in ${\cal H}^{(+)}$ flow to minus infinity as $\lambda \to \infty$ and thus are the only states surviving in the low energy sector where we measure the energies above the vacuum. Hence in this case we flow to the $(+)$ component. In the $\lambda<0$ case the situation is the opposite with the energies in the ${\cal H}^{(+)}$ going to plus infinity and 
only the states in ${\cal H}^{(-)}$ surviving in the far infrared. The RG operator in the first case is the operator 
\be
\hat P_{+}^{\rm ir, uv} : {\cal H}^{(+)}\oplus {\cal H}^{(-)} \to {\cal H}^{(+)}  \, , \quad \hat P_{+}^{\rm ir, uv}:(v_{+}, v_{-}) \mapsto v_{+} 
\ee 
with the conjugate 
\be
\hat P_{+}^{\rm uv, ir} :  {\cal H}^{(+)} \to {\cal H}^{(+)}\oplus {\cal H}^{(-)} \, , \quad \hat P_{+}^{\rm uv, ir}:v_{+}  \mapsto (v_{+}, 0) \, .
\ee
Clearly we have 
\be
\hat P_{+}^{\rm uv, ir} \hat P_{+}^{\rm ir, uv} = 1_{+} 
\ee
and
\be
 \hat P_{+}^{\rm ir, uv} \hat P_{+}^{\rm uv, ir} = 1_{\rm ir} \, . 
\ee
Our conjecture 3 holds trivially in this case because there is no infrared theory operator along which the flow arrives at the fixed 
point. In this respect the situation is similar to massive flows in the bulk. 
For the negative coupling $\lambda$ all formulae above are valid if one replaces everywhere $+$ by $-$.
The flows triggered by boundary identity operators are also similar to the bulk RG flows in topological QFTs which were recently 
considered in \cite{Roggenkamp1}, \cite{Roggenkamp2}. 

\subsection{$\psi_{1,3}$ flows } \label{sec:psi13}
Such flows can start on one of the three boundary conditions: $(0+)$, $(-0)$, $(d)$. We assume that the perturbing $\psi_{1,3}$ field is normalised so that the identity field appears with coefficient 1 it its OPE with itself.  

We are going to compare the variational energies using formula (\ref{Emin2}) and assuming that the RG operators are primary.  The OPE coefficient that enters (\ref{Emin2})  can be expressed using \cite{Runkel} in terms of minimal model fusion matrices as 
\be
D_{\psi \hat \psi}^{\hat \psi}= \frac{F_{{\rm uv}, \hat \psi}\left[ \begin{array}{cc} {\rm uv}&{\rm ir}\\
 \psi^{\rm uv} & \hat \psi
\end{array} \right]}{\sqrt{F_{{\rm uv},1}\left[ \begin{array}{cc} {\rm uv}&{\rm uv}\\
\psi^{\rm uv} & \psi^{\rm uv} 
\end{array} \right]}}
\ee
where one needs to enter the corresponding Kac table labels for each symbolic entry in the fusion matrices. 
The fusion matrices can be calculated\footnote{The author thanks I. Runkel for sharing his Mathematica code for calculating fusion matrices.} using the recurrence based routine from \cite{Runkel} or the explicit formula 
presented in \cite{Runkel_PhD}.

It is convenient to factorise the expression in (\ref{Emin2}) as 
\be \label{e_factor}
E_{\rm min}(\hat \psi) = e(\hat \psi) |\lambda|^{\frac{1}{1-\Delta^{\rm uv}}}  \, .
\ee
If  $\lambda D_{\psi\hat\psi}^{\hat \psi}>0$  formulae  (\ref{Emin2}), (\ref{e_factor}) define $e(\hat \psi)<0$ while when 
$\lambda D_{\psi\hat\psi}^{\hat \psi}<0$ we set $e(\hat \psi)=0$. 
As we intend to compare the variational energies for fixed $\lambda$ and $\Delta^{\rm uv}$ it suffices to compare the dimensionless factors  $e(\hat \psi)$.

Tables \ref{VAR:table1} and \ref{VAR:table2} list the factors $e(\hat \psi)$   for the $\psi_{1,3}$ perturbation with positive and negative coupling $\lambda$ respectively. Zero energy means that either 
$D_{\psi \hat \psi}^{\hat \psi}<0$ or that it equals zero. Note that we put all possible Cardy boundary conditions as tentative 
IR fixed points regardless of their value of the g-factor in order to provide  additional testing to the method.  
The coincident values of the variational energies are due to the spin reversal symmetry and the duality induced by the $X_{\sigma'}$ 
defect. 

The lowest values of the variational energy give the same end points as presented on the diagram on Figure \ref{space_of_flows}. 
This includes the flow $(d)\to (+)\oplus (-)$ for which we get an exactly degenerate value of the variational energy. 
We remarked in section \ref{boundary_var_main} that when such a degeneration happens there may be corrections from 
the off-diagonal terms in the Hamiltonian which would compete with the constant terms from the spectator boundary condition if 
$\Delta_{1,2}<1$ where $\Delta_{1,2}$ is the lowest weight of the boundary condition changing fields linking the two infrared boundary conditions. We note that there are no such fields in this case\footnote{This is actually the case for all $\psi_{1,3}$ flows in the A-series minimal models}. 

\begin{center}
\begin{table}
\centering
\begin{tabular}{|c|>{\centering\arraybackslash}p{4.5cm}|c|c|}
\hline 
\diagbox{IR}{UV}  & (d) & (0+)& (-0) \\
\hline 
\rule{0pt}{4ex}  \rule[-3ex]{0pt}{0pt}  (d) &$e(\hat 1)=e(\hat \epsilon'')=e(\hat \epsilon')=0$; $e(\hat \epsilon)=-3.63$&
$e(\hat \sigma')=e(\hat \sigma)=0$ &$e(\hat \sigma')=0$; $e(\hat \sigma)=-2.80$  \\
\hline 
\rule{0pt}{4ex}  \rule[-3ex]{0pt}{0pt}  (0+) &$e(\hat \sigma)=e(\hat \sigma')=0$ &$e(\hat 1)=e(\hat \epsilon')=0$& 
$e(\hat \epsilon) =e(\hat \epsilon'')=0$ \\
\hline 
\rule{0pt}{4ex}  \rule[-3ex]{0pt}{0pt}  (-0) &$e(\hat \sigma)=e(\hat \sigma')=0$  &$e(\hat \epsilon'')=0$; $e(\hat \epsilon)=-3.63$&
$e(\hat 1)=0$; $e(\hat \epsilon')=-0.24$\\
\hline
\rule{0pt}{4ex}  \rule[-3ex]{0pt}{0pt}  (0) &$e(\hat \epsilon)=0$, $e(\hat \epsilon')=-0.82$&$e(\hat \sigma)= \textcolor{red}{-9.32}$&
$e(\hat \sigma)=0$ \\
\hline
\rule{0pt}{4ex}  \rule[-3ex]{0pt}{0pt} (+) & $e(\hat \sigma) =\textcolor{red}{-9.32}$&$e(\hat \epsilon)=0$& $e(\hat \epsilon')=0$\\
\hline 
\rule{0pt}{4ex}  \rule[-3ex]{0pt}{0pt} (-) &$e(\hat \sigma) =\textcolor{red}{-9.32}$  &$e(\hat \epsilon')=-0.82$&$e(\hat \epsilon)=\textcolor{red}{-12.11}$\\
\hline 
\end{tabular}
\caption{Variational energy factors for $\psi_{1,3}$ flows with positive coupling given to the second decimal place. Here the columns are labeled by  the UV boundary conditions while the rows correspond to tentative IR fixed points. The smallest values are highlighted in red. }
\label{VAR:table1}
\end{table}
\end{center}

\begin{center}
\begin{table}
\centering
\begin{tabular}{|c|>{\centering\arraybackslash}p{4.5cm}|c|c|}
\hline 
\diagbox{IR}{UV}  & (d) & (0+)& (-0) \\
\hline 
\rule{0pt}{4ex}  \rule[-3ex]{0pt}{0pt}  (d) &$e(\hat 1)=e(\hat \epsilon'')=e(\hat \epsilon)=0$; $e(\hat \epsilon')=-0.24$&
$e(\hat \sigma')=0$; $e(\hat \sigma)=-2.80$ &$e(\hat \sigma')=e(\hat \sigma)=0$  \\
\hline 
\rule{0pt}{4ex}  \rule[-3ex]{0pt}{0pt}  (0+) &$e(\hat \sigma')=0$; $e(\hat \sigma)=-2.80$ &$e(\hat 1)=0$; $e(\hat \epsilon')=-0.24$& 
$e(\hat \epsilon'') =0$; $e(\hat \epsilon)=-3.63$ \\
\hline 
\rule{0pt}{4ex}  \rule[-3ex]{0pt}{0pt}  (-0) &$e(\hat \sigma')=0$; $e(\hat \sigma)=-2.80$  &$e(\hat \epsilon'')=e(\hat \epsilon)=0$&
$e(\hat 1)=e(\hat \epsilon')=0$\\
\hline
\rule{0pt}{4ex}  \rule[-3ex]{0pt}{0pt}  (0) &$e(\hat \epsilon')=0$, $e(\hat \epsilon)=\textcolor{red}{-12.11}$&$e(\hat \sigma)=0$&
$e(\hat \sigma)=\textcolor{red}{-9.32}$ \\
\hline
\rule{0pt}{4ex}  \rule[-3ex]{0pt}{0pt} (+) & $e(\hat \sigma) =0$&$e(\hat \epsilon)=\textcolor{red}{-12.11}$& $e(\hat \epsilon')= -0.82$\\
\hline 
\rule{0pt}{4ex}  \rule[-3ex]{0pt}{0pt} (-) &$e(\hat \sigma) =0$  &$e(\hat \epsilon')=0$&$e(\hat \epsilon)=0$\\
\hline 
\end{tabular}
\caption{Variational energy factors for $\psi_{1,3}$ flows with negative coupling given to the second decimal place. Here the columns are labeled by  the UV boundary conditions while the rows correspond to tentative IR fixed points. The smallest values are highlighted in red.}
\label{VAR:table2}
\end{table}
\end{center}

\newpage
Besides confirming the blue arrows on the diagram on Figure \ref{space_of_flows} we also obtain the RG operators for these 
flows which we summarise in the following diagrams where we put the RG operator above the arrow designating the flow. 
\be
(+)\oplus (-)\,  \mathop{\xleftarrow{\hspace*{1.5cm}}}^{\mathlarger{\hat \sigma \oplus \hat \sigma}}\,  (d)\, \mathop{\xrightarrow{\hspace*{1.5cm}}}^{\mathlarger{\hat \epsilon}} \,  (0) 
\ee
\be
(0)\,  \mathop{\xleftarrow{\hspace*{1.5cm}}}^{\mathlarger{\hat \sigma}}\,  (0+)\, \mathop{\xrightarrow{\hspace*{1.5cm}}}^{\mathlarger{\hat \epsilon}} \,  (+) 
\ee
\be
(-)\,  \mathop{\xleftarrow{\hspace*{1.5cm}}}^{\mathlarger{\hat \epsilon}}\,  (-0)\, \mathop{\xrightarrow{\hspace*{1.5cm}}}^{\mathlarger{\hat \sigma}} \,  (0) 
\ee
Here the left arrows correspond to the flows with $\lambda>0$ and the right arrows to the flows with $\lambda<0$. 
We observe that these assignments satisfy our general conjecture 2 as well as the mapping of RG flows  induced by the $X_{\sigma'}$ topological defect. 


\subsection{$\psi_{1,2}$ flows } \label{sec:psi12}
The $\psi_{1,2}$ boundary field is only present on the $(d)$ boundary condition. The factors $e(\hat \psi)$ 
for the variational energy are presented in Table \ref{VAR:table3}. The smallest energies correspond to the two 
flows which are drawn in red on Figure \ref{space_of_flows}. 
 We  also represent these two flows on the diagram below indicating the RG 
operators above the arrows. 
\be
(-)\,  \mathop{\xleftarrow{\hspace*{1.5cm}}}^{\mathlarger{\hat   \sigma}}\,  (d)\, \mathop{\xrightarrow{\hspace*{1.5cm}}}^{\mathlarger{\hat \sigma}} \,  (+) 
\ee

\begin{center}
\begin{table}[H]
\centering
\begin{tabular}{|c|c|c|}
\hline 
\diagbox{IR}{UV}  & (d),\,  $\lambda>0$ & (d),\, $\lambda<0$ \\
\hline 
\rule{0pt}{4ex}  \rule[-3ex]{0pt}{0pt}  (d) &$e(\hat 1)=e(\hat \epsilon)=e(\hat \epsilon')=e(\hat \epsilon'')=0$&$e(\hat 1)=e(\hat \epsilon)=e(\hat \epsilon')=e(\hat \epsilon'')=0$  \\
\hline 
\rule{0pt}{4ex}  \rule[-3ex]{0pt}{0pt}  (0+) &$e(\hat \sigma')=e(\hat \sigma)=0$& $e(\hat \sigma')=0$; $e(\hat \sigma)=-0.69$\\
\hline 
\rule{0pt}{4ex}  \rule[-3ex]{0pt}{0pt}  (-0) &$e(\hat \sigma')=0$; $e(\hat \sigma)=-0.69$&$e(\hat \sigma')=e(\hat \sigma)=0$\\
\hline
\rule{0pt}{4ex}  \rule[-3ex]{0pt}{0pt}  (0) &$e(\hat \epsilon)=e(\hat \epsilon')=0$&$e(\hat \epsilon)=e(\hat \epsilon')=0$ \\
\hline
\rule{0pt}{4ex}  \rule[-3ex]{0pt}{0pt} (+) &$e(\hat \sigma)=0$ &$e(\hat \sigma)=\textcolor{red}{-1.17}$\\
\hline 
\rule{0pt}{4ex}  \rule[-3ex]{0pt}{0pt} (-) &$e(\hat \sigma)=\textcolor{red}{-1.17}$ &$e(\hat \sigma)=0$\\
\hline 
\end{tabular}
\caption{Variational energy factors for $\psi_{1,2}$ flows  given to the second decimal place. The smallest values are highlighted in red. }
\label{VAR:table3}
\end{table}
\end{center}

Regarding our conjecture 3 we observe that the leading irrelevant operator for the $(d)\to~(+)\oplus~(-)$ flow is believed to be the boundary condition changing operator 
$(\epsilon'')^{[+,-]}$. This operator is  present in the OPE $\hat \psi^{[\rm ir, uv]} \hat \psi^{[\rm uv, ir]}  $ of the RG operator with itself with our identification of its components  
as the primaries: $\hat \sigma^{[d,+]}$,   $\hat \sigma^{[d,-]}$. The leading irrelevant 
operator for the $(d)\to (0)$ flow is the stress energy tensor which  is present in the relevant OPE. 
 While the leading IR operators for the pure $\psi_{1,2}$ flows are not known they can be only composites of 
the  stress-energy tensor which are always present in the OPE of interest.

\subsection{Superpositions of trial states} \label{sec:offdiag}
As we discussed in section \ref{boundary_var_main} the off-diagonal matrix elements of the Hamiltonian  between 
trial states built upon different IR BCFTs are suppressed by powers of $\tau/L$. However there may be $L$-independent corrections in 
the average of the perturbation for the same IR BCFT but different RG operators. We find that this is precisely the 
case for the trial states linking $(d)$ with $(0)$ and $(d)$ with $(0+)$ or $(-0)$.  In the first case we have two trial states: 
\be
\hat \epsilon^{[d, 0]}(-\tau)|0\rangle_{(0)} \, , \qquad (\hat \epsilon')^{[d, 0]}(-\tau)|0\rangle_{(0)}
\ee
while in the second: 
\be
\hat \sigma^{[d, 0+]}(-\tau)|0\rangle_{(0+)} \, , \qquad (\hat \sigma')^{[d, 0+]}(-\tau)|0\rangle_{(0+)} \, 
\ee
and similarly for $(0-)$.
The perturbing operators $\psi_{1,3}$ and $\psi_{1,2}$ each have off-diagonal matrix elements in the above states. 
We can thus introduce  more general trial states
\be \label{new_trial0}
|\theta, \tau\rangle_{(0)} = \cos(\theta) \frac{(2\tau)^{\hat \Delta_{\epsilon}}}{\sqrt{C^{1}_{\hat \epsilon \hat \epsilon}}} \hat \epsilon^{[d,0]}(-\tau) |0\rangle_{(0)}  + \sin(\theta) \frac{(2\tau)^{\hat \Delta'_{\epsilon}}}{\sqrt{C^{1}_{\hat \epsilon' \hat \epsilon'}}} (\hat \epsilon')^{[d,0]}(-\tau) |0\rangle_{(0)}  \, , 
\ee
\be \label{new_trial0+}
|\varphi, \tau\rangle_{(0+)} = \cos(\varphi) \frac{(2\tau)^{\hat \Delta_{\sigma}}}{\sqrt{C^{1}_{\hat \sigma \hat \sigma}}} 
\hat \sigma^{[d,0+]}(-\tau) |0\rangle_{(0+)}  + \sin(\varphi) \frac{(2\tau)^{\hat \Delta'_{\sigma}}}{\sqrt{C^{1}_{\hat \sigma' \hat \sigma'}}} (\hat \sigma')^{[d,0+]}(-\tau) |0\rangle_{(0+)} 
\ee
where 
\be
\hat \Delta_{\epsilon} = \Delta_{1,2}=\frac{1}{10} \, , \quad \hat \Delta'_{\epsilon}  =\Delta_{1,3}= \frac{3}{5}\, , \quad 
\hat \Delta_{\sigma}=\Delta_{2,2}=\frac{3}{80}\, , \quad \hat \Delta_{\sigma}'=\Delta_{2,1}=\frac{7}{16} \, .
\ee
The states (\ref{new_trial0}), (\ref{new_trial0+})  are normalised up to finite size suppressed corrections. 
As far as the components of these trial vectors are concerned the contribution from the operator of larger dimension: $\hat \epsilon'$ 
or $\hat \sigma'$ 
is suppressed because it contains  an extra positive power of $\tau$  that at the minimum of energy is proportional 
to a negative power of a coupling. Thus the terms with the higher dimension operator  can be considered as an additional  correction to the RG operators $\hat \epsilon$, $\hat \sigma$ in addition 
to the corrections coming from the $\tau$-shift (which are a series in derivatives). 

The variational energy averages in the above states  receive contributions from the off-diagonal matrix elements which at the leading order are expressed via OPE coefficients. The general form of these contributions can be written as 
\be
{}_{\rm ir} \langle 0|  (\hat \psi')^{[0,d]}(-\tau) \psi_{i}(0)    \hat \psi^{[d,0]}(-\tau) |0\rangle_{\rm ir} \sim 
\frac{C_{i \hat \psi}^{\hat \psi'} C_{\hat \psi' \hat \psi'}^{1}}{2^{\hat \Delta_{\psi} + \hat \Delta'_{\psi}- \Delta_{i}}\tau^{\hat \Delta_{\psi} + \hat \Delta'_{\psi} + \Delta_{i}}} \, .
\ee
For completeness we calculate the variational energies in the states (\ref{new_trial0}), (\ref{new_trial0+}) for the generic perturbation 
\be
H = H_{0} - \lambda_{1,3} \psi_{1,3} - \lambda_{1,2}\psi_{1,2} \, .
\ee
It is convenient to rescale the couplings as
\bea
\nu_{12}&=&\lambda_{12}\,  2^{1/10}  \left(F_{(2,2),(1,1)}\left[ \begin{array}{cc} (2,2)&(2,2)\\
(1,2) & (1,2)
\end{array} \right]\right)^{-1/2} \, , \nonumber \\
\nu_{13}&=& \lambda_{13}\,  2^{3/5}\left(F_{(2,2),(1,1)}\left[ \begin{array}{cc} (2,2)&(2,2)\\
(1,3) & (1,3)
\end{array} \right]\right)^{-1/2} \, .
\eea 
We find the following averages 
\be
E^{(+)}(\tau) =  \frac{3}{80 \tau} -  \frac{\nu_{13}}{\tau^{3/5}}  +\frac{\nu_{12}}{\tau^{0.1}}  \, , 
\ee
\be
E^{(0)}(\tau, \theta) = \frac{1}{20\tau}(7-5\cos(2\theta)) + \frac{2\nu_{13} \cos(2\theta)}{\tau^{3/5}} 
- \frac{\nu_{12}\sqrt{2/3}\sin(2\theta)}{\tau^{0.1}}  \, , 
\ee
\bea
&& E^{(0+)}(\tau, \varphi) = \frac{1}{\tau}\left( \frac{3}{80}\cos^2(\varphi) + \frac{7}{16}\sin^2(\varphi)\right)  
+ \cos^2(\varphi)\frac{\sqrt{5}-1}{2}\left( \frac{\nu_{13}}{\tau^{3/5}} + \frac{\nu_{12}}{\tau^{0.1}}\right)  \nonumber \\
&&- \sin(2\varphi)\left( \frac{b \nu_{12}}{\tau^{0.1}} + \frac{a \nu_{13}}{\tau^{3/5}} \right)
\eea
where 
\bea
&&a = F_{(2,2),(2,1)}\left[ \begin{array}{cc} (2,2)&(1,3)\\
(1,3) & (2,2)
\end{array} \right] \left( \frac{F_{(2,2),(1,1)}\left[ \begin{array}{cc} (1,3)&(1,3)\\
(2,1) & (2,1)
\end{array} \right]}{F_{(2,2),(1,1)}\left[ \begin{array}{cc} (1,3)&(1,3)\\
(2,2) & (2,2)
\end{array} \right]}  \right)^{1/2}  \nonumber \\
&& = -\frac{15\Gamma(6/5)\sqrt{(3-\sqrt{5})\Gamma(4/5)}}{2\Gamma^{3/2}(2/5)} \approx - 1.9658
\eea
and 
\bea
&& b = F_{(2,2),(2,1)}\left[ \begin{array}{cc} (2,2)&(1,3)\\
(1,2) & (2,2)
\end{array} \right] \left( \frac{F_{(2,2),(1,1)}\left[ \begin{array}{cc} (1,3)&(1,3)\\
(2,1) & (2,1)
\end{array} \right]}{F_{(2,2),(1,1)}\left[ \begin{array}{cc} (1,3)&(1,3)\\
(2,2) & (2,2)
\end{array} \right]}  \right)^{1/2}  \nonumber \\
&& = \frac{2^{\frac{1}{20}}((5-\sqrt{5})\pi)^{3/4}\sqrt{-3\Gamma(-3/10)}}{25\Gamma(7/10)\Gamma(7/5)}\approx 0.6552
\eea
For completeness we also included above $E^{(+)}(\tau)$ -- the variational energy linking $(d)$ and $(+)$ that does not receive 
any additional terms because there is a unique boundary condition changing primary.

To minimise the energy one can either minimise the above expressions in both $\tau$ and $\theta$ or find the eigenvalues of the corresponding  $2\times 2$ matrix 
and minimise them in $\tau$. Here we focus on the pure $\psi_{1,3}$ and pure $\psi_{1,2}$ perturbations to compare with the 
results obtained in sections \ref{sec:psi13}, \ref{sec:psi12}.

 As the new ansatze   (\ref{new_trial0}), (\ref{new_trial0+}) contain the old trial vectors 
 the new variational energies will be lower in comparison to the ones tabulated in Tables \ref{VAR:table1}, \ref{VAR:table2}, \ref{VAR:table3}. We define the dimensionless factors for the new trial energies similarly to  (\ref{e_factor}) as 
 \be
 E_{\rm min}^{(0)} = e(\hat \epsilon, \hat \epsilon') |\lambda|^{\frac{1}{1-\Delta^{\rm uv}}} \, , \quad 
 E_{\rm min}^{(0+)} = e(\hat \sigma, \hat \sigma') |\lambda|^{\frac{1}{1-\Delta^{\rm uv}}}
 \ee
 where for $\lambda$ we take $\lambda_{13}$   for pure $\psi_{1,3}$ flows and $\lambda_{12}$ for pure $\psi_{1,2}$ flows.
If there is no local minimum  we set the corresponding  factor to zero.
   For the pure $\psi_{1,3}$ flows we obtain the same values for $e(\hat \epsilon, \hat \epsilon')$ as before while 
   \be
    e(\hat \sigma, \hat \sigma') = -2.39 \, , \quad \mbox{for } \lambda_{13}> 0\, ,  \qquad 
     e(\hat \sigma, \hat \sigma') = \textcolor{red}{-12.64} \, , \quad \mbox{for } \lambda_{13}< 0
   \ee
   with two decimal places retained.
 The corresponding factors for $E_{\rm min}^{(-0)}$ are the same.  The value of $ e(\hat \sigma, \hat \sigma')$ 
for the negative coupling  has changed by a large amount from $e(\sigma)=-2.80$. Moreover, it is now lower than 
the value for the correct endpoint $(0)$ which is $e(\hat \epsilon)=-12.11$. This presents a problem. We do know 
from TBA and TCSA numerics that  the correct end point for this flow is $(0)$. We also know from the TCSA results presented 
in section \ref{sec:TCSA} that $\hat \epsilon$ is the correct RG operator. It may be that further local corrections to the interface operator 
are needed to be taken into account, e.g, the $(L_{-2}\hat \psi)$ descendant field, or that non-local corrections are present. 
We plan to investigate this further in future work. 

For the pure $\psi_{1,2}$ flows we obtain $e(\hat \epsilon, \hat \epsilon')=-0.73$ for $ \lambda_{12}\ne 0$ and 
\be
  e(\hat \sigma, \hat \sigma') = -0.35 \, , \quad \mbox{for } \lambda_{12}> 0\, ,  \qquad 
     e(\hat \sigma, \hat \sigma') = -1.06 \, , \quad \mbox{for } \lambda_{12}< 0 \, .
\ee
(The two values are swapped for $(-0)$.) Comparing these new values to the ones in Table \ref{VAR:table3} we see that the 
winning trial state remains the same. 

It would be very interesting to locate the black lines on the diagramme on Figure \ref{space_of_flows}. We have done this numerically using TCSA 
approach and will present the results elsewhere. 
As far as the variational method is concerned we feel that one needs first 
to understand the above mentioned problem with the $(d)\rightarrow (0)$ flow and the trial state (\ref{new_trial0+}) before trying to apply the  method to the genuine two-coupling flows. However, even without doing the variational calculations, we can say that if the flow exists the corresponding RG operator must be the operator $\hat \sigma^{[0+,d]}$. This is because there are only two primaries 
linking the two fixed points: $\sigma$ and $\sigma'$, and the $\sigma'$ primary OPE with itself only produces the identity and the $\epsilon''$ fields.

\subsection{The eigenvector equation} \label{sec:eigenvector}
Some insight into the general structure of the vacuum state can be gained by looking at the eigenvector equation. We discussed 
such equations in section \ref{section_states}. The eigenvector equation in the perturbed theory can be written as 
\be \label{id_3_copy}
\lim_{\epsilon \to\, + \,0}  \Bigl[H_{0} -   \lambda \psi(\epsilon) + \mbox{Counterterms}  \Bigr] \hat \psi^{[0,\lambda]}(0) |{\cal E}_{I}\rangle_{\lambda} =  {\cal E}_{I} \hat \psi^{[0,\lambda]}(0)|{\cal E}_{I}\rangle_{\lambda}    \, .
\ee
In the variational method we try to approximate the vacuum state $\hat \psi^{[0,\lambda]}(0) |0\rangle_{\lambda}$ by a 
state in the IR BCFT. This approximation can be written as 
\be \label{eigv_exp}
\hat \psi^{[0,\lambda]}(0) |0\rangle_{\lambda} = \hat \psi^{[\rm uv, ir]}(0) |0\rangle_{\rm ir}  + C 
|\lambda|^{\frac{1}{\Delta^{\rm uv}-1}}\partial_{\tau}  \hat \psi^{[\rm uv, ir]}(0) |0\rangle_{\rm ir} + \dots 
\ee
where $C$ is a constant and the omitted terms all contain higher powers of the inverse correlation length and may contain local operators such as higher 
Virasoro descendants of the RG operator as well as non-local operators such as Virasoro modes (that act on $ |0\rangle_{\rm ir}$ and thus give  different states in the IR BCFT on which the local boundary condition changing operators act). When we act on such a state 
by the perturbed Hamiltonian $H_{0} -   \lambda \psi(0)$ the divergences come from collision of $\psi$ with the local operators present  in the above expansion. Assuming the divergences correspond to the divergence in the vacuum energy ${\cal E}_{0}$ (as can be demonstrated perturbatively) they should be cancelled by  counterterms proportional to the eigenvector  itself. This may be possible only if 
$\hat \psi^{[\rm uv, ir]}$ is a primary. If it wasn't then the most singular terms in the OPE with $\psi$ would come with the primary underlying $\hat \psi^{[\rm uv, ir]}$ which would not be in the eigenvector itself. 

The eigenvector equation also tells us that if the OPE of $\psi$ with $\hat \psi^{[\rm uv, ir]}$ contains other singular terms proportional to primary fields  then the corresponding primaries must also be present in the expansion (\ref{eigv_exp}). This is what we tried to incorporate 
in the extended variational ansatze (\ref{new_trial0}), (\ref{new_trial0+}). We also note that the RG operators identified for the pure 
$\psi_{1,3}$ and pure $\psi_{1,2}$ flows in sections \ref{sec:id}, \ref{sec:psi13}, \ref{sec:psi12} are all closed under the action of the relevant perturbing operator that is 
the OPE contains only terms from the same primary tower. In general the requirement that the OPE of the perturbing operator 
$\psi$, which is a primary or a linear combination of different primaries, with  the RG operator must contain the RG operator itself 
is essentially equivalent to our conjecture 2. The insight based on the eigenvector equation even suggests a stronger version of that conjecture: the term in the OPE 
\be
\psi_{i}(0) \hat \psi^{[\rm uv, ir]}(-\tau) \sim \frac{D_{i}}{\tau^{\Delta_{i}^{\rm uv}}}  \hat \psi^{[\rm uv, ir]}(-\tau) + \dots 
\ee
containing the RG operator must be the most singular term. 

\section{Some numerical results} \label{sec:TCSA}
\setcounter{equation}{0}

The truncated conformal space approach (TCSA) is a numerical method put forward in \cite{YZ1}, \cite{YZ2} which allows one to find approximately 
the   spectrum of perturbed 2D CFTs. It was adapted to boundary RG flows in \cite{DPTW}.  The eigenvalues and their degeneracies in particular allow one to identify the IR fixed point of the flow. Here we investigate the boundary flows in TIM focusing  on the numerical eigenvectors for the vacuum and the first excited state. We choose the spectator boundary condition on the strip to be $(-)$ that corresponds to the identity primary state.  The state space of the UV BCFT then has a single Virasoro tower of states. 
Let $|0\rangle_{\rm uv}$ and $|0\rangle_{\rm ir}$ denote the vacuum states on the strip in the UV and the IR state spaces respectively. 
Assuming that $|0\rangle_{\rm uv}$  is a primary of positive weight, which with the $(-)$ spectator is true when the UV BCFT has a relevant operator and thus must be one of  $(d)$, $(0+)$, $(-0)$,  
the image of the IR vacuum in the UV state space can be expanded in components of increasing weights as 
\be
\hat \psi^{\rm [uv, ir]}(0) |0\rangle_{\rm ir} = C_{0} |0\rangle_{\rm uv} + C_{1}L_{-1}|0\rangle_{\rm uv} + D_1 L_{-1}^2|0\rangle_{\rm uv} 
+ D_{2}L_{-2}|0\rangle_{\rm uv} + \dots 
\ee
While the normalisations of these states depend on the normalisation of the RG operator, the following ratios of the lowest 
components are independent of the overall normalisation\footnote{Of course one can also consider $D_{1}/C_{0}$, it is just a matter of convenience that dictated our choice of the ratios for which we present results.}:
\be
\Gamma^{1}_{\rm vac} = \frac{C_{1}}{C_{0}} \, , \qquad \Gamma^{2}_{\rm vac} = \frac{D_{2}}{C_{0}} 
\ee
Given the RG operator we can calculate the values of these ratios using the mapping of the RG field  from the half plane to the strip. 
To find the matrix elements of  $\hat \psi^{\rm [ir, uv]}(0)$ we note that 
the conformal transformations that preserve the insertion point  are generated by 
\be
l_{n} = L_{n} - L_{0} \, .
\ee
For a primary $\hat \psi^{\rm [ir, uv]}$ inserted at  $z=1$ on the upper half plane the commutation 
relations with the generators $l_{n}$ are 
\be \label{Vir_gluing}
l_{n}^{\rm ir} \hat \psi^{\rm [ir, uv]}(1) -  \hat \psi^{\rm [ir, uv]}(1) l_{n}^{\rm uv} = n \hat \Delta   \hat \psi^{\rm [ir, uv]}(1)  \, .
\ee
This equation can be interpreted as a gluing condition for the UV and IR generators preserving the position of the interface. 
The term on the right hand side of (\ref{Vir_gluing}) means that the generators are glued up to a central element\footnote{This neat  interpretation holds only when the RG operator is a primary which we believe is always the case.}. 
Using (\ref{Vir_gluing}) we find 
\be
\Gamma^{1}_{\rm vac} =\frac{\Delta^{\rm uv}_{0}- \Delta^{\rm ir}_{0} + \hat \Delta}{2\Delta^{\rm uv}_{0}} \, , 
\ee
\be
\Gamma^{2}_{\rm vac} = -3\frac{(\hat \Delta - \Delta^{\rm ir}_{0} + \Delta^{\rm uv}_{0})(\hat \Delta - \Delta^{\rm ir}_{0}+ \frac{1}{3}(1-\Delta^{\rm uv}_{0})) - \frac{2}{3}\hat \Delta (2\Delta^{\rm uv}_{0} +1)}{c(1+2\Delta^{\rm uv}_{0})+2\Delta^{\rm uv}_{0}(17\Delta^{\rm uv}_{0}-5)}
\ee
where $c$ is the central charge of the bulk CFT which is equal to $7/10$ for TIM. 

We can also consider ratios of components in the excited states. If $|0\rangle_{\rm ir}$ is a primary with a positive weight then 
the first excited state in the IR theory is $|1\rangle_{\rm ir}=L_{-1}|0\rangle_{\rm ir}$ and its image in the UV theory is 
\be 
 \hat \psi^{\rm [uv, ir]}(0) |1\rangle_{\rm ir} = C_{0}^{(1)} |0\rangle_{\rm uv} + C_{1}^{(1)}L_{-1}|0\rangle_{\rm uv} +  \dots 
\ee
We define a ratio
\be
\Gamma^{1}_{\rm 1} = \frac{C_{1}^{(1)}}{C_{0}^{(1)}} 
\ee
for which the theoretical value is 
\be
\Gamma^{1}_{\rm 1}= \Gamma^{1}_{\rm vac} + \frac{1}{2\Delta^{\rm uv}_{0}}\left( \frac{2\Delta^{\rm ir}_{0}}{\Delta^{\rm ir}_{0} + \hat \Delta - \Delta^{\rm uv}_{0}}-1\right) \, .
\ee
In the case when $|0\rangle_{\rm ir}$ has zero weight, which happens e.g. for a flow into the $(-)$  boundary condition, 
the first excited state is $|1\rangle_{\rm ir}=L_{-2}|0\rangle_{\rm ir}$ and the predicted value of the ratio is 
\be
\Gamma^{1}_{\rm 1}= \frac{\hat \Delta + \Delta^{\rm uv}_{0}  -2}{2\Delta^{\rm uv}_{0}} \, . 
\ee

\begin{center}
\begin{figure}[h!]

\begin{minipage}[b]{0.5\linewidth}
\centering
\includegraphics[scale=0.8]{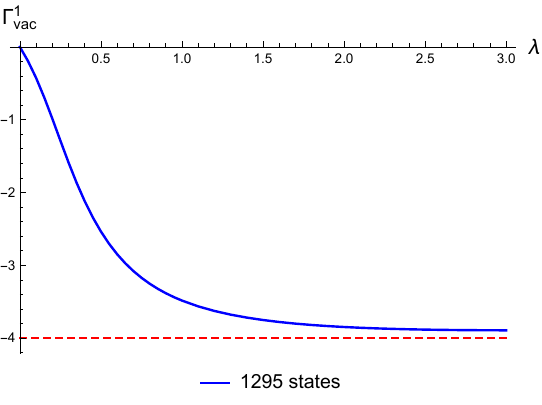}
\end{minipage}%
\begin{minipage}[b]{0.5\linewidth}
\centering
\includegraphics[scale=0.8]{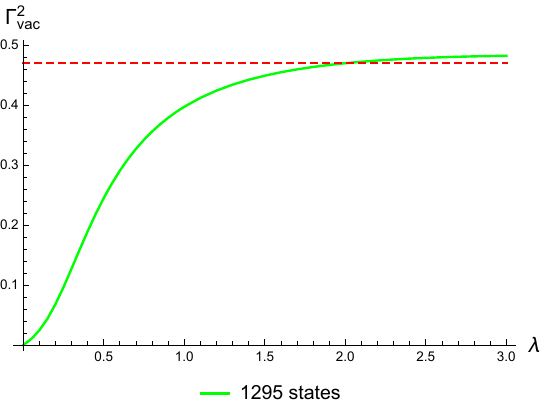}
\end{minipage}
\caption{The component ratios in the vacuum state for the $(d)\rightarrow (0)$ flow. The red dashed lines mark 
the predicted values: $\Gamma_{\rm vac}^{1}=-4$, $\Gamma_{\rm vac}^{2}=\frac{640}{1361}\approx 0.47$. The dimension of the truncated state space is indicated in the legend.}
\label{Ratios_pic1}
\end{figure}
\end{center}

\begin{center}
\begin{figure}[h!]
\begin{minipage}[b]{0.5\linewidth}
\centering
\includegraphics[scale=0.8]{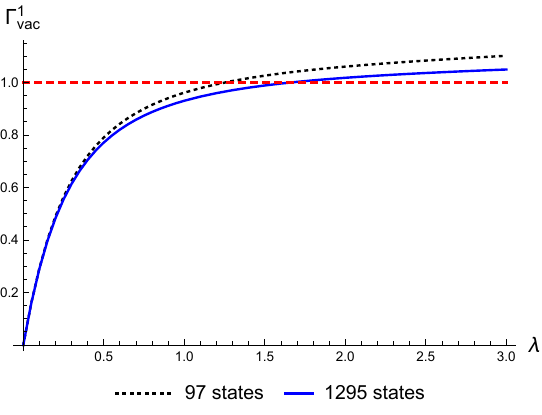}
\end{minipage}%
\begin{minipage}[b]{0.5\linewidth}
\centering
\includegraphics[scale=0.8]{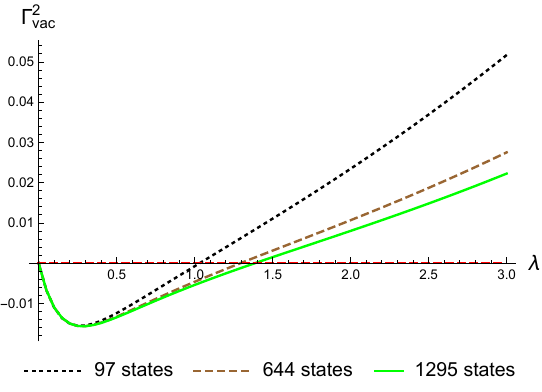}
\end{minipage}
\caption{The component ratios in the vacuum state for the $(d)\rightarrow (+)\oplus (-)$ flow. The red dashed lines mark 
the predicted values: $\Gamma_{\rm vac}^{1}=1$, $\Gamma_{\rm vac}^{2}=0$. }
\label{Ratios_pic2}
\end{figure}
\end{center}

\begin{center}
\begin{figure}[h!]

\begin{minipage}[b]{0.5\linewidth}
\centering
\includegraphics[scale=0.8]{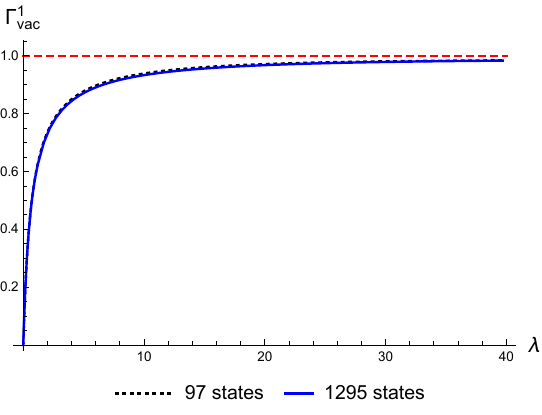}
\end{minipage}%
\begin{minipage}[b]{0.5\linewidth}
\centering
\includegraphics[scale=0.8]{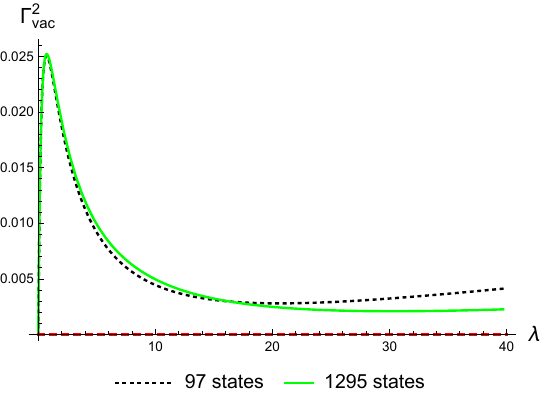}
\end{minipage}
\caption{The component ratios in the vacuum state for the $(d)\rightarrow  (-)$ flow. The red dashed lines mark 
the predicted values: $\Gamma_{\rm vac}^{1}=1$, $\Gamma_{\rm vac}^{2}=0$. }
\label{Ratios_pic3}
\end{figure}
\end{center}

In TCSA the infinite dimensional state space of a BCFT is truncated to a finite dimensional subspace of states with the conformal 
weight being less than a cutoff value. Empirically one finds that the lowest energy eigenstates are the best approximated in this scheme. Moreover the lowest weight components of the numerical eigenvectors are the most reliable ones. Hence the above ratios are the
  observables we hope are best approximated numerically. On the other hand the theoretical values of the  ratios change significantly when we change the tentative RG operators. Hence the numerical values can be used to confirm the RG operators of particular flows. 
  A similar strategy was used in \cite{AK_Ising} to identify the conformal boundary states giving the RG interfaces for massive bulk flows. 
  For the boundary flows the RG operator should be independent of the choice of the spectator that can be also tested numerically. 
  The plots of the vacuum ratios versus the coupling for 3 different boundary flows in TIM are presented on Figures \ref{Ratios_pic1},  \ref{Ratios_pic2}, 
   \ref{Ratios_pic3}.

As in the bulk case, where similar ratios were discussed in \cite{AK_Ising}, we observe a rather slow power-like convergence in the coupling. At  very large coupling one expects the truncation errors to be large, however, empirically TCSA tends to have bounded errors even for very large couplings. On the plots we can see that for some flows the ratios get closer to the predicted values in the asymptotic region for example for the $(d)\rightarrow (0)$ flow, while for the $\Gamma_{\rm vac}^{2}$ ratio in the $(d)\rightarrow (+)\oplus(-)$ flow
the best fit happens near a local minimum after which the ratio diverges linearly from the theoretical value. This fits well with the observation that there is a "flow beyond" present in the second case but not in the first one. The TCSA numerics in the second flow behaves as if the theory flows back to the UV fixed point. This phenomenon for boundary flows was first observed in \cite{Gerard2}  and discussed more recently in \cite{withDermot}. For the $\psi_{1,2}$ flows the asymptotic behaviour is monotonic with the approch to theoretical values being even slower than for the $\psi_{1,3}$ flows. 
The dimensions of the truncated state spaces are indicated in the legend for each plot. For all quantities we see the movement towards the theoretical values for increased size of the truncated state space. For the $\psi_{1,2}$ flows this improvement is only visible for very large values of the coupling.

\begin{center}
\begin{figure}[h!]
\begin{minipage}[b]{0.5\linewidth}
\centering
\includegraphics[scale=0.8]{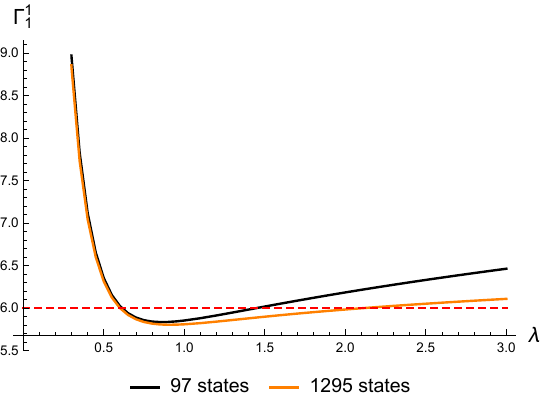}
\end{minipage}%
\begin{minipage}[b]{0.5\linewidth}
\centering
\includegraphics[scale=0.8]{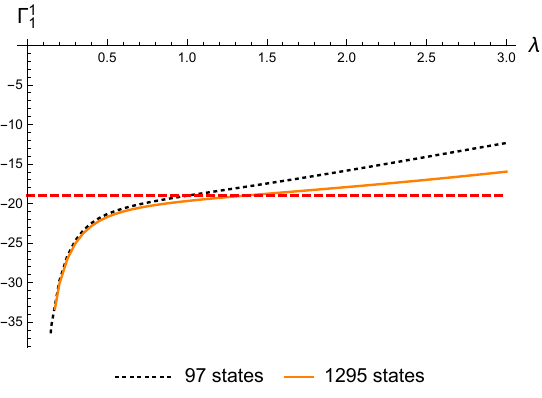}
\end{minipage}
\caption{The component ratio $\Gamma_{1}^{1}$ of the first excited state for the $(d)\rightarrow  (0)$ (left) and $(d)\rightarrow~(+)\oplus(-)$ (right) flows. The red dashed lines mark 
the predicted values: $\Gamma_{1}^{1}=6$ and  $\Gamma_{1}^{1}=-19$. For the second flow the first excited IR state is the $\epsilon''$ primary. }
\label{Ratios_pic4}
\end{figure}
\end{center}

\begin{center}
\begin{figure}[h!]
\centering
\includegraphics[scale=0.8]{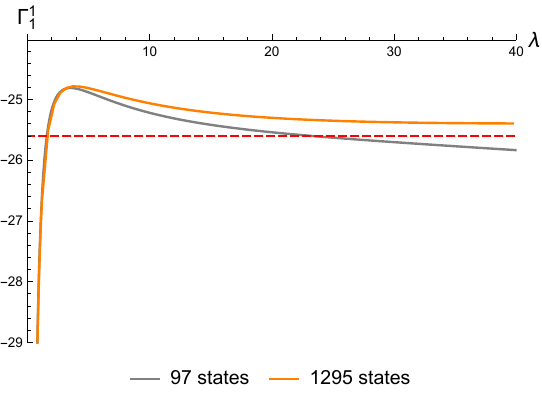}
\caption{The component ratio $\Gamma_{1}^{1}$ of the first excited state for the $(d)\rightarrow  (-)$  flow. The red dashed line marks 
the predicted value   $\Gamma_{1}^{1}=-\frac{77}{3}\approx -25.6$. }
\label{Ratios_pic5}
\end{figure}
\end{center}

On Figures \ref{Ratios_pic4}, \ref{Ratios_pic5} we present plots of the ratio of the first two components in the first excited energy eigenvector. 
They have similar features to the plots of the vacuum ratios.

\section{Concluding remarks} \label{sec:conclusions}
\setcounter{equation}{0}

Our main non-perturbative tool for using RG operators has been the variational method we developed in section 
\ref{sec:VAR}. The method assumes a particular ansatz for the perturbed theory vacuum state based on the RG 
operator. We think that to make further progress with RG operators the method needs to be developed further.  
The main technical advantage of the method is that its key ingredient is a local operator and we could use OPE to derive variational 
energies. It seems to be important to understand to what extent such local terms capture the true vacuum and how one can 
systematically improve such an ansatz. Such general understanding should clarify the puzzle with the superpositions we 
pointed at in  section \ref{sec:offdiag}. The same remarks go for the variational method of \cite{Cardy_var} where a similar puzzle was also  noted for the case of  bulk massive flows in the Ising field theory. Provided that we understand better the variational method and know how to improve it, it would be interesting to use it to gain insight into the special critical lines like the ones 
designated by the black arrows on the diagram on Figure \ref{space_of_flows}. Such critical lines appear to be important features of the  spaces of RG flows. We hope to return to these questions in future work. 

Another set-up in  which it would be interesting to consider the RG operators is that of  spliced flows. Suppose we have pairwise flows between 3 fixed points as on Figure \ref{spliced_flow}. We assume further that there is a continuous family of flows between 
${\rm BCFT_{1}}$ 
and ${\rm BCFT_{3}}$ such that in a limit one obtains a spliced flow that is a concatenation of flow from ${\rm BCFT_{1}}$ to 
${\rm BCFT_{2}}$ with the flow from ${\rm BCFT_{2}}$ to ${\rm BCFT_{3}}$. Such a spliced flow is not a real RG flow but it does belong to the boundary of the space of true flows (see e.g. \cite{Cohen_Morse} for a nice mathematical exposition of such matters). 
Let  $\hat \psi^{[2,1]}$ and $\hat \psi^{[3,2]}$ be the RG operator for each component flow in the spliced flow. Assume further that 
all flows in the family of flows between  ${\rm BCFT_{1}}$ 
and ${\rm BCFT_{3}}$ have the same RG operator: $\hat \psi^{[3,1]}$. (For the theories along these flows near the IR fixed point the interface operators can differ in the subleading components as in our discussion of superpositions in section \ref{sec:offdiag}.)

\begin{center}
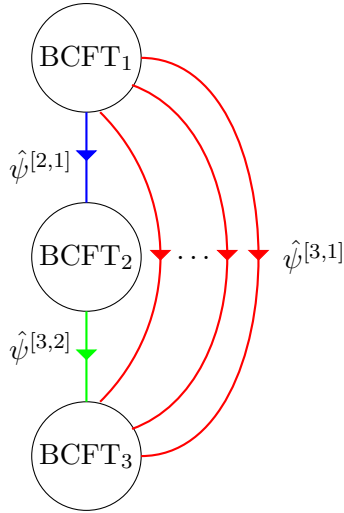
\begin{figure}[H]
\centering
\begin{tikzpicture}[scale=1.2]
\draw (0,2.2) circle [radius=0.6] node {${\rm BCFT_{1}}$};
\draw (0,0) circle [radius=0.6] node {${\rm BCFT_{2}}$};
\draw (0,-2.2) circle [radius=0.6] node {${\rm BCFT_{3}}$};
\begin{scope}[thick, every node/.style={sloped,allow upside down}]
\draw[blue] (0,1.6) -- node {\midarrow} (0,0.6);
\draw[green] (0,-0.6) -- node {\midarrow} (0,-1.6); 
\draw[red] (0.6,2.2) to[out=0,in=0] node {\midarrow} (0.6,-2.2);
\draw[red] (0.5,1.9) to[out=-20,in=20] node {\midarrow} (0.5,-1.9);
\draw[red] (0.15,1.6) to[out=-45,in=45] node {\midarrow} (0.15,-1.6);
\end{scope}
\draw (1.2,0) node {$\dots$};
\draw (-0.5,1) node {$\hat \psi^{[2,1]}$};
\draw (-0.5,-1) node {$\hat \psi^{[3,2]}$};
\draw (2.5,0) node {$\hat \psi^{[3,1]}$};
\end{tikzpicture}
\caption{A space of RG flows between 3 fixed points. The associated RG operators are put next to the arrows. 
 Here we assume that there is a family of flows between ${\rm BCFT_{1}}$ and ${\rm BCFT_{3}}$ for which the spliced flow 
 that passes through ${\rm BCFT_{2}}$ appears as a singular limit.  }
\label{spliced_flow}
\end{figure}
\end{center}

It is tempting to conjecture then that given such a setup the OPE of the first two RG operators must contain the third:
\be \label{splice_rule}
\hat \psi^{[3,2]}(0) \hat \psi^{[2,1]}(0) \sim  C_{32,21}^{31}  \tau^{\hat \Delta_{31} - \hat \Delta_{32}-\hat \Delta_{21}}\hat \psi^{[3,1]}(0) + \dots \qquad C_{32,21}^{31}  \ne 0 \, .
\ee
The situation depicted on Figure \ref{spliced_flow}  is realised in the space of boundary flows in TIM with the role of  ${\rm BCFT_{1}}$ 
played by $(d)$, the role  of ${\rm BCFT_{2}}$ by $(0+)$ and that of ${\rm BCFT_{3}}$ by $(0)$. As argued in the end of 
section \ref{sec:offdiag} the RG operator for the flow from $(d)$ to $(0+)$ (which we were able to locate numerically using TCSA) 
must be $\hat \sigma^{[0+,d]}$. For the other two flows we found the RG operators in section \ref{sec:psi13}, they are 
$\hat \epsilon^{[0,d]}$ and $\hat \sigma^{[0,0+]}$ and the rule (\ref{splice_rule}) indeed holds. It would be interesting to investigate this conjecture further. 

Although we did discuss the bulk RG interfaces in section \ref{sec_op_map} the focus of the paper is on the boundary RG interfaces. 
Our conjectures 2 and 3 can be generalised to the bulk case as follows. Instead of fusing the RG interface line with itself, that would be the direct analogue of the OPEs in conjectures 2 and 3, we can place the perturbing operator close to the RG interface and perform the 
bulk to boundary  OPE. It is natural then to require, especially using the intuition from the eigenvector equation considered in section 
\ref{sec:eigenvector}, that this OPE contains the identity operator. Equivalently we can say that the UV operator must have a non-vanishing one-point function in the presence of the RG interface. A similar condition holds for the massive flows as follows from Cardy's variational method in which the interaction term comes with the one point function of the perturbing operator in the conformal boundary condition giving the vacuum state. We can also formulate the same condition for the leading IR operator bringing it close to the interface on the IR side. 
The principle obstacle in making these conjectures useful in the bulk case is our poor knowledge of 
conformal interfaces.   It is possible though that in the context of topological or supersymmetric QFTs (in two dimensions or higher) these conjectures  can lead to some interesting insights. The RG interfaces for such theories have been studied in \cite{Gaiotto_N2}, 
\cite{Roggenkamp1}, \cite{Roggenkamp2}. 

\begin{center}
{\bf \Large Acknowledgements}
\end{center}
  The TCSA numerical results presented in this paper have been obtained using Wolfram's Mathematica package (version 10.2.0.0).
  I thank Ingo Runkel for sharing his Mathematica code for calculating fusion matrices. I also thank Matthew Buican, Marco Meineri, Cornelius Schmidt-Colinet and Gerard Watts for stimulating discussions.

\end{document}